\documentclass[aps,prd,reprint,superscriptaddress,nofootinbib,longbibliography]{revtex4-2}
\usepackage[T1]{fontenc}
\usepackage{amsmath,amssymb,bm,graphicx,booktabs,xcolor,tikz}
\usepackage{listings}
\usepackage{needspace}
\usetikzlibrary{arrows.meta,positioning,calc}
\usepackage[colorlinks=true,linkcolor=blue!55!black,citecolor=blue!55!black,urlcolor=blue!55!black]{hyperref}
\definecolor{rose}{HTML}{F6ECEE}
\definecolor{cream}{HTML}{F8EED2}
\definecolor{mint}{HTML}{E7F0EA}
\definecolor{ice}{HTML}{E7EDF3}
\newcommand{\dd}{\mathrm{d}}
\newcommand{\as}{a_s}

\begin{document}
\title{Infrared Subtraction with Artificial Intelligence}
\author{Wenjie He}
\email{202531101011@mail.bnu.edu.cn}
\affiliation{School of Physics and Astronomy, Beijing Normal University, and Key Laboratory of Multiscale Spin Physics (Beijing Normal University), Ministry of Education, Beijing 100875, China}
\author{Xiaohui Liu}
\email{xiliu@bnu.edu.cn}
\affiliation{School of Physics and Astronomy, Beijing Normal University, and Key Laboratory of Multiscale Spin Physics (Beijing Normal University), Ministry of Education, Beijing 100875, China}
\affiliation{Southern Center for Nuclear Science Theory (SCNT), Institute of Modern Physics, Chinese Academy of Science, Huizhou 516000, China}
\author{Yandong Liu}
\email{ydliu@bnu.edu.cn}
\affiliation{School of Physics and Astronomy, Beijing Normal University, and Key Laboratory of Multiscale Spin Physics (Beijing Normal University), Ministry of Education, Beijing 100875, China}
\author{Zhan Wang}
\email{zhanwang@mail.bnu.edu.cn}
\affiliation{School of Physics and Astronomy, Beijing Normal University, and Key Laboratory of Multiscale Spin Physics (Beijing Normal University), Ministry of Education, Beijing 100875, China}
\date{September 27, 2026}
\begin{abstract}
We present AI-developed local infrared subtraction, building on projection to Born and effective-field-theory (EFT) matching. The framework separates an integrable radiation term from a finite contribution at Born kinematics, referred to as the Born contact. The contact is determined using the EFT singular distribution in a resolution observable such as $N$-jettiness $\tau_N$. Under human physics guidance, a large language model (LLM) develops two implementations. One uses a neural network for phase space projection and fits the contact by matching to EFT cumulants. The other uses an analytic construction that keeps the Born momenta fixed while integrating over radiation. It combines the EFT $\delta(\tau_N)$ coefficient with finite four-dimensional radiation integrals to calculate the contact term directly. This gives a local subtraction formula without a slicing parameter, while reusing existing lower-order radiation calculations and EFT singular predictions. As a demonstration, we reconstruct the full next-to-leading order (NLO) correction for massless three- and four-jet production in electron--positron annihilation. The attempt to the next-to-next-to-leading order (NNLO) di-jet production is also made by recursively using the NLO P2B construction with machine-learning controls to reduce the variance of the contact integral. The tested differential predictions are in good agreement with \texttt{EERAD3}. The numerical calculation and projection-network training use the CPU of a 2020 Apple M1 MacBook, without GPU acceleration, illustrating the feasibility of the construction with modest local computing resources. The appendices develop an extension of the local subtraction to three-jet NNLO, giving explicit radiation maps and a proposed contact formula. We also show how to integrate over NNLO radiation while keeping the Born momenta fixed, for any number of massless final-state jets. Our results show how AI could help higher-order calculations by constructing infrared subtraction and improving its numerical integration.
\end{abstract}
\maketitle

\section{Introduction}
Real-radiation corrections are essential to precision QCD predictions. Additional partons generate resolved jets, redistribute recoil, and change event shapes and the acceptance of experimental cuts. Their contribution must therefore be included with its measurement dependence when extracting the strong coupling, constraining parton distributions, or comparing collider data with the Standard Model. The same radiation becomes singular when a parton is soft or two partons are collinear. Infrared safety ensures cancellation with virtual corrections in the physical cross section, but does not make the separate phase-space integrals finite~\cite{Kinoshita:1962ur,Lee:1964is,Ellis:1996mzs}. Turning this cancellation into a differential numerical prediction remains a central problem in higher-order calculations.

The treatment of unresolved radiation has developed alongside perturbative jet physics, from infrared-safe jet cross sections in the 1970s~\cite{Sterman:1977wj} and calculations of jet structure in electron--positron annihilation in the early 1980s~\cite{Ellis:1980wv} to general prescriptions based on universal soft and collinear factorization. The Frixione--Kunszt--Signer and Catani--Seymour methods established such prescriptions at next-to-leading order (NLO) in the 1990s~\cite{Frixione:1995ms,Catani:1996vz}, opening the way to automated real-radiation subtraction~\cite{Frederix:2009yq}. At next-to-next-to-leading order (NNLO), simultaneous unresolved emissions introduce overlapping singularities. Sector decomposition provided a way to disentangle these singularities in real-radiation integrals~\cite{Anastasiou:2003gr}; antenna, sector-improved residue, nested soft-collinear, and local analytic sector subtraction developed complementary organizations of their cancellation~\cite{Gehrmann-DeRidder:2005btv,Czakon:2010td,Caola:2017dug,Magnea:2018hab}.

Phase-space slicing developed in parallel, with invariant-mass cutoffs and separate soft and collinear cutoffs used to isolate unresolved radiation~\cite{Giele:1993dj,Harris:2001sx}. At NNLO, the $q_T$ method exploited the singular small-recoil distribution for color-singlet production~\cite{Catani:2007vq}. Fully differential top-quark decay was subsequently calculated at NNLO using the invariant mass of the hadronic final state as a slicing variable~\cite{Gao:2012ja}. A factorization theorem in soft-collinear effective theory (SCET)~\cite{Bauer:2000ew,Bauer:2000yr,Bauer:2001ct,Bauer:2001yt,Beneke:2002ph} determines the contribution below the mass cut, while the region above it is obtained from $t\to Wb+\mathrm{jet}$ at NLO. This demonstrated how integrated singular information could replace the construction of counterterms for double-unresolved radiation. The definition and factorization of $N$-jettiness~\cite{Stewart:2010tn,Jouttenus:2011wh} provided a resolution variable for processes with several hard directions; its use in NNLO calculations extended this strategy to jet production~\cite{Boughezal:2015dva,Gaunt:2015pea}. Recently, recoil-free winner-take-all jet axes have also extended $q_T$-based slicing to jet final states~\cite{Fu:2024fgj}.

These developments have made NLO calculations broadly automated~\cite{Alwall:2014hca,Sherpa:2019gpd} and NNLO predictions available for a growing range of collider observables. The public \texttt{NNLOJET} framework covers jet production in lepton and hadron collisions~\cite{NNLOJET:2025rno}, while recent antenna and local-sector developments pursue greater generality and automation~\cite{Marcoli:2026zub,Bertolotti:2025clg}.

At N$^3$LO, several complementary constructions have produced differential predictions. Projection to Born has been applied to jet production in deep-inelastic scattering~\cite{Currie:2018fgr,Gehrmann:2018odt}, Higgs decay to bottom quarks in combination with $N$-jettiness slicing~\cite{Mondini:2019gid}, and gluon-fusion Higgs production~\cite{Chen:2021isd}. Independently, $q_T$ subtraction combined with NNLO antenna subtraction yielded the Higgs rapidity distribution, with an approximation for then-unknown third-order collinear terms~\cite{Cieri:2018oms}, followed by neutral- and charged-current Drell--Yan distributions using the complete third-order small-$q_T$ ingredients~\cite{Chen:2021vtu,Chen:2022lwc}. In these calculations, the small-$q_T$ factorization formula supplies the unresolved N$^3$LO contribution, while antenna subtraction treats color-singlet-plus-jet production at NNLO in the resolved region. Predictions with fiducial cuts have also been obtained through transverse-momentum subtraction and resummation in several implementations~\cite{Billis:2021ecs,Camarda:2021ict,Chen:2022cgv,Neumann:2022lft}.

More recently, antenna subtraction has been extended to the N$^3$LO unresolved limits themselves and applied directly to $e^+e^-\to2$ jets, yielding the two-jet rate and leading-jet energy distribution with singlet contributions neglected~\cite{Chen:2025kez}. Fully differential Higgs-pair production has also reached N$^3$LO in the heavy-top limit using $q_T$ slicing~\cite{Chen:2026zmi}. Further progress includes the three-loop zero-jettiness soft function~\cite{Chen:2022yre,Baranowski:2024ene,Baranowski:2024vxg} and a transverse-momentum subtraction construction with a winner-take-all axis~\cite{Larkoski:2014uqa} for semi-inclusive deep-inelastic scattering at N$^3$LO~\cite{Dong:2026uvw}. Despite this progress, extending this reach to more external legs and higher orders still requires substantial analytic work and numerical integration. Moreover, achieving the broad automation established at NLO remains a challenge at higher orders, where the subtraction terms, momentum maps, and integrated contributions often require dedicated analytic construction even when the amplitudes are available.

To illustrate these difficulties, we compare subtraction and slicing at NLO for an $N$-parton Born process with no incoming colored partons. We denote an $m$-parton configuration by $\Phi_m$, its phase-space measure by $\dd\Phi_m$, and the value of an infrared-safe measurement on that configuration by $J_m$. We write $R$ for the real contribution and $V$ for the ultraviolet-renormalized virtual contribution. With the Born term omitted, local subtraction expresses the NLO correction as~\cite{Frixione:1995ms,Catani:1996vz}
\begin{equation}
\begin{aligned}
 \delta\sigma_{\mathrm{NLO}}[J]
 &=\int\dd\Phi_{N+1}\big[RJ_{N+1}-S\widetilde J_N\big]\\
 &\quad+\int\dd\Phi_N\left[V+\int\dd\Phi_{\mathrm{rad}}S\right]J_N.
\end{aligned}
\label{eq:introsubtraction}
\end{equation}
Here the subtraction map $\widehat m:\Phi_{N+1}\to\Phi_N$ takes a real-emission configuration to an $N$-parton Born configuration while preserving on-shell kinematics and total four-momentum. It approaches the corresponding Born limit when radiation becomes unresolved. We evaluate the projected measurement as $\widetilde J_N=J_N(\widehat m\Phi_{N+1})$. The counterterm $S$ reproduces $R$ in each single-unresolved limit, making the first line finite. We integrate $S$ over the unresolved radiation at fixed mapped Born momenta, including the phase-space Jacobian; sums over counterterms and maps are implicit. The resulting poles cancel those of $V$ in dimensional regularization, so the second line is also finite in four dimensions.

Slicing instead introduces a dimensionless resolution variable $\tau$ that vanishes on Born kinematics and resolves all relevant unresolved limits. With a cut $t_c>0$, the same correction takes the form
\begin{equation}
\begin{aligned}
 \delta\sigma_{\mathrm{NLO}}[J]
 &=\int\dd\Phi_N\,\Sigma_{N,\mathrm{sing}}^{(1)}(\Phi_N;t_c)J_N\\
 &\quad+\int_{\tau>t_c}\dd\Phi_{N+1}\,RJ_{N+1}
       +\Delta_{\mathrm{pow}}[J;t_c].
\end{aligned}
\label{eq:introslicing}
\end{equation}
The quantity $\Sigma_{N,\mathrm{sing}}^{(1)}$ is the NLO singular cumulant below the cut, differential in Born kinematics. It combines virtual and unresolved real contributions and can be obtained from a factorization theorem~\cite{Gao:2012ja,Gaunt:2015pea}. The second term is a finite real-emission integral above the cut. The remainder $\Delta_{\mathrm{pow}}$ contains the power-suppressed contributions, including the difference between the real and Born measurements below the cut, and vanishes as $t_c\to0$. Dropping it defines the finite-cut slicing approximation.

The two formulas show how the difficulties noted above arise in different parts of the calculation. Subtraction cancels the real singularity near each unresolved configuration and preserves the full measurement dependence, avoiding a neglected below-cut contribution. This can improve numerical convergence, but constructing the counterterms, their momentum maps, and their integrated forms becomes demanding as the multiplicity and perturbative order increase. Slicing based on factorization uses an integrated singular prediction calculated in effective field theory, reducing the need to construct local counterterms. The resolved contribution reuses a calculation with one additional jet at one lower perturbative order. However, its numerical cost grows when a small cut is needed to suppress power corrections. This is because lowering the cut exposes more singular regions of the resolved integral and produces large logarithmic contributions of opposite sign above and below the cut, whose cancellation requires increasingly precise numerical integration.

In this work, we use a large language model (LLM) to advance multijet infrared subtraction, building on the established connection between projection to Born and EFT singular predictions~\cite{Cacciari:2015jma,Ebert:2019zkb,Campbell:2024hjq}. Human guidance sets the framework and physical constraints, while the LLM devises momentum maps and contact formulas and implements the numerical calculations. In one implementation, it constructs the phase-space projection with a neural network and fits the Born contact through EFT matching. This tests whether the expressive power of neural networks can accommodate the kinematic and infrared constraints of subtraction. Furthermore, the LLM extends the same framework to \emph{local subtraction}. It devises analytic maps that keep the Born momenta fixed during radiation integration and derives a formula that calculates the contact directly from the EFT $\delta(\tau)$ coefficient and finite four-dimensional radiation integrals. The resulting construction evaluates both radiation and contact in four spacetime dimensions over the full phase space, without a small slicing parameter, while recycling existing lower-order radiation calculations and EFT singular predictions. We demonstrate both implementations for three-jet production at NLO and extend the local subtraction to four jets. At NNLO, we calculate the two-jet contact recursively from our NLO P2B construction. The LLM also designs a machine-learning strategy that reduces the variance of the contact integral. The appendices also develop the three-jet NNLO extension, with explicit radiation maps at fixed Born momenta and a proposed local contact formula.

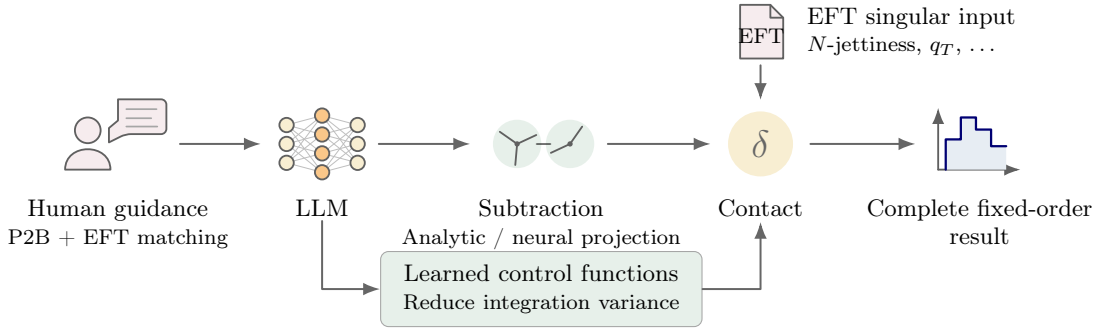
\begin{figure*}[t]
\centering
\begin{tikzpicture}[font=\small,>=Latex,
 line cap=round,line join=round,
 arr/.style={->,draw=gray!80!black,line width=.75pt},
 icon/.style={draw=gray!75!black,line width=.7pt},
 label/.style={align=center,anchor=north},
 person/.pic={
   \draw[icon,fill=rose] (0,.18) circle (.17);
   \draw[icon,fill=rose] (-.3,-.34) .. controls (-.3,.02) and (.3,.02) .. (.3,-.34) -- cycle;
 }]
\pic at (-.2,0) {person};
\draw[icon,fill=rose] (.08,.21) -- (.2,.31) -- (.2,.59)
  [rounded corners=2pt] -- (.84,.59) -- (.84,.13) -- (.3,.13)
  [sharp corners] -- (.08,.02) -- cycle;
\foreach \y/\end in {.46/.71,.34/.64,.23/.72}
  \draw[gray!70,line width=.65pt] (.31,\y)--(\end,\y);
\node[label] at (.2,-.62) {Human guidance\\{\footnotesize P2B + EFT matching}};
\foreach \i in {1,2,3}{\foreach \j in {1,2,3,4}{
  \draw[gray!50,line width=.35pt] (2.43,{.5-.25*\i})--(2.9,{.625-.25*\j});
  \draw[gray!50,line width=.35pt] (2.9,{.625-.25*\j})--(3.37,{.5-.25*\i});
}}
\foreach \x in {2.43,3.37}{\foreach \i in {1,2,3}{
  \draw[icon,fill=cream] (\x,{.5-.25*\i}) circle (.095);
}}
\foreach \j in {1,2,3,4}{\draw[icon,fill=cream!65!orange] (2.9,{.625-.25*\j}) circle (.095);}
\node[label] at (2.9,-.62) {LLM};
\fill[mint] (5.48,0) circle (.32);
\fill[mint] (6.18,0) circle (.32);
\foreach \ang in {25,145,265}{\draw[icon] (5.48,0)--++(\ang:.28);}
\foreach \ang in {55,205}{\draw[icon] (6.18,0)--++(\ang:.28);}
\fill[gray!75!black] (5.48,0) circle (.035);
\fill[gray!75!black] (6.18,0) circle (.035);
\node at (5.83,0) {$-$};
\node[label] at (5.8,-.62) {Subtraction\\{\footnotesize Analytic / neural projection}};
\fill[cream] (8.7,0) circle (.43);
\node[font=\Large,text=gray!80!black] at (8.7,0) {$\delta$};
\node[label] at (8.7,-.62) {Contact};
\draw[icon,fill=rose] (8.43,1.13)--(8.43,1.84)--(8.83,1.84)
  --(9,1.67)--(9,1.13)--cycle;
\draw[icon] (8.83,1.84)--(8.83,1.67)--(9,1.67);
\node[font=\footnotesize] at (8.71,1.43) {EFT};
\node[align=left,anchor=west] at (9.2,1.49)
  {EFT singular input\\{\footnotesize $N$-jettiness, $q_T$, $\ldots$}};
\draw[arr] (8.7,1.04)--(8.7,.58);
\draw[icon,->] (11.05,-.34)--(12.14,-.34);
\draw[icon,->] (11.05,-.34)--(11.05,.48);
\fill[ice] (11.15,-.34)--(11.15,.06)--(11.35,.06)--(11.35,.35)
  --(11.55,.35)--(11.55,.19)--(11.75,.19)--(11.75,-.03)
  --(11.95,-.03)--(11.95,-.34)--cycle;
\draw[blue!45!black,line width=.85pt] (11.15,-.34)--(11.15,.06)
  --(11.35,.06)--(11.35,.35)--(11.55,.35)--(11.55,.19)
  --(11.75,.19)--(11.75,-.03)--(11.95,-.03);
\node[label] at (11.6,-.62) {Complete fixed-order\\result};
\draw[arr] (1.02,0)--(2.09,0);
\draw[arr] (3.65,0)--(4.86,0);
\draw[arr] (6.68,0)--(8.04,0);
\draw[arr] (9.36,0)--(10.75,0);
\node[draw=gray!70,rounded corners=3pt,fill=mint,align=center,
  text width=39mm,inner sep=5pt] (control) at (5.8,-1.92)
  {Learned control functions\\{\footnotesize Reduce integration variance}};
\draw[arr] (2.9,-1.02)--(2.9,-1.92)--(control.west);
\draw[arr] (control.east)--(8.7,-1.92)--(8.7,-1.03);
\end{tikzpicture}
\caption{Roles of human physics guidance, the LLM, machine learning and EFT input in the subtraction construction. Humans choose the P2B + EFT matching framework and specify the physical constraints. The LLM develops neural and analytic projections, contact formulas and their numerical implementation. EFT supplies the Born-dependent singular distribution. The contact is fitted through matching or calculated from finite radiation integrals and the EFT $\delta(\tau_N)$ coefficient. In the LLM-developed local subtraction, learned control functions reduce the integration variance, with their known integrals added back to preserve the result. The radiation term and Born contact together give the fixed-order prediction.}
\label{fig:ai-role}
\end{figure*}

\section{P2B + EFT matching for multijet production}
Projection to Born (P2B) separates an integrable radiation contribution from a finite Born contribution by evaluating the higher-multiplicity input on the difference between real and projected measurements~\cite{Cacciari:2015jma}. Its connection to differential $N$-jettiness subtraction provides an EFT determination of the Born contribution~\cite{Gaunt:2015pea,Ebert:2019zkb}. We refer to this organization as \emph{P2B + EFT matching}.

The P2B-improved $\tau_0$ construction in Ref.~\cite{Campbell:2024hjq} treats color-singlet production. It evaluates the difference between real and Born-projected measurements locally. The remaining contribution, differential in the projected Born variables, is the Born contact in our notation. This contact is still obtained by slicing, so a small cutoff is needed to suppress the residual power corrections. A ``slice-to-subtraction'' construction in NLO shower matching was explicitly developed for processes with two colored Born
 legs~\cite{vanBeekveld:2025lpz}. It combines a slicing cumulant below a cut with an integrated shower counterterm above it, analytically taking the cut to zero to obtain a finite Born-local NLO weight.

Building on these developments, we use AI to construct multijet infrared subtraction in two ways. First, we follow the established P2B + EFT matching strategy, using the expressive power of neural networks to construct the phase-space projection and determining the Born contact by direct fitting through EFT matching. This tests the ability of neural networks to handle the kinematic and infrared constraints of subtraction. Furthermore, we extend the framework to local subtraction. Our analytic construction provides explicit projections and inverse radiation maps for any number of massless final-state jets, in contrast to the two-colored-leg construction given in Ref.~\cite{vanBeekveld:2025lpz}. At fixed Born momenta, the contact is calculated from the EFT $\delta(\tau_N)$ coefficient and finite four-dimensional radiation integrals, using finite rescalings of the full real contribution. Both radiation and contact are then evaluated over the full phase space without a small slicing parameter. We demonstrate both implementations for three jets at NLO, and local subtraction for four jets at NLO and two jets at NNLO. Appendices~\ref{app:nnlo-fixed-born}--\ref{app:local-contact} develop the extension to three-jet NNLO.

Both constructions share the following P2B + EFT matching framework. For a process whose Born term starts at $\alpha_s^p$, let $\Sigma_N^{(k)}$ denote the coefficient at order $\alpha_s^{p+k}$, with $k\ge1$, and $\mathcal I_{N+1}^{(k-1)}$ the coefficient at the same order in the calculation with one additional jet. Let $\widehat{M}$ denote the P2B projection. It leaves a Born configuration unchanged, $\widehat{M}\Phi_N=\Phi_N$. We define $J_{\widehat{M}}(\Phi)=J_N(\widehat{M}\Phi)$ and $\Delta J=J-J_{\widehat{M}}$. Input multiplicities and branch or sector labels are displayed when needed.
At any such order, the construction has the form
\begin{equation}
\begin{aligned}
 \Sigma_N^{(k)}[J]
 &=\mathcal I_{N+1}^{(k-1)}[\Delta J]\\
 &\quad+\int\dd\Phi_N\,c_N^{(k)}(\Phi_N)J_N(\Phi_N).
\end{aligned}
\label{eq:p2bframework}
\end{equation}
This organization recycles a calculation with one additional jet at one lower perturbative order and an EFT singular prediction at the target order, preserving an advantage of slicing. The radiation calculation uses its existing internal subtraction, and $\widehat{M}$ must preserve its unresolved limits so that $\Delta J$ admits local cancellation and an integrable radiation term. The finite density $c_N^{(k)}$ depends on the projection prescription and contains the remaining Born contribution in the same normalization. 

To determine this contact, choose a dimensionless resolution observable $\tau_N$ that vanishes on Born configurations, and weight functions $w_a(\Phi_N)$ on Born phase space. For $t>0$, define $W_{a,t}(\Phi)=w_a(\widehat{M}\Phi)\Theta[t-\tau_N(\Phi)]$ and its projected measurement $W_{a,t,\widehat{M}}(\Phi)=w_a(\widehat{M}\Phi)$. The matching relation is
\begin{equation}
\begin{aligned}
 \int\dd\Phi_N\,c_N^{(k)}w_a
 =\lim_{t\to0}\big\{&\Sigma_{N,\mathrm{sing}}^{(k)}[W_{a,t}]\\
 &-\mathcal I_{N+1}^{(k-1)}[W_{a,t}-W_{a,t,\widehat{M}}]\big\}.
\end{aligned}
\label{eq:frameworkcontact}
\end{equation}
Here $\Sigma_{N,\mathrm{sing}}^{(k)}$ is the EFT singular cumulant at the same order, including its Born dependence and the finite coefficient of $\delta(\tau_N)$ in the singular distribution. The subtracted term is the radiation contribution in Eq.~\eqref{eq:p2bframework}, evaluated with $J=W_{a,t}$. Equation~\eqref{eq:frameworkcontact} thus determines a weighted moment of the contact for each choice of $w_a$. Weights localized in different Born regions, or chosen from a function basis, probe its Born dependence. A neural-network or other parametrization can therefore be fitted by requiring its weighted integrals to reproduce these moments, as illustrated by the three-jet NLO fit in Sec.~\ref{sec:numerical}.

For the analytic implementation, we parametrize the higher-multiplicity phase space by Born coordinates and radiation variables $\xi$. The map $G(\Phi_N,\xi)$ generates radiation that projects back to the prescribed Born configuration, $\widehat{M}G(\Phi_N,\xi)=\Phi_N$. Holding the Born momenta fixed, we integrate only over $\xi$, so the matching relation determines the contact at each Born point. Appendix~\ref{app:fixed-born} gives the analytic NLO projection and its inverse for any number of massless final-state jets.

Appendix~\ref{app:local-contact} derives a direct contact calculation that avoids this small-cut limit. At NLO, we add the finite part of the radiation integral at fixed Born momenta to the EFT $\delta(\tau_N)$ coefficient. Denote this coefficient by $H_1(b)$ at a Born point $b$, and let $\mathcal Z_1(\zeta;b)$ be the tree-level radiation integral weighted by $\tau_N^\zeta$, with the same projection prescription. The contact is then
\begin{equation}
 c_N^{(1)}(b)=H_1(b)+\mathcal A_1(b),\qquad
 \mathcal A_1(b)=[\zeta^0]\mathcal Z_1(\zeta;b).
 \label{eq:framework-finite-contact}
\end{equation}
Here $[\zeta^0]$ selects the constant term in the Laurent expansion of $\mathcal Z_1$ around $\zeta=0$; this constant is $\mathcal A_1$. To calculate it, we combine radiation contributions evaluated at coordinates related by fixed rescalings, so that their common soft and collinear terms cancel before integration. We also include the finite contributions from the changes in integration range. The result is a sum of convergent integrals in four spacetime dimensions, retaining the nonsingular radiation over the full phase space. No small numerical value of $\zeta$ or slicing cutoff is required. Together with Eq.~\eqref{eq:p2bframework}, this gives a \emph{local subtraction formula at NLO} while reusing existing radiation and EFT results.

Once the projection and contact are fixed, Eq.~\eqref{eq:p2bframework} predicts different observables without further matching. The numerical examples compare the two implementations for three jets at NLO and demonstrate local subtraction for four jets. They also demonstrate recursive local contact extraction at NNLO in the simpler two-jet case, where the observables require only one contact coefficient. Appendices~\ref{app:nnlo-fixed-born}--\ref{app:local-contact} propose the extension of the local subtraction to three-jet NNLO, combining radiation integration at fixed Born momenta with a finite-difference representation of the contact.

\section{The role of artificial intelligence}
Machine learning supports event reconstruction, simulation, and analysis in high-energy physics~\cite{Albertsson:2018maf}, while LLM agents have begun to develop analyses and coordinate research and simulation workflows~\cite{Diefenbacher:2025zzn,Menzo:2025cim,Qiu:2026iby,Gao:2026cpd,Aether:Project}. In perturbative QCD, applications include machine learning for loop-integral differential equations~\cite{Calisto:2023vmm,Liu:2025tje}, AI-assisted integration-by-parts reduction~\cite{vonHippel:2025okr,Song:2025pwy}, factorization-aware matrix-element emulation~\cite{Maitre:2021uaa}, and transformer-based modeling of non-global logarithms~\cite{Li:2026yrp} and Balitsky--Kovchegov evolution~\cite{Gao:2025dkn}. LLM-driven bootstrapping has also reached the nine-loop six-gluon maximally helicity-violating amplitude in planar $\mathcal{N}=4$ super-Yang--Mills theory~\cite{Claude:2026NineLoops}. Work related to infrared subtraction includes GAN generation of subtracted event samples using prescribed collinear counterterms~\cite{Butter:2019eyo} and neural importance sampling within NNLO subtraction~\cite{Janssen:2025zke}.

These developments motivate using AI to devise the subtraction itself. The projection must satisfy kinematic constraints and reproduce the soft and collinear limits, but its form away from these limits is not uniquely fixed. Recent LLM-guided searches for mathematical constructions illustrate the ability to devise algorithms under explicit constraints~\cite{Ellenberg:2025umt,Ulam:2026Jacobian,OpenAI:2026NavierStokes}. Here the LLM devises momentum maps that preserve on-shell kinematics, four-momentum conservation, particle symmetries, and unresolved limits. The analytic construction gives explicit radiation maps at fixed Born momenta for massless final-state jets of any multiplicity, while the neural network parametrizes the projection away from unresolved limits. The LLM also implements both constructions, including training for the neural network. Human physics guidance specifies the constraints, while the LLM develops and refines the construction.

Within this construction, neural networks provide the capacity to represent functions of many variables. Universal approximation on compact domains~\cite{Cybenko:1989iql} motivates using them to parametrize the projection away from soft and collinear limits. We test whether this flexibility can be combined with exact kinematic and infrared constraints to produce an integrable radiation term. A neural parametrization could also represent the signed Born contact; the numerical fit presented here uses a spline basis. Machine learning has a further role after local subtraction has made the radiation integrand integrable. A function learned from integration samples can be subtracted from the exact weight and its known integral added back, reducing fluctuations without changing the prediction. We realize this strategy with fitted Gaussian and polynomial control functions in the two-jet NNLO calculation of Sec.~\ref{sec:twojet-results}.

As illustrated in Fig.~\ref{fig:ai-role}, humans choose the framework and specify the physical constraints. The LLM devises the projections and contact formulas and implements the numerical calculation, including the learned controls for integration. For the local NLO calculation, it also assembles the EFT $\delta(\tau_N)$ coefficient from established hard, jet and soft contributions. We used GPT-5.6 and GPT-6 for this construction and implementation. The neural implementation tests a flexible representation of the projection. The LLM derives radiation maps at fixed Born momenta and finite contact integrals, yielding the local subtraction detailed in Appendices~\ref{app:fixed-born} and~\ref{app:local-contact}. The two-jet NNLO example combines recursive contact extraction with learned variance reduction. Appendices~\ref{app:nnlo-fixed-born}--\ref{app:local-contact} develop the extension to three-jet NNLO.

\begin{table*}[t]
\caption{Physical inputs and construction restrictions for AI-developed P2B subtraction. The exclusions concern the new subtraction being constructed. The lower-order radiation input may use its existing subtraction, as indicated by the dash.}
\label{tab:construction-inputs}
\centering
\footnotesize
\renewcommand{\arraystretch}{1.15}
\begin{tabular}{@{}p{0.14\textwidth}p{0.43\textwidth}@{\hspace{\dimexpr2\tabcolsep+6pt\relax}}p{\dimexpr0.37\textwidth-6pt\relax}@{}}
\toprule
Ingredient & Input or permitted use & Excluded from the new construction \\
\midrule
Projection & Born information, parton kinematics and identities, universal infrared limits and geometric constraints. & Reuse of local counterterms or momentum maps from existing subtraction schemes. \\[3pt]
Radiation & The complete calculation with one additional jet at one lower perturbative order, including its internal subtraction. & \makebox[\linewidth][c]{---} \\[3pt]
Contact & EFT singular predictions at fixed Born kinematics, including the coefficient of $\delta(\tau_N)$, after fixing the projection. & Separate virtual or integrated-counterterm contributions as contact-fitting targets. \\[3pt]
Prediction & A common fixed projection and contact; independent reference comparisons. & Observable-dependent changes to the fixed projection or contact. \\
\bottomrule
\end{tabular}
\end{table*}

\section{NLO workflow}
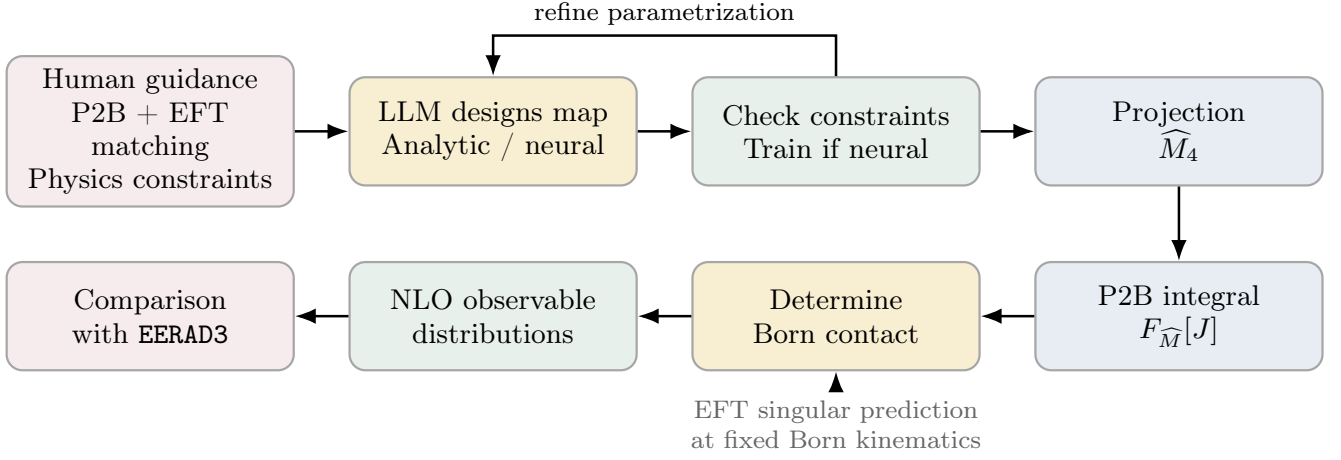
\begin{figure*}[t]
\centering
\resizebox{.98\textwidth}{!}{\begin{tikzpicture}[font=\small,>=Latex,
 box/.style={draw=gray!70,rounded corners=5pt,line width=.7pt,align=center,minimum height=12mm,text width=29mm,inner sep=4pt},
 arr/.style={->,line width=.8pt}]
\node[box,fill=rose] (brief) at (0,0) {Human guidance\\P2B + EFT matching\\Physics constraints};
\node[box,fill=cream] (llm) at (3.8,0) {LLM designs map\\Analytic / neural};
\node[box,fill=mint] (code) at (7.6,0) {Check constraints\\Train if neural};
\node[box,fill=ice] (freeze) at (11.4,0) {Projection\\$\widehat{M}_4$};
\draw[arr] (brief)--(llm);\draw[arr] (llm)--(code);\draw[arr] (code)--(freeze);
\draw[arr] (code.north) -- ++(0,.45) -| node[pos=.25,above,font=\footnotesize]{refine parametrization} (llm.north);
\node[box,fill=ice] (resp) at (11.4,-2.05) {P2B integral\\$F_{\widehat{M}}[J]$};
\node[box,fill=cream] (contact) at (7.6,-2.05) {Determine\\Born contact};
\node[box,fill=mint] (predict) at (3.8,-2.05) {NLO observable\\distributions};
\node[box,fill=rose] (check) at (0,-2.05) {Comparison\\with \texttt{EERAD3}};
\draw[arr] (freeze)--(resp);\draw[arr] (resp)--(contact);\draw[arr] (contact)--(predict);\draw[arr] (predict)--(check);
\node[align=center,font=\footnotesize,text=gray!80!black] (oracle) at (7.6,-3.25) {EFT singular prediction\\at fixed Born kinematics};
\draw[arr] (oracle)--(contact);
\end{tikzpicture}}
\caption{Common workflow for the two NLO implementations. Humans specify the framework and physical constraints. The LLM develops a projection and checks its kinematic and infrared properties, including training for the neural option. With the projection fixed, the contact is fitted through EFT matching or calculated directly using finite radiation integrals and the EFT $\delta(\tau_3)$ coefficient. Each implementation uses its own common projection and contact for all final distributions.}
\label{fig:workflow}
\end{figure*}

In both NLO implementations, we first construct a projection satisfying the physical constraints. We then fix the projection, determine its contact from EFT input and predict several observables using the same map and contact. To test the LLM's ability to carry out this construction, we required it to develop the maps and contact calculation from the given constraints, without importing local counterterms or momentum maps from existing subtraction schemes. Table~\ref{tab:construction-inputs} summarizes the inputs and restrictions, and Fig.~\ref{fig:workflow} shows the sequence.

\begin{figure*}[t]
\centering
\resizebox{.98\textwidth}{!}{\begin{tikzpicture}[font=\footnotesize,>=Latex,
 box/.style={draw=gray!70,rounded corners=4pt,line width=.65pt,align=center,text width=29mm,minimum height=12mm,inner sep=4pt},
 arr/.style={->,line width=.75pt}]
\node[anchor=west,font=\small\bfseries] at (-1.6,.95) {(a) Projection module};
\node[box,fill=rose] (input) at (0,0) {Four momenta, labels\\$E_i/Q$, $s_{ij}/Q^2$};
\node[box,fill=mint] (feat) at (3.8,0) {Branch selection\\10 kinematic inputs\\+ 6 label entries};
\node[box,fill=cream,minimum height=19mm] (net) at (7.6,0) {};
\node at (7.6,.65) {Mapper MLP};
\foreach \x/\lid/\count/\col in {6.8/one/2/white,7.35/two/3/orange!65,7.9/three/3/orange!65,8.4/four/2/red!60}{
  \foreach \i in {1,...,\count}{\node[circle,draw,fill=\col,inner sep=1.4pt] (n\lid-\i) at (\x,{.15-(\i-1)*.23}) {};}}
\foreach \a in {1,2}{\foreach \b in {1,2,3}{\draw[thin] (none-\a)--(ntwo-\b);}}
\foreach \a in {1,2,3}{\foreach \b in {1,2,3}{\draw[thin] (ntwo-\a)--(nthree-\b);}}
\foreach \a in {1,2,3}{\foreach \b in {1,2}{\draw[thin] (nthree-\a)--(nfour-\b);}}
\node at (7.6,-.64) {$16\to32\to32\to11$};
\node[box,fill=ice] (shell) at (11.4,0) {Shape + orientation\\Constrained decoder\\One state $\widetilde\Phi_3$};
\draw[arr] (input)--(feat);\draw[arr] (feat)--(net);\draw[arr] (net)--(shell);
\node[align=center,text=gray!80!black] at (7.6,-1.27) {Analytic limiting behaviour; symmetry constraints; geometric training criterion};
\draw[densely dotted,gray] (-1.6,-1.65)--(13,-1.65);
\node[anchor=west,font=\small\bfseries] at (-1.6,-2.04) {(b) Real-emission integral};
\node[box,fill=rose] (sample) at (0,-2.95) {Integration sample\\$\Phi_4$, exact $R_4$};
\node[box,fill=ice] (pair) at (3.8,-2.95) {Projection\\$(\Phi_4,\widetilde\Phi_3)$};
\node[box,fill=mint] (measure) at (7.6,-2.95) {Measure both states\\with specified support};
\node[box,fill=cream] (weight) at (11.4,-2.95) {Integrate weighted\\measurement difference};
\draw[arr] (sample)--(pair);\draw[arr] (pair)--(measure);\draw[arr] (measure)--(weight);
\draw[densely dotted,gray] (-1.6,-3.9)--(13,-3.9);
\node[anchor=west,font=\small\bfseries] at (-1.6,-4.29) {(c) Contact module};
\node[box,fill=rose] (mom) at (0,-5.22) {EFT singular\\cumulants weighted\\by $w_a(b)$};
\node[box,fill=mint] (diff) at (3.8,-5.22) {EFT minus P2B\\$t\to0$ limit};
\node[box,fill=cream] (fit) at (7.6,-5.22) {Contact as a function\\of Born coordinates};
\node[box,fill=ice] (out) at (11.4,-5.22) {Born contact $c_{\widehat{M}}(b)$\\with uncertainty};
\draw[arr] (mom)--(diff);\draw[arr] (diff)--(fit);\draw[arr] (fit)--(out);
\node[text=gray!80!black] at (7.6,-6.25) {The same projection and contact enter all observable distributions};
\draw[densely dotted,gray] (-1.6,-6.65)--(13,-6.65);
\node[anchor=west,font=\small\bfseries] at (-1.6,-7.04) {(d) Learned integration controls for the local NNLO contact};
\node[box,fill=rose] (local) at (0,-7.95) {Radiation integrand\\for the contact};
\node[box,fill=mint] (learn) at (3.8,-7.95) {Learn a control from\\separate samples};
\node[box,fill=cream] (residual) at (7.6,-7.95) {Integrate exact\\integrand\\minus control};
\node[box,fill=ice] (restore) at (11.4,-7.95) {Add the known\\control integral};
\draw[arr] (local)--(learn);\draw[arr] (learn)--(residual);\draw[arr] (residual)--(restore);
\node[text=gray!80!black] at (7.6,-8.98) {Control held fixed during independent integration; unchanged physical prediction};
\end{tikzpicture}}
\caption{Modules for projection, radiation integration and contact determination. Panels (a)--(c) describe the neural implementation at NLO. The projection network has two hidden layers with sigmoid linear unit (SiLU) activations. Its 11 outputs comprise three shape parameters and eight coefficients for two orientation vectors. The real-emission integral uses the exact matrix element. Contact fitting uses EFT singular cumulants weighted by functions of the Born variables, in the limit $t\to0$. Here $\widetilde\Phi_3=\widehat{M}_4\Phi_4$. Panel (d) shows the additional use of machine learning in local two-jet NNLO contact integration. A control function is trained on separate samples, then held fixed while the exact residual is integrated and the known control integral restored.}
\label{fig:modules}
\end{figure*}

Panels (a)--(c) of Fig.~\ref{fig:modules} give the three modules of the neural implementation. The projection module selects a branch labelled by a soft parton or collinear pair. Here $Q$ is the centre-of-mass energy, $p_i$ is a parton four-momentum, $E_i=p_i^0$, and $s_{ij}=2p_i\cdot p_j$, with $p^2=(p^0)^2-|\mathbf p|^2$. Four energy fractions and six pair invariants, together with six learned components encoding this label, enter a multilayer perceptron (MLP). Its outputs determine a single projected state through a constrained parametrization. The network is trained on kinematic information; pointwise real weights and reference NLO distributions do not enter its training or selection. The second module integrates the exact real coefficient against the difference between real and projected measurements. The contact module fits a function of Born coordinates to the $t\to0$ difference between EFT singular cumulants and the P2B integral. Panel (d) illustrates the learned variance reduction used for the local NNLO contact below.

In the analytic implementation, the map allows radiation to be generated while the Born momenta remain fixed. The radiation term uses the real and projected measurements as before, while the contact is calculated from the finite radiation contribution $\mathcal A_1$ and the EFT $\delta(\tau_3)$ coefficient $H_1$. The next section describes the two constructions. All numerical calculations presented here, including projection-network training and contact extraction, use the CPU of a 2020 Apple M1 MacBook, without GPU acceleration.

\section{NLO construction and contact extraction}
\label{sec:nlo}
We consider a pure virtual-photon current, $n_f=5$ massless quark flavours, and $\mu_R=Q$ with the $\overline{\mathrm{MS}}$ coupling. We work in the centre-of-mass frame with flavour-blind, rotationally invariant measurements. With $\as=\alpha_s(Q)/(2\pi)$ and the photon-mediated two-parton Born cross section $\sigma_0$, the three-jet expansion is
\begin{equation}
 \frac{\sigma[J]}{\sigma_0}=\as A[J]+\as^2 B[J]+\mathcal O(\as^3).
 \label{eq:normalization}
\end{equation}
The displayed three-jet NLO results give the coefficient $B$ in this expansion. All coefficient densities below have the corresponding powers of $\as$ and the common factor $\sigma_0$ removed and use Lorentz-invariant phase-space measures. For the rotationally invariant measurements used here, we average the kernels over the overall event orientation. The real channels are $q\bar qgg$ and four-quark production, with the flavour sums, interference terms, and correlated azimuthal samples supplied by \texttt{EERAD3}~\cite{Gehrmann-DeRidder:2014hxk}. These samples pair configurations differing by a $\pi/2$ azimuthal rotation of a daughter pair in its rest frame.

\subsection{Neural projection and contact fitting}

Let $\Gamma_m^0$ select JADE jets~\cite{JADE:1986kta} with E-scheme recombination and strict $y_{23}>0.02$; no $y_{34}$ cut is applied. Here $y_{23}$ and $y_{34}$ denote the $3\to2$ and $4\to3$ jet transition values. The neural projection $\widehat{M}_4:\Phi_4\to\Phi_3$ is evaluated only for $\Gamma_4^0=1$. Its projected counterevent contributes only when $\Gamma_3^0(\widehat{M}_4\Phi_4)=1$. For a measurement $J$ supported within this three-jet region, define
\begin{equation}
\begin{aligned}
 F_{\widehat{M}}[J]=\int_{\Gamma_4^0=1}\dd\Phi_4\,R_4(\Phi_4)
 \big[ J_4(\Phi_4)\hspace{22mm}\\[-2pt]
 {}-\Gamma_3^0(\widehat{M}_4\Phi_4)J_3(\widehat{M}_4\Phi_4)\big].
\end{aligned}
\label{eq:nlofinite}
\end{equation}
Here $R_4\dd\Phi_4$ includes the normalization of the real contribution to $B$. All additional physical cuts are evaluated on each state separately.

In a single-unresolved limit with a resolved three-parton limiting configuration, the map approaches the corresponding energy flow. The real and projected weights therefore cancel their leading singular behaviour before integration, giving an integrable $F_{\widehat{M}}[J]$ for a fixed infrared-safe measurement without a slicing boundary. At finite resolution, the map changes the finite subtraction prescription. In this normalization, Eq.~\eqref{eq:p2bframework} at $k=1$ becomes
\begin{equation}
 B[J]=F_{\widehat{M}}[J]+\int\dd\Phi_3(b)\,c_{\widehat{M}}(b)J_3(b).
 \label{eq:nlomaster}
\end{equation}
Here $b$ denotes the Born momenta, ordered by decreasing energy, and $c_{\widehat M}$ is a density relative to $\dd\Phi_3(b)$, with the parton-label assignments summed. The contact $c_{\widehat{M}}$ contains the finite combination of virtual and projected real contributions. It depends on the map and support prescription, while their sum in Eq.~\eqref{eq:nlomaster} does not. The fit parametrizes its ratio to the Born density, as specified in Sec.~\ref{sec:nlo-fitted-results}.

For the projection entering Eq.~\eqref{eq:nlomaster}, we parametrize the shape and orientation of a massless three-parton state. An analytic reference is obtained by removing the selected soft parton or combining the selected collinear pair. We subtract the mean of the three spatial momenta, assign massless energies, and rescale to total energy $Q$. This recovers the Born state in the corresponding unresolved limit; an equal-energy configuration is used if the reference degenerates away from that limit. The network deforms this reference through
\begin{equation}
 \begin{aligned}
 L_i&=\ln u_i^{\mathrm{anchor}}+\frac{2d}{1+d}\tanh\ell_i,\\
 u_i&=\frac{e^{L_i}}{\sum_j e^{L_j}},\qquad
 E_i=\frac Q2(1-u_i).
 \end{aligned}
 \label{eq:nlodecoder}
\end{equation}
Here $u_i^{\mathrm{anchor}}=1-2E_i^{\mathrm{anchor}}/Q$ specifies the reference shape, and $\ell_i$ are network outputs. We use $d=2E_s/Q$ for a soft parton of energy $E_s$ and $d=(1-\cos\theta_{ij})/2$ for a collinear pair of opening angle $\theta_{ij}$, so that the deformation vanishes in the corresponding unresolved limit. The orientation is constructed covariantly, with the required particle-exchange symmetries, and all three momenta remain on shell and conserve total four-momentum. The geometric training criterion contains neither the four-parton matrix element nor the NLO cumulants. Since Eq.~\eqref{eq:nlofinite} is integrated over the original four-parton phase space, no Jacobian for the projection is required.

With the projection fixed, we determine the contact through its moments in Born coordinates. We use geometric $N$-jettiness~\cite{Stewart:2010tn,Jouttenus:2011wh} with reference directions chosen by global minimization,
\begin{equation}
 \begin{aligned}
 \tau_N(\Phi_m)&=\frac1Q\min_{\{\hat{\mathbf n}_a\}}
 \sum_{k=1}^{m}\min_{a=1,\ldots,N} n_a\cdot p_k,\\
 \tau_3(\Phi_4)&=\min_{i<j}\frac{E_i+E_j-|\mathbf p_i+\mathbf p_j|}{Q}.
 \end{aligned}
 \label{eq:tau3}
\end{equation}
Here $n_a=(1,\hat{\mathbf n}_a)$ with $|\hat{\mathbf n}_a|=1$, so $n_a\cdot p_k=E_k-\hat{\mathbf n}_a\cdot\mathbf p_k$. For four massless partons, the minimizing partition consists of one pair and two singletons. Their axes align with the respective momentum sums, yielding the second line. On three massless partons, $\tau_3=0$.
For $N=3$, the measurement $W_{a,t}$ of Eq.~\eqref{eq:frameworkcontact} has Born value $w_a(b)$. Its contact moment in the present normalization is
\begin{equation}
 C_{\widehat{M},a}=\int\dd\Phi_3(b)\,c_{\widehat{M}}(b)w_a(b).
 \label{eq:contact}
\end{equation}

Since three-jettiness vanishes on every projected Born state, the contact contributes $C_{\widehat{M},a}\delta(\tau_3)$ to the Born-weighted distribution. The matching relation in Eq.~\eqref{eq:frameworkcontact} reads
\begin{equation}
 C_{\widehat{M},a}=\lim_{t\to0}\big\{B_{\mathrm{sing}}[W_{a,t}]
                         -F_{\widehat{M}}[W_{a,t}]\big\}.
\label{eq:endpointmatching}
\end{equation}
Here $B_{\mathrm{sing}}$ is the leading-power EFT prediction for the same measurement $W_{a,t}$ and phase-space cuts, including its $\delta(\tau_3)$ term and plus distributions~\cite{Gaunt:2015pea}. The logarithms of $t$ cancel between $B_{\mathrm{sing}}[W_{a,t}]$ and $F_{\widehat{M}}[W_{a,t}]$, while the nonsingular cumulant vanishes as $t\to0$. The remaining constant determines the contact moment. This requires the finite EFT $\delta(\tau_3)$ coefficient, including the hard contribution, as well as the logarithmic terms. We determine the Born-dependent contact by fitting its weighted integrals to these moments. Section~\ref{sec:nlo-fitted-results} gives the parametrization and numerical results.

\subsection{Analytic projection and local contact}
\label{sec:nlo-local-method}

The second implementation calculates the contact directly at each Born point. The analytic map in Appendix~\ref{app:fixed-born} merges a selected pair and rescales the remaining momenta, preserving the mass shell, total four-momentum and single-unresolved limits. Its inverse generates the four-parton state from the prescribed Born momenta and three radiation variables. This defines the common projection and phase-space measure needed for the radiation term and the contact.

At fixed Born kinematics, we evaluate $\mathcal A_1$ using the finite four-dimensional integrals in Eq.~\eqref{eq:contact-nlo-interior} and add the EFT $\delta(\tau_3)$ coefficient $H_1$ to obtain $c_{\widehat M}=H_1+\mathcal A_1$. Appendix~\ref{app:local-contact} derives these weights and the compensations that restore the full integration domain. The resulting local subtraction formula evaluates both radiation and contact without a small slicing parameter and includes the exact nonsingular contributions. Section~\ref{sec:nlo-local-results} compares its three-jet predictions with the same distributions used for the neural implementation and presents the extension to four jets. The extension of the local subtraction to three-jet NNLO is proposed in Appendices~\ref{app:nnlo-fixed-born}--\ref{app:local-contact}.

\begin{samepage}
\section{Two-jet NNLO example}
\label{sec:nnlo}
The two-jet case provides a simple test of the P2B organization at NNLO. The radiation term in Eq.~\eqref{eq:p2bframework} is obtained from the NLO correction to three-jet production. We reuse this complete NLO calculation, including its internal subtraction, and apply P2B to remove the remaining singularity associated with the approach to two-parton Born kinematics~\cite{Cacciari:2015jma,Currie:2018fgr}.\par
\end{samepage}

The radiation term preserves the full infrared cancellation within the NLO input, including its integrated contributions. The measurement $\Delta J$ is evaluated on every real configuration and every subtraction configuration. In a single-unresolved limit it must approach the corresponding lower-multiplicity measurement difference; on reaching the two-parton Born configuration it vanishes. These limits must be preserved, with sufficient suppression for the remaining integration to be finite. No extra-jet generation cut may remove a nonzero contribution to $\Delta J$. A completed NLO subtraction, including one constructed with AI, can therefore be reused at the measurement level.

We use the same massless, pure-photon theory as at NLO. For measurements of thrust~\cite{Farhi:1977sg}, the parton energy--energy correlation (EEC)~\cite{Basham:1978bw}, and the two highest jet energies, the Born value $j_B$ is independent of the orientation of the back-to-back pair. The Born contact therefore reduces to one scalar. Writing the coefficient of $\sigma_0\as^2$ as $\mathcal B_2$, Eq.~\eqref{eq:p2bframework} becomes
\begin{equation}
 \mathcal B_2[J]=\mathcal R_3[J-j_B]+c_2j_B,
 \label{eq:nnlotwojet}
\end{equation}
where $\mathcal R_3$ is the three-jet NLO correction in this normalization. For these rotationally invariant measurements the three-to-two projection is trivial, since only the Born value $j_B$ is needed. The LLM therefore implemented this step without a neural network. For this two-direction Born configuration, geometric two-jettiness with back-to-back thrust axes is $\tau_2=1-T$. A Born-dependent or axis-sensitive measurement requires a nontrivial projection.

The scalar contact $c_2$ contains the finite combination of double-virtual and Born-projected radiation contributions. We calculate it directly as $c_2=H_2+\mathcal A_2$, recursively using the NLO P2B construction as described in Appendix~\ref{app:nnlo-contact-extension}. Here $H_2$ is the coefficient of $\delta(\tau_2)$ in the NNLO EFT thrust distribution, with the hard, jet and soft contributions combined~\cite{Becher:2008cf,Kelley:2011ng}, and $\mathcal A_2$ is the finite radiation contribution. Both the radiation term and the contact are evaluated by local subtraction without a slicing cut. Section~\ref{sec:twojet-results} presents the numerical results and the learned controls used to stabilize the contact integral. Appendices~\ref{app:nnlo-fixed-born} and~\ref{app:local-contact} develop the extension to a Born-dependent three-jet NNLO contact.

\section{Numerical results}
\label{sec:numerical}
\subsection{Three-jet NLO with a fitted contact}
\label{sec:nlo-fitted-results}
\begin{figure*}[tp]
\centering
\includegraphics[width=\textwidth]{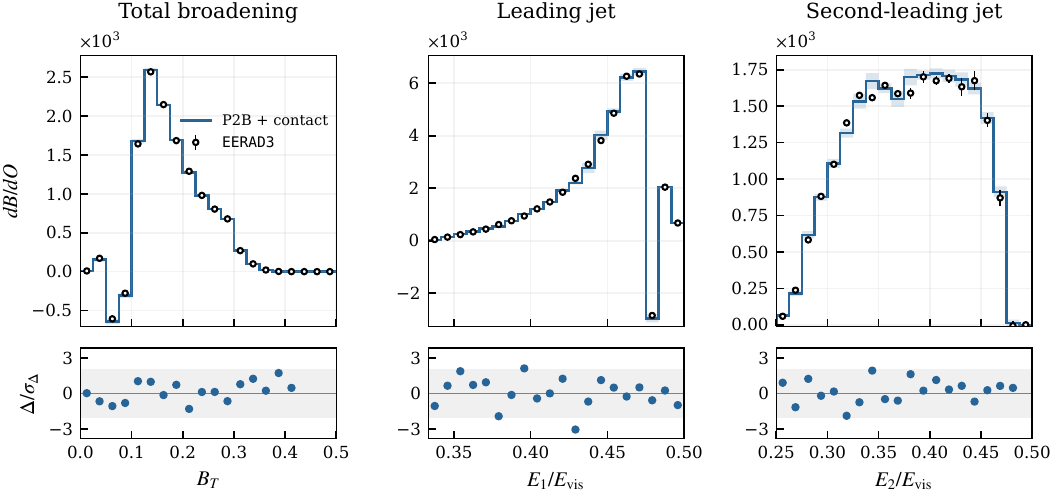}
\caption{Complete three-jet NLO coefficient distributions with strict JADE/E $y_{23}>0.04$ and no $y_{34}$ cut. The projection and the 127-parameter contact are common to all panels. Blue bands combine P2B Monte Carlo and contact bootstrap errors; black points show \texttt{EERAD3} with Monte Carlo errors. In all distribution plots, $O$ denotes the observable on the horizontal axis. Lower panels show $\Delta/\sigma_\Delta$, where $\Delta$ is the difference from the reference and $\sigma_\Delta$ its combined statistical error. The same event sample contributes to different bins, and the fitted contact introduces additional correlations.}
\label{fig:nlo}
\end{figure*}
The matching prescription in Eq.~\eqref{eq:endpointmatching} uses EFT singular cumulants in the limit $t\to0$. For simplicity in this numerical test, we use full reference cumulants from \texttt{EERAD3} at $t=0.004,0.025,0.09$ to constrain the contact fit. At these finite thresholds, the exact identity $C_{\widehat{M},a}=B[W_{a,t}]-F_{\widehat{M}}[W_{a,t}]$ includes the nonsingular contribution and requires no small-$t$ approximation. These Monte Carlo cumulants carry statistical errors and require weighted integrations to resolve the Born dependence, making contact extraction more demanding than with an analytic EFT cumulant evaluated at chosen Born points~\cite{Jouttenus:2011wh,Gaunt:2015pea}.

For the fit, we use the ordered energy fractions $x_i=2E_i/Q$, with $x_1\ge x_2\ge x_3$ and $x_3=2-x_1-x_2$. After integrating over the overall orientation, $\dd\Phi_3=Q^2\dd x_1\dd x_2/(128\pi^3)$. Let $B_3(b)$ be the three-parton Born coefficient and $\overline B_3(b)=\sum_{\pi\in S_3}B_3(b_\pi)$ its sum over the six assignments of $q,\bar q,g$ to these momenta. We expand the dimensionless ratio $c_{\widehat M}(b)/\overline B_3(b)$ in bicubic splines with 127 independent coefficients and $C^2$ continuity at $x_1=x_2=2/3$. The fit integrals therefore use the Born weight $\dd\Phi_3\,\overline B_3$. The coefficients may have either sign, as required for the finite virtual contribution. The statistical uncertainty is estimated from 128 bootstrap replicas of the cumulant fit. The resulting contact is common to all distributions in Fig.~\ref{fig:nlo}.

We reconstruct the full three-jet NLO coefficient $B$ by combining the subtracted real-emission integral with the fitted Born contact, including the finite virtual contribution. The contact is determined in the region $y_{23}>0.02$. Figure~\ref{fig:nlo} presents the resulting distributions with the tighter cut $y_{23}>0.04$, using the same projection and contact throughout. Total broadening $B_T$~\cite{Catani:1992jc} is measured on the original partons; the ordered jet energies follow exclusive three-jet recombination, with total visible energy $E_{\mathrm{vis}}=Q$. For this comparison, the \texttt{EERAD3} reference is evaluated with its own subtraction prescription, applying the cuts separately to real events and counterevents. The comparison therefore probes both the observable dependence and the change in the three-jet cut.

Figure~\ref{fig:nlo} shows good agreement with \texttt{EERAD3}, reproducing the full NLO coefficient with a single fitted contact and no observable-dependent refit. For broadening and leading energy, the integrated NLO coefficients are $286.43\pm1.00$ and $287.75\pm1.25$, respectively, compared with $287.09\pm0.28$ and $287.72\pm1.21$ from \texttt{EERAD3}. For second-leading energy we obtain $288.33\pm0.92$, compared with $285.50\pm0.76$, a difference of 2.36 combined standard deviations. These normalizations agree within two combined standard errors after accounting for the common contact. The \texttt{EERAD3} second-leading-jet reference lies slightly below its other determinations, also by less than two combined standard errors, consistent with fluctuations between independent reference runs. The largest binwise differences are 1.71, 3.04 and 1.91 combined standard deviations for broadening, leading and second-leading energy, respectively.

For this numerical study, we deliberately restricted the computational budget to a single M1 laptop with limited running time, keeping the integration statistics modest and the contact parametrization to 127 coefficients. At fixed sampling, narrow bins have fewer contributing events, while away from unresolved limits real and projected configurations can populate neighbouring bins. Such fluctuations are common to histogrammed subtraction calculations. \texttt{EERAD3} reduces the sampling cost with adaptive \texttt{VEGAS} grids~\cite{Lepage:1977sw} and phase-space parametrizations adapted to unresolved limits~\cite{Gehrmann-DeRidder:2014hxk}. Our calculation also uses adaptive integration, but neither the geometric projection training nor the sampling was optimized for individual distributions.

Sampling adapted to the bins and a projection optimized for variance can improve precision per event, while more precise cumulants weighted by functions of the Born variables can constrain a richer contact representation. These improvements offer a direct route to higher precision for the reconstructed NLO distributions.

\subsection{Three- and four-jet NLO with a local contact}
\label{sec:nlo-local-results}
We next calculate the same three-jet NLO distributions using the analytic local subtraction of Sec.~\ref{sec:nlo-local-method}. The map in Appendix~\ref{app:fixed-born} generates radiation at fixed Born momenta and defines a common projection and measure for the radiation term and the contact. At each Born point, Eq.~\eqref{eq:contact-nlo-direct} gives $c_{\widehat M}=H_1+\mathcal A_1$.

The LLM obtains $H_1$ by expanding the three-jettiness factorization formula~\cite{Jouttenus:2011wh} to one loop and collecting the coefficient of $\delta(\tau_3)$. It derives the hard contribution from the finite three-parton virtual interference~\cite{Gehrmann-DeRidder:2005btv} supplied by \texttt{EERAD3}, converting it to the $\overline{\mathrm{MS}}$ hard-function convention. It then combines this with the one-loop quark and gluon jet functions~\cite{Becher:2006qw,Becher:2010pd} and the three-jettiness soft function, including its finite geometry-dependent terms~\cite{Jouttenus:2011wh}. All terms use $\mu_R=Q$, the geometric jettiness normalization with reference scales $Q_i=2E_i$, and the dimensionless $\tau_3$ with the plus-distribution convention of Eq.~\eqref{eq:contact-nlo-eft-spectrum}. The LLM also includes the finite $\delta(\tau_3)$ terms generated when expressing the jet and soft distributions in this common variable.

The finite radiation contribution $\mathcal A_1$ is calculated directly with the weights of Eq.~\eqref{eq:contact-nlo-interior}, including all finite-interval compensations. Adding $H_1$ completes the contact at each Born point. Both the radiation term and contact are integrated in four dimensions, including all nonsingular contributions. This gives a \emph{local subtraction calculation} without a small slicing parameter; its inputs are the tree matrix elements and the EFT singular prediction. No contact fit or reference cumulant enters this prediction.

\begin{figure*}[tp]
\centering
\includegraphics[width=\textwidth]{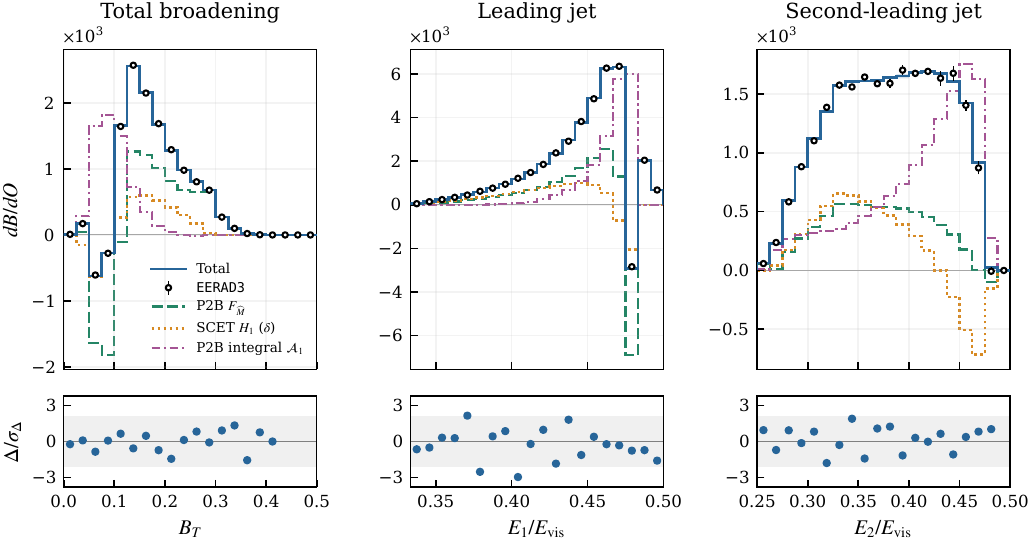}
\caption{Three-jet NLO coefficient distributions from the analytic local subtraction with a directly calculated contact, using the same observables and physical cuts as Fig.~\ref{fig:nlo}. Green, orange and purple curves show the P2B radiation term $F_{\widehat M}$, the Born-weighted EFT $\delta(\tau_3)$ coefficient $H_1$, and the Born-weighted finite radiation integral $\mathcal A_1$, respectively. Their sum is the blue result; its band includes the correlation between radiation and contact. Black points reproduce the \texttt{EERAD3} reference of Fig.~\ref{fig:nlo}. Lower panels compare the total with the reference in combined standard errors, with grey bands at $\pm2$. The calculation uses the maps of Appendix~\ref{app:fixed-born} and the local contact weights of Appendix~\ref{app:local-contact}, over the full radiation phase space.}
\label{fig:local-contact-nlo}
\end{figure*}

Figure~\ref{fig:local-contact-nlo} separates the P2B radiation contribution, labelled $F_{\widehat M}$, from the Born-weighted $H_1$ and $\mathcal A_1$ terms. Their sum agrees well with \texttt{EERAD3}. The calculation uses $3.36\times10^8$ samples and gives a common integrated coefficient $B=286.680\pm0.277$, with a relative statistical error of $0.10\%$. The largest binwise differences from \texttt{EERAD3} are $1.56$, $2.96$ and $1.90$ combined standard errors. These results demonstrate complete multijet NLO predictions from the AI-constructed analytic projection and direct local contact calculation.

\begin{figure*}[p]
\centering
\includegraphics[width=\textwidth]{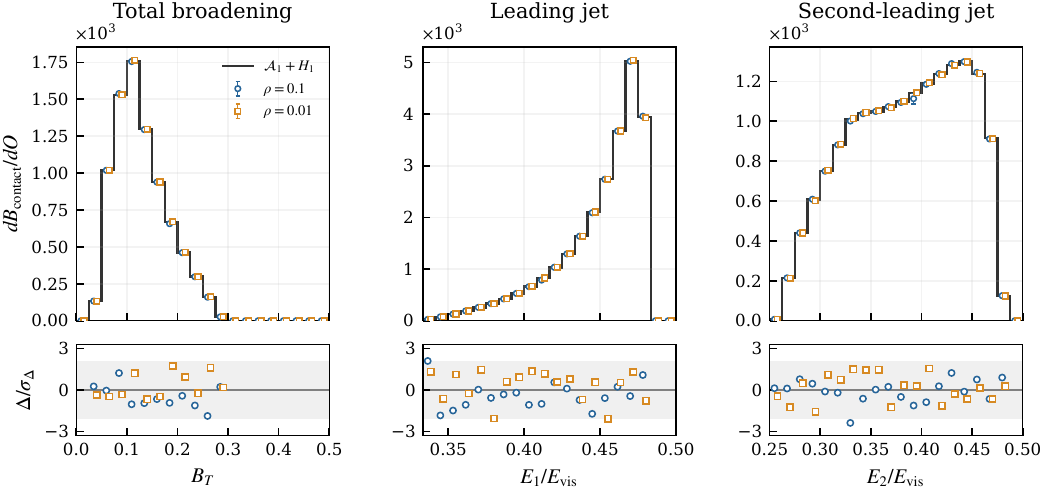}
\caption{Comparison of three contact determinations for the analytic projection. Each curve shows the Born integral in Eq.~\eqref{eq:nlomaster}, denoted $B_{\mathrm{contact}}$, for the analytic projection and the cuts of Fig.~\ref{fig:nlo}. The black histogram uses the finite radiation integral $\mathcal A_1$ in Eq.~\eqref{eq:contact-nlo-boundary} plus the EFT $\delta(\tau_3)$ coefficient $H_1$. Blue circles and orange squares use Eq.~\eqref{eq:contact-nlo-interior} for $\mathcal A_1$, with $\rho=0.1$ and $0.01$. Error bars and the black band denote one standard error; markers are displaced within each bin for visibility. Lower panels show the differences from the black result in standard errors, including shared-sample correlations; grey bands indicate $\pm2$. As defined in Appendix~\ref{app:local-contact}, $\rho$ is a free rescaling parameter, not a cutoff.}
\label{fig:contact-methods}
\vspace{0.8em}
\includegraphics[width=\textwidth]{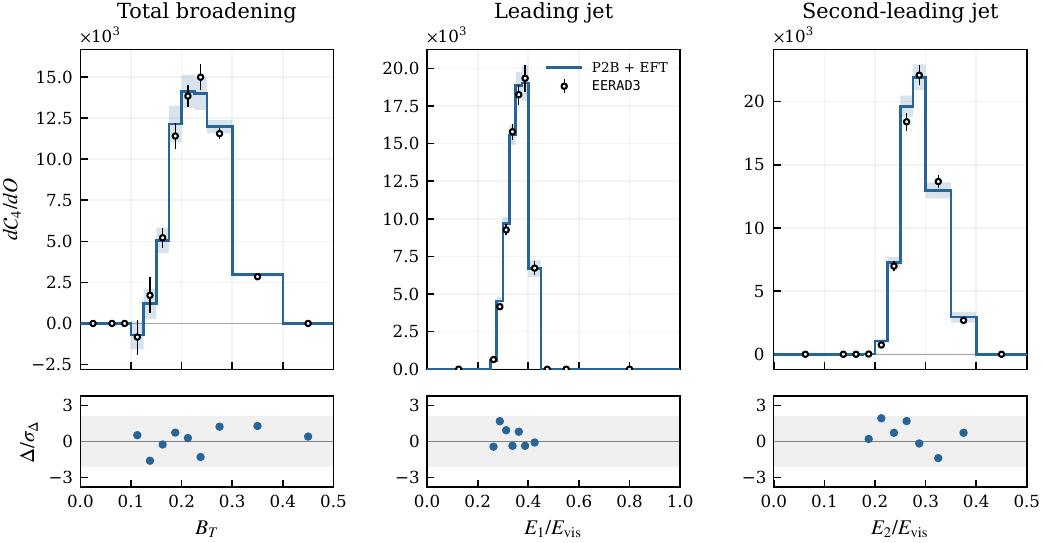}
\caption{Four-jet NLO coefficient distributions from the analytic local subtraction with a directly calculated contact, with strict JADE/E $y_{34}>0.04$. Here $\mathcal C_4$ is the coefficient of $\as^3$ in $\sigma[J]/\sigma_0$. Total broadening is measured on the partons, and the jet energies follow exclusive four-jet recombination with $E_{\mathrm{vis}}=Q$. Blue curves and bands show the P2B + EFT prediction and one standard error; black points show the \texttt{EERAD3} antenna reference with Monte Carlo errors. Lower panels show the differences in standard errors, including shared-sample correlations, with grey bands at $\pm2$.}
\label{fig:four-jet-nlo-local}
\end{figure*}

Figure~\ref{fig:contact-methods} compares the contact $\mathcal A_1+H_1$ obtained with finite rescalings $\rho=0.1$ and $0.01$ and by direct evaluation of the P2B integral $\mathcal A_1$ in Eq.~\eqref{eq:contact-nlo-boundary}. The same EFT $\delta(\tau_3)$ coefficient $H_1$ is added in all three cases. Here $\rho\in(0,1)$ is a freely chosen rescaling parameter, not a cutoff. The exact compensations in Appendix~\ref{app:local-contact} restore the full phase space for any such choice. The two rescalings use independent samples, while Eq.~\eqref{eq:contact-nlo-boundary} is evaluated on those same samples. The differences shown below the distributions account for their correlations. All three determinations agree, with the largest binwise difference from the direct integral below $2.4$ standard errors. This supports both the equivalence of the two contact formulas and the independence of the extracted contact from the free parameter $\rho$.

We further show four-jet NLO predictions in Fig.~\ref{fig:four-jet-nlo-local}, using the analytic projection and local contact with four-parton Born kinematics. The contact combines the EFT $\delta(\tau_4)$ coefficient, calculated directly by the LLM, with finite radiation integrals, without a small slicing parameter. All distributions agree well with \texttt{EERAD3}.\footnote{The LLM identified \texttt{7**s12} as a typo in the four-parton one-loop $n_f$ contribution of \texttt{EERAD3}~2.0.0. Its symbolic checks show that replacing it by \texttt{7*s12} restores the expected scaling with the invariants and the symmetry under simultaneous quark--antiquark and gluon exchange. The correction also makes the leading $qg$ and $\bar qg$ collinear residues vanish, as required for this $n_f$ term. We use this LLM-suggested matrix-element modification here and for $K_4$ in Fig.~\ref{fig:fixed-born-nnlo}.}

\subsection{Two-jet NNLO}
\label{sec:twojet-results}
We now turn to the two-jet NNLO results, comparing a directly calculated contact with one determined by cumulant matching. The local construction recursively uses our NLO P2B contact, as described in Appendix~\ref{app:nnlo-contact-extension}, and requires no slicing cut. As in the NLO examples, the LLM constructs the EFT input $H_2$ by combining the known hard and quark-jet contributions with the analytic two-loop hemisphere-soft constant~\cite{Becher:2008cf,Kelley:2011ng}. It collects the coefficient of $\delta(\tau_2)$ in the NNLO singular thrust distribution, using the plus-distribution convention of Appendix~\ref{app:local-contact}.

\begin{figure*}[tp]
\centering
\includegraphics[width=\textwidth]{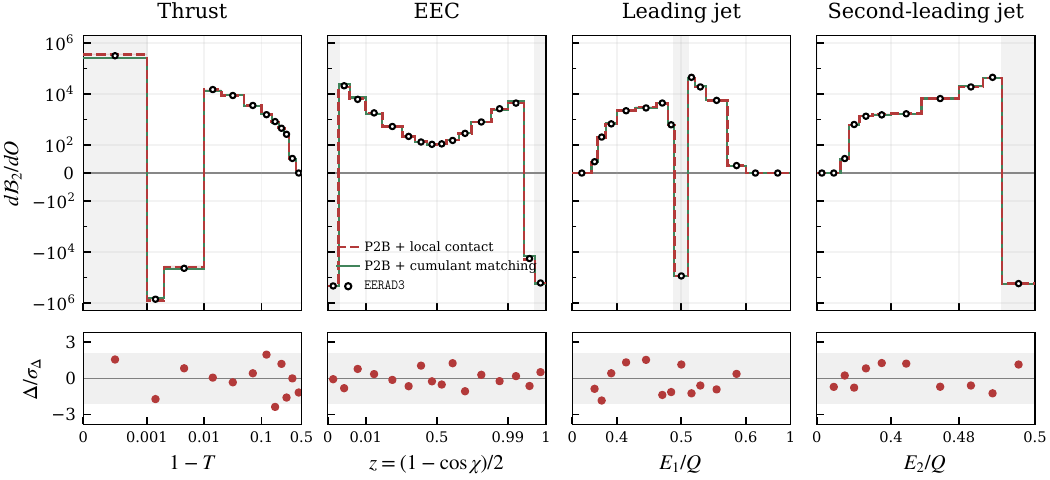}
\caption{Two-jet NNLO coefficient distributions $\mathcal B_2$. Red dashed curves use local subtraction with the directly calculated contact. The red bands combine P2B and contact errors, retaining the common contact correlation between Born-supported bins. Green solid curves show the cumulant-matched calculation, and black points the \texttt{EERAD3} reference. Lower panels compare the red prediction with the reference in combined statistical errors; grey bands indicate $\pm2$. Shaded upper-panel bins contain Born support. There are no event-selection cuts; EEC includes all ordered parton pairs and self correlations, and JADE/E clustering at $y_{\rm cut}=0.04$ includes two-, three-, and four-jet events.}
\label{fig:nnlo}
\end{figure*}

Local cancellation makes the contact integral finite, but signed contributions can still produce substantial Monte Carlo fluctuations. We reduce these fluctuations with control functions learned from previous integration samples. Write the radiation contribution to the contact as $\mathcal A_2=\int\dd\mu\,\mathcal W_2$, where $\dd\mu$ includes all radiation variables and channel sums. Let $q$ be a normalized sampling density on this domain, so that $\mathcal W_2/q$ is the Monte Carlo weight. For a learned function $h_\theta$ with fitted parameters $\theta$ and known integral $I_\theta$, we use the identity
\begin{equation}
 \mathcal A_2=I_\theta+
 \left\langle\frac{\mathcal W_2-h_\theta}{q}\right\rangle_q,
 \qquad I_\theta=\int\dd\mu\,h_\theta.
 \label{eq:learned-contact-control}
\end{equation}
The brackets denote an average over $q\,\dd\mu$. Gaussian and polynomial control functions are fitted to the signed integrand $\mathcal W_2$ using separate samples, then held fixed during integration. Their known integrals are added back, so the learned approximation affects the variance without changing the contact. No inclusive cross section is used to train or normalize the prediction.

The contact shown in Fig.~\ref{fig:nnlo} is calculated directly with about 22 CPU-hours on the M1 laptop, excluding training. The machine-learning controls designed by the LLM are used to reduce the estimated variance of the integration by about $80\%$. Further optimization to stabilize the integration is needed and is left to future work~\footnote{The results shown here use the numerical contact evaluation. The LLM has also obtained a closed-form result by directly evaluating the recursive P2B contact relation in Eq.~\eqref{eq:contact-twojet-recursion} of Appendix~\ref{app:nnlo-contact-extension}, taking the complete three-jet NLO calculation and the EFT coefficients as known inputs. The radiation phase-space integrations were kept strictly four-dimensional. In the normalization of Eq.~\eqref{eq:nnlotwojet}, with $a_s=\alpha_s(Q)/(2\pi)$ and $\mu_R=Q$, the result is
\begin{eqnarray}
   \begin{aligned}
   c_2={}&-\frac{3}{8}C_F^2
   +C_FC_A\left(\frac{123}{8}-11\zeta_3\right)\\
   &+C_FT_Rn_f\left(-\frac{11}{2}+4\zeta_3\right).
   \end{aligned}
   \end{eqnarray}
 No known NNLO total cross section~\cite{Celmaster:1979xr} or pre-integrated NNLO subtraction result was used to infer the answer, fix integration constants, or normalize the result. The calculation required resolving overlapping soft and collinear limits, preserving the prescribed order of finite-part extraction, and analytically retaining all finite boundary compensations. This provides a direct validation of our P2B subtraction framework and a concrete demonstration of an LLM completing a nontrivial analytic NNLO contact calculation under specified physical constraints.}. The P2B contribution $\mathcal R_3[J-j_B]$ is obtained following Appendix~\ref{app:nnlo-contact-extension}.

For comparison, the green curves use the same contact-matching strategy as the neural NLO implementation. As a proof of concept, that calculation uses a full reference thrust cumulant at $\tau_2\leq0.3$, giving $c_2=5.538\pm0.329$. Both implementations use one contact for thrust, EEC, and the two jet energies, with no observable-dependent adjustment.

For an EEC bin $\ell$, the measurement is $J_\ell=\sum_{i,j}(E_iE_j/Q^2)\mathbf{1}_\ell(z_{ij})$, where $\mathbf{1}_\ell$ is its indicator, $z_{ij}=(1-\cos\chi_{ij})/2$, and $\chi_{ij}$ is the opening angle. The sum includes $i=j$. At Born level, this measurement gives weight $1/2$ in each of the bins containing $z=0$ and $z=1$. Thrust has unit Born weight in the first bin, while both jet energies have unit weight in the bin containing $E/Q=1/2$. Jet energies are ranked after JADE/E clustering at $y_{\rm cut}=0.04$, including all two-, three-, and four-jet events. These distributions give the NNLO coefficient, including the bins with Born support.

The reference uses \texttt{EERAD3} and restores the Born contribution with the known NNLO correction to the total cross section~\cite{Gehrmann-DeRidder:2009fgd}. Figure~\ref{fig:nnlo} shows good agreement with the local-contact prediction, including the bins with Born support. The largest binwise differences are $2.36$, $1.26$, $1.83$ and $1.28$ combined standard errors for thrust, EEC, leading and second-leading jet energy, respectively. The agreement across these observables demonstrates the feasibility of recursive local contact extraction at NNLO. Similar to the NLO examples, larger samples and improved variance reduction provide a direct route to higher precision.

Together, these results demonstrate two complementary uses of AI. Neural projections and fitted contacts reconstruct the full three-jet NLO coefficient, while analytic maps and direct contact integrals give local predictions for three and four jets at NLO and two jets at NNLO. The NNLO calculation also shows how machine learning can improve the integration of the constructed subtraction. Appendices~\ref{app:nnlo-fixed-born}--\ref{app:local-contact} develop the three-jet NNLO extension and test its radiation contribution.

\section{Conclusions and outlook}
We have developed infrared subtraction for multijet production with AI within the P2B + EFT matching framework. Human guidance specifies the framework and physical constraints, while the LLM devises the projections and contact formulas and implements the numerical calculation. The first implementation uses the expressive power of a neural network to construct a phase-space projection and determines its Born contact by fitting. Furthermore, the analytic implementation returns a \emph{local subtraction formula}. Radiation is integrated at fixed Born momenta, and the contact is calculated directly from the EFT $\delta(\tau_N)$ coefficient and finite four-dimensional radiation integrals. Both radiation and contact are evaluated over the full phase space without a small slicing parameter. The construction thus combines local subtraction with slicing's advantage of recycling lower-order radiation calculations and EFT singular predictions.

The numerical examples reconstruct the complete three-jet NLO coefficient in both implementations and the four-jet NLO coefficient with local subtraction, giving good agreement with \texttt{EERAD3}. The neural network implementation determines the contact through cumulant matching, while the analytic local calculations use the EFT $\delta(\tau_N)$ coefficient and the directly evaluated radiation contribution $\mathcal A_1$. At NNLO, recursive use of our NLO P2B construction gives the two-jet contact directly. Gaussian and polynomial machine-learning controls designed by the LLM reduce the variance of its finite radiation integral while preserving its value through exact integral compensation. The resulting predictions agree well with \texttt{EERAD3}. Machine learning therefore serves both to represent the subtraction ingredients and to improve their numerical integration. The calculations use the CPU of a 2020 M1 MacBook, without GPU acceleration, demonstrating that the construction is practical with modest local resources.

For three jets at NNLO, Appendices~\ref{app:nnlo-fixed-born}--\ref{app:local-contact} give explicit inverse maps and paired radiation weights that allow all Born momenta to remain fixed, together with a proposed finite-difference representation of the Born contact. The existing NLO matrix elements and subtraction terms are reused, while their phase-space integrations are reorganized at fixed Born momenta. Numerical checks test the inverse maps, radiation cancellation and integration at fixed Born kinematics. Appendix~\ref{app:njet-fixed-born} extends this construction to any number of massless final-state jets and gives explicit maps and measures for radiation integration at fixed Born momenta. A complete study of the NNLO extension, including the numerical determination of $\mathcal A_2$, its combination with the EFT coefficient of $\delta(\tau_3)$, and complete three-jet NNLO predictions, is left for future work.

Further progress should focus on efficient evaluation of the local contact integrals at NNLO and higher perturbative orders. This calls for phase-space maps that preserve local cancellation and importance sampling adapted to the complete subtracted weight. LLM-designed machine-learning controls provide a complementary way to reduce variance while preserving the integral. Together with larger samples and parallel evaluation, these developments could make local subtraction practical for more demanding multijet calculations. Our results show how AI could help higher-order calculations by constructing infrared subtraction and making its numerical evaluation more efficient.

\begin{acknowledgments}
We thank Xuan Chen for helpful discussions. 
This work is supported by the National Natural Science Foundation of China under Grant Nos.~12625505 and 12547109 and by the Fundamental Research Funds for the Central Universities, Beijing Normal University. X.~L. would like to thank the Erwin-Schr\"{o}dinger International Institute for Mathematics and Physics at the University of Vienna for partial support during the Programme `New Paradigms for Harnessing Quantum Field Theory at Colliders', July 27--August 28, 2026.
\end{acknowledgments}

\bibliographystyle{apsrev4-2}
\bibliography{references}

@article{Ebert:2019zkb,
    author = "Ebert, Markus A. and Tackmann, Frank J.",
    title = "{Impact of isolation and fiducial cuts on q$_{T}$ and N-jettiness subtractions}",
    eprint = "1911.08486",
    archivePrefix = "arXiv",
    primaryClass = "hep-ph",
    reportNumber = "DESY 19-199, DESY-19-199, MIT-CTP/5158",
    doi = "10.1007/JHEP03(2020)158",
    journal = "JHEP",
    volume = "03",
    pages = "158",
    year = "2020"
}

@article{Campbell:2024hjq,
    author = "Campbell, John and Neumann, Tobias and Vita, Gherardo",
    title = "{Projection-to-Born-improved subtractions at NNLO}",
    eprint = "2408.05265",
    archivePrefix = "arXiv",
    primaryClass = "hep-ph",
    reportNumber = "CERN-TH-2024-102, FERMILAB-PUB-24-0402-T",
    doi = "10.1007/JHEP05(2025)172",
    journal = "JHEP",
    volume = "05",
    pages = "172",
    year = "2025"
}

@article{NNLOJET:2025rno,
    author = "Huss, Alexander and others",
    collaboration = "NNLOJET",
    title = "{NNLOJET: A parton-level event generator for jet cross sections at NNLO QCD accuracy}",
    eprint = "2503.22804",
    archivePrefix = "arXiv",
    primaryClass = "hep-ph",
    doi = "10.21468/SciPostPhysCodeb.69",
    journal = "SciPost Phys. Codeb.",
    volume = "69",
    pages = "1",
    year = "2026"
}

@misc{Dong:2026uvw,
    author = "Dong, Liang and Fang, Shen and Gao, Jun and Li, Hai Tao and Shao, Ding Yu and Zhu, Hua Xing and Zhu, Yu Jiao",
    title = "{Two-Dimensional Transverse-Momentum Subtraction and Semi-Inclusive Deep-Inelastic Scattering at N$^3$LO in QCD}",
    eprint = "2603.29673",
    archivePrefix = "arXiv",
    primaryClass = "hep-ph",
    month = "3",
    year = "2026"
}

@article{Frixione:1995ms,
    author = "Frixione, S. and Kunszt, Z. and Signer, A.",
    title = "{Three jet cross-sections to next-to-leading order}",
    eprint = "hep-ph/9512328",
    archivePrefix = "arXiv",
    doi = "10.1016/0550-3213(96)00110-1",
    journal = "Nucl. Phys. B",
    volume = "467",
    pages = "399--442",
    year = "1996"
}

@article{Cacciari:2015jma,
    author = "Cacciari, Matteo and Dreyer, Fr{\'e}d{\'e}ric A. and Karlberg, Alexander and Salam, Gavin P. and Zanderighi, Giulia",
    title = "{Fully Differential Vector-Boson-Fusion Higgs Production at Next-to-Next-to-Leading Order}",
    eprint = "1506.02660",
    archivePrefix = "arXiv",
    primaryClass = "hep-ph",
    doi = "10.1103/PhysRevLett.115.082002",
    journal = "Phys. Rev. Lett.",
    volume = "115",
    number = "8",
    pages = "082002",
    year = "2015",
    note = "[Erratum: Phys.Rev.Lett. 120, 139901 (2018)]"
}

@article{Catani:2007vq,
    author = "Catani, Stefano and Grazzini, Massimiliano",
    title = "{An NNLO subtraction formalism in hadron collisions and its application to Higgs boson production at the LHC}",
    eprint = "hep-ph/0703012",
    archivePrefix = "arXiv",
    doi = "10.1103/PhysRevLett.98.222002",
    journal = "Phys. Rev. Lett.",
    volume = "98",
    pages = "222002",
    year = "2007"
}

@article{Catani:1996vz,
    author = "Catani, S. and Seymour, M. H.",
    title = "{A General algorithm for calculating jet cross-sections in NLO QCD}",
    eprint = "hep-ph/9605323",
    archivePrefix = "arXiv",
    doi = "10.1016/S0550-3213(96)00589-5",
    journal = "Nucl. Phys. B",
    volume = "485",
    pages = "291--419",
    year = "1997",
    note = "[Erratum: Nucl.Phys.B 510, 503--504 (1998)]"
}

@article{Gehrmann-DeRidder:2014hxk,
    author = "Gehrmann-De Ridder, A. and Gehrmann, T. and Glover, E. W. N. and Heinrich, G.",
    title = "{EERAD3: Event shapes and jet rates in electron-positron annihilation at order $\alpha_s^3$}",
    eprint = "1402.4140",
    archivePrefix = "arXiv",
    primaryClass = "hep-ph",
    doi = "10.1016/j.cpc.2014.07.024",
    journal = "Comput. Phys. Commun.",
    volume = "185",
    pages = "3331",
    year = "2014"
}

@article{Gehrmann-DeRidder:2005btv,
    author = "Gehrmann-De Ridder, A. and Gehrmann, T. and Glover, E. W. Nigel",
    title = "{Antenna subtraction at NNLO}",
    eprint = "hep-ph/0505111",
    archivePrefix = "arXiv",
    doi = "10.1088/1126-6708/2005/09/056",
    journal = "JHEP",
    volume = "2005",
    number = "09",
    pages = "056",
    year = "2005"
}

@misc{Marcoli:2026zub,
    author = "Marcoli, Matteo",
    title = "{Recent progress in antenna subtraction at NNLO and N$^3$LO}",
    booktitle = "{17th International Symposium on Radiative Corrections: Applications of Quantum Field Theory to Phenomenolog}",
    eprint = "2603.05747",
    archivePrefix = "arXiv",
    primaryClass = "hep-ph",
    month = "3",
    year = "2026"
}

@article{Caola:2017dug,
    author = {Caola, Fabrizio and Melnikov, Kirill and R{\"o}ntsch, Raoul},
    title = "{Nested soft-collinear subtractions in NNLO QCD computations}",
    eprint = "1702.01352",
    archivePrefix = "arXiv",
    primaryClass = "hep-ph",
    doi = "10.1140/epjc/s10052-017-4774-0",
    journal = "Eur. Phys. J. C",
    volume = "77",
    number = "4",
    pages = "248",
    year = "2017"
}

@misc{Bertolotti:2025clg,
    author = "Bertolotti, Gloria and Limatola, Giovanni and Torrielli, Paolo and Uccirati, Sandro",
    title = "{Advances in Local Analytic Sector Subtraction: massive NLO and elements of NNLO automation}",
    eprint = "2503.14629",
    archivePrefix = "arXiv",
    primaryClass = "hep-ph",
    month = "3",
    year = "2025"
}

@article{Boughezal:2015dva,
    author = "Boughezal, Radja and Focke, Christfried and Liu, Xiaohui and Petriello, Frank",
    title = "{$W$-boson production in association with a jet at next-to-next-to-leading order in perturbative QCD}",
    eprint = "1504.02131",
    archivePrefix = "arXiv",
    primaryClass = "hep-ph",
    doi = "10.1103/PhysRevLett.115.062002",
    journal = "Phys. Rev. Lett.",
    volume = "115",
    number = "6",
    pages = "062002",
    year = "2015"
}

@article{Czakon:2010td,
    author = "Czakon, M.",
    title = "{A novel subtraction scheme for double-real radiation at NNLO}",
    eprint = "1005.0274",
    archivePrefix = "arXiv",
    primaryClass = "hep-ph",
    doi = "10.1016/j.physletb.2010.08.036",
    journal = "Phys. Lett. B",
    volume = "693",
    pages = "259--268",
    year = "2010"
}

@article{Frederix:2009yq,
    author = "Frederix, Rikkert and Frixione, Stefano and Maltoni, Fabio and Stelzer, Tim",
    title = "{Automation of next-to-leading order computations in QCD: The FKS subtraction}",
    eprint = "0908.4272",
    archivePrefix = "arXiv",
    primaryClass = "hep-ph",
    doi = "10.1088/1126-6708/2009/10/003",
    journal = "JHEP",
    volume = "2009",
    number = "10",
    pages = "003",
    year = "2009"
}

@article{Magnea:2018hab,
    author = "Magnea, L. and Maina, E. and Pelliccioli, G. and Signorile-Signorile, C. and Torrielli, P. and Uccirati, S.",
    title = "{Local analytic sector subtraction at NNLO}",
    eprint = "1806.09570",
    archivePrefix = "arXiv",
    primaryClass = "hep-ph",
    doi = "10.1007/JHEP12(2018)107",
    journal = "JHEP",
    volume = "2018",
    number = "12",
    pages = "107",
    year = "2018",
    note = "[Erratum: JHEP 06, 013 (2019)]"
}

@article{Gaunt:2015pea,
    author = "Gaunt, Jonathan and Stahlhofen, Maximilian and Tackmann, Frank J. and Walsh, Jonathan R.",
    title = "{N-jettiness Subtractions for NNLO QCD Calculations}",
    eprint = "1505.04794",
    archivePrefix = "arXiv",
    primaryClass = "hep-ph",
    doi = "10.1007/JHEP09(2015)058",
    journal = "JHEP",
    volume = "2015",
    number = "09",
    pages = "058",
    year = "2015"
}

@book{Ellis:1996mzs,
author={Ellis, R. Keith and Stirling, W. James and Webber, Bryan R.},
title={QCD and Collider Physics},
publisher={Cambridge University Press},
year={1996},
doi={10.1017/CBO9780511628788},
note={Chap. 3, pp. 58 and 72}
}

@article{Maitre:2021uaa,
author={Ma{\^i}tre, Daniel and Truong, Henry},
title={A factorisation-aware Matrix element emulator},
eprint={2107.06625},
archivePrefix={arXiv},
primaryClass={hep-ph},
doi={10.1007/JHEP11(2021)066},
journal={JHEP},
volume={2021},
number={11},
pages={066},
year={2021}
}

@article{Currie:2018fgr,
author={Currie, J. and Gehrmann, T. and Glover, E. W. N. and Huss, A. and Niehues, J. and Vogt, A.},
title={N$^3$LO Corrections to Jet Production in Deep Inelastic Scattering using the Projection-to-Born Method},
eprint={1803.09973},
archivePrefix={arXiv},
primaryClass={hep-ph},
doi={10.1007/JHEP05(2018)209},
journal={JHEP},
volume={2018},
number={05},
pages={209},
year={2018}
}

@article{Ellis:1980wv,
author={Ellis, R. Keith and Ross, D. A. and Terrano, A. E.},
title={The Perturbative Calculation of Jet Structure in $e^+e^-$ Annihilation},
journal={Nucl. Phys. B},
volume={178},
pages={421--456},
year={1981},
doi={10.1016/0550-3213(81)90165-6}
}

@article{Giele:1993dj,
    author = "Giele, W. T. and Glover, E. W. Nigel and Kosower, David A.",
    title = "{Higher order corrections to jet cross-sections in hadron colliders}",
    eprint = "hep-ph/9302225",
    archivePrefix = "arXiv",
    reportNumber = "FERMILAB-PUB-92-230-T, DTP-92-64, CERN-TH-6750-92",
    doi = "10.1016/0550-3213(93)90365-V",
    journal = "Nucl. Phys. B",
    volume = "403",
    pages = "633--670",
    year = "1993"
}

@article{Harris:2001sx,
    author = "Harris, B. W. and Owens, J. F.",
    title = "{The Two cutoff phase space slicing method}",
    eprint = "hep-ph/0102128",
    archivePrefix = "arXiv",
    reportNumber = "ANL-HEP-PR-00-044, FSU-HEP-010130",
    doi = "10.1103/PhysRevD.65.094032",
    journal = "Phys. Rev. D",
    volume = "65",
    pages = "094032",
    year = "2002"
}

@article{Anastasiou:2003gr,
    author = "Anastasiou, Charalampos and Melnikov, Kirill and Petriello, Frank",
    title = "{A new method for real radiation at NNLO}",
    eprint = "hep-ph/0311311",
    archivePrefix = "arXiv",
    reportNumber = "SLAC-PUB-10252, UH-511-1039-03",
    doi = "10.1103/PhysRevD.69.076010",
    journal = "Phys. Rev. D",
    volume = "69",
    pages = "076010",
    year = "2004"
}

@article{Gao:2012ja,
    author = "Gao, Jun and Li, Chong Sheng and Zhu, Hua Xing",
    title = "{Top Quark Decay at Next-to-Next-to Leading Order in QCD}",
    eprint = "1210.2808",
    archivePrefix = "arXiv",
    primaryClass = "hep-ph",
    reportNumber = "SLAC-PUB-15260",
    doi = "10.1103/PhysRevLett.110.042001",
    journal = "Phys. Rev. Lett.",
    volume = "110",
    number = "4",
    pages = "042001",
    year = "2013"
}

@article{Stewart:2010tn,
    author = "Stewart, Iain W. and Tackmann, Frank J. and Waalewijn, Wouter J.",
    title = "{N-Jettiness: An Inclusive Event Shape to Veto Jets}",
    eprint = "1004.2489",
    archivePrefix = "arXiv",
    primaryClass = "hep-ph",
    reportNumber = "MIT-CTP-4139",
    doi = "10.1103/PhysRevLett.105.092002",
    journal = "Phys. Rev. Lett.",
    volume = "105",
    pages = "092002",
    year = "2010"
}

@article{Camarda:2021ict,
    author = "Camarda, Stefano and Cieri, Leandro and Ferrera, Giancarlo",
    title = "{Drell{\textendash}Yan lepton-pair production: qT resummation at N3LL accuracy and fiducial cross sections at N3LO}",
    eprint = "2103.04974",
    archivePrefix = "arXiv",
    primaryClass = "hep-ph",
    doi = "10.1103/PhysRevD.104.L111503",
    journal = "Phys. Rev. D",
    volume = "104",
    number = "11",
    pages = "L111503",
    year = "2021"
}

@article{Chen:2022lwc,
    author = "Chen, Xuan and Gehrmann, Thomas and Glover, Nigel and Huss, Alexander and Yang, Tong-Zhi and Zhu, Hua Xing",
    title = "{Transverse mass distribution and charge asymmetry in W boson production to third order in QCD}",
    eprint = "2205.11426",
    archivePrefix = "arXiv",
    primaryClass = "hep-ph",
    reportNumber = "IPPP/22/32, ZU-TH 19/22, P3H-22-055, KA-TP-15-2022, CERN-TH-2022-083",
    doi = "10.1016/j.physletb.2023.137876",
    journal = "Phys. Lett. B",
    volume = "840",
    pages = "137876",
    year = "2023"
}

@article{Chen:2021isd,
    author = "Chen, X. and Gehrmann, T. and Glover, E. W. N. and Huss, A. and Mistlberger, B. and Pelloni, A.",
    title = "{Fully Differential Higgs Boson Production to Third Order in QCD}",
    eprint = "2102.07607",
    archivePrefix = "arXiv",
    primaryClass = "hep-ph",
    reportNumber = "CERN-TH-2021-021, IPPP/20/80, KA-TP-25-2021, SLAC-PUB-17588, NIKHEF 2021-005,
  SLAC-PUB-17588, ZU-TH 07/21",
    doi = "10.1103/PhysRevLett.127.072002",
    journal = "Phys. Rev. Lett.",
    volume = "127",
    number = "7",
    pages = "072002",
    year = "2021"
}

@article{Cieri:2018oms,
    author = "Cieri, Leandro and Chen, Xuan and Gehrmann, Thomas and Glover, E. W. N. and Huss, Alexander",
    title = "{Higgs boson production at the LHC using the $q_T$ subtraction formalism at N$^3$LO QCD}",
    eprint = "1807.11501",
    archivePrefix = "arXiv",
    primaryClass = "hep-ph",
    reportNumber = "ZU-TH 27/18, IPPP/18/67, CERN-TH-2018-167",
    doi = "10.1007/JHEP02(2019)096",
    journal = "JHEP",
    volume = "2019",
    number = "02",
    pages = "096",
    year = "2019"
}

@article{Chen:2025kez,
    author = "Chen, Xuan and Jakub{\v{c}}{\'\i}k, Petr and Marcoli, Matteo and Stagnitto, Giovanni",
    title = "{Jet production at electron-positron colliders at next-to-next-to-next-to-leading order in QCD}",
    eprint = "2505.10618",
    archivePrefix = "arXiv",
    primaryClass = "hep-ph",
    reportNumber = "IPPP/25/27, ZU-TH 36/25",
    doi = "10.1016/j.physletb.2025.139804",
    journal = "Phys. Lett. B",
    volume = "869",
    pages = "139804",
    year = "2025"
}

@article{Gehrmann:2018odt,
    author = "Gehrmann, T. and Huss, A. and Niehues, J. and Vogt, A. and Walker, D. M.",
    title = "{Jet production in charged-current deep-inelastic scattering to third order in QCD}",
    eprint = "1812.06104",
    archivePrefix = "arXiv",
    primaryClass = "hep-ph",
    reportNumber = "CERN-TH-2018-272, IPPP/18/108, ZU-TH 45/18, LTH 1188, IPPP-18-108, ZU-TH-45-18, LTH-1188",
    doi = "10.1016/j.physletb.2019.03.003",
    journal = "Phys. Lett. B",
    volume = "792",
    pages = "182--186",
    year = "2019"
}

@article{Chen:2026zmi,
    author = "Chen, Xuan and Dai, Yuesheng and Li, Hai Tao and Li, Shi-Yuan and Shao, Hua-Sheng and Wang, Jian",
    title = "{Fully differential Higgs boson pair production at N$^{3}$LO with top quark mass effects}",
    eprint = "2601.19990",
    archivePrefix = "arXiv",
    primaryClass = "hep-ph",
    doi = "10.1007/JHEP06(2026)005",
    journal = "JHEP",
    volume = "2026",
    number = "06",
    pages = "005",
    year = "2026"
}

@article{Neumann:2022lft,
    author = "Neumann, Tobias and Campbell, John",
    title = "{Fiducial Drell-Yan production at the LHC improved by transverse-momentum resummation at N4LLp+N3LO}",
    eprint = "2207.07056",
    archivePrefix = "arXiv",
    primaryClass = "hep-ph",
    reportNumber = "FERMILAB-PUB-22-528-T",
    doi = "10.1103/PhysRevD.107.L011506",
    journal = "Phys. Rev. D",
    volume = "107",
    number = "1",
    pages = "L011506",
    year = "2023"
}

@article{Billis:2021ecs,
    author = "Billis, Georgios and Dehnadi, Bahman and Ebert, Markus A. and Michel, Johannes K. L. and Tackmann, Frank J.",
    title = "{Higgs pT Spectrum and Total Cross Section with Fiducial Cuts at Third Resummed and Fixed Order in QCD}",
    eprint = "2102.08039",
    archivePrefix = "arXiv",
    primaryClass = "hep-ph",
    reportNumber = "DESY 21-022, DESY-21-022, MPP-2021-16, MIT-CTP 5266",
    doi = "10.1103/PhysRevLett.127.072001",
    journal = "Phys. Rev. Lett.",
    volume = "127",
    number = "7",
    pages = "072001",
    year = "2021"
}

@article{Chen:2021vtu,
    author = "Chen, Xuan and Gehrmann, Thomas and Glover, Nigel and Huss, Alexander and Yang, Tong-Zhi and Zhu, Hua Xing",
    title = "{Dilepton Rapidity Distribution in Drell-Yan Production to Third Order in QCD}",
    eprint = "2107.09085",
    archivePrefix = "arXiv",
    primaryClass = "hep-ph",
    reportNumber = "KA-TP-17-2021, ZU-TH 33/21, CERN-TH-2021-110, IPPP/21/13",
    doi = "10.1103/PhysRevLett.128.052001",
    journal = "Phys. Rev. Lett.",
    volume = "128",
    number = "5",
    pages = "052001",
    year = "2022"
}

@article{Mondini:2019gid,
    author = "Mondini, Roberto and Schiavi, Matthew and Williams, Ciaran",
    title = "{N$^{3}$LO predictions for the decay of the Higgs boson to bottom quarks}",
    eprint = "1904.08960",
    archivePrefix = "arXiv",
    primaryClass = "hep-ph",
    doi = "10.1007/JHEP06(2019)079",
    journal = "JHEP",
    volume = "2019",
    number = "06",
    pages = "079",
    year = "2019"
}

@article{Chen:2022cgv,
    author = "Chen, Xuan and Gehrmann, Thomas and Glover, E. W. N. and Huss, Alexander and Monni, Pier Francesco and Re, Emanuele and Rottoli, Luca and Torrielli, Paolo",
    title = "{Third-Order Fiducial Predictions for Drell-Yan Production at the LHC}",
    eprint = "2203.01565",
    archivePrefix = "arXiv",
    primaryClass = "hep-ph",
    reportNumber = "CERN-TH-2022-023, IPPP/22/08, LAPTH-009/22, KA-TP-03-2022,
  P3H-22-022, ZU-TH 07/22",
    doi = "10.1103/PhysRevLett.128.252001",
    journal = "Phys. Rev. Lett.",
    volume = "128",
    number = "25",
    pages = "252001",
    year = "2022"
}

@article{Baranowski:2024vxg,
    author = "Baranowski, Daniel and Delto, Maximilian and Melnikov, Kirill and Pikelner, Andrey and Wang, Chen-Yu",
    title = "{Zero-Jettiness Soft Function to Third Order in Perturbative QCD}",
    eprint = "2409.11042",
    archivePrefix = "arXiv",
    primaryClass = "hep-ph",
    reportNumber = "TTP24-033,P3H-24-061,TUM-HEP-1526/24,MPP-2024-179,ZU-TH 45/24",
    doi = "10.1103/PhysRevLett.134.191902",
    journal = "Phys. Rev. Lett.",
    volume = "134",
    number = "19",
    pages = "191902",
    year = "2025"
}

@article{Albertsson:2018maf,
    author = "Albertsson, Kim and others",
    title = "{Machine Learning in High Energy Physics Community White Paper}",
    eprint = "1807.02876",
    archivePrefix = "arXiv",
    primaryClass = "physics.comp-ph",
    reportNumber = "FERMILAB-PUB-18-318-CD-DI-PPD",
    doi = "10.1088/1742-6596/1085/2/022008",
    journal = "J. Phys. Conf. Ser.",
    volume = "1085",
    number = "2",
    pages = "022008",
    year = "2018"
}

@misc{Diefenbacher:2025zzn,
    author = {Diefenbacher, Sascha and Hallin, Anna and Kasieczka, Gregor and Kr{\"a}mer, Michael and Lauscher, Anne and Lukas, Tim},
    title = "{Agents of Discovery}",
    eprint = "2509.08535",
    archivePrefix = "arXiv",
    primaryClass = "hep-ph",
    month = "9",
    year = "2025"
}

@misc{Menzo:2025cim,
    author = {Menzo, Tony and Roman, Alexander and Gleyzer, Sergei and Matchev, Konstantin and Fleming, George T. and H{\"o}che, Stefan and Mrenna, Stephen and Shyamsundar, Prasanth},
    title = "{HEPTAPOD: Orchestrating High Energy Physics Workflows Towards Autonomous Agency}",
    eprint = "2512.15867",
    archivePrefix = "arXiv",
    primaryClass = "hep-ph",
    reportNumber = "FERMILAB-PUB-25-0923-CSAID-ETD-T",
    month = "12",
    year = "2025"
}

@article{Calisto:2023vmm,
    author = "Calisto, Francesco and Moodie, Ryan and Zoia, Simone",
    title = "{Learning Feynman integrals from differential equations with neural networks}",
    eprint = "2312.02067",
    archivePrefix = "arXiv",
    primaryClass = "hep-ph",
    reportNumber = "CERN-TH-2023-225",
    doi = "10.1007/JHEP07(2024)124",
    journal = "JHEP",
    volume = "2024",
    number = "07",
    pages = "124",
    year = "2024"
}

@article{vonHippel:2025okr,
    author = "von Hippel, Matt and Wilhelm, Matthias",
    title = "{Refining Integration-by-Parts Reduction of Feynman Integrals with Machine Learning}",
    eprint = "2502.05121",
    archivePrefix = "arXiv",
    primaryClass = "hep-th",
    doi = "10.1007/JHEP05(2025)185",
    journal = "JHEP",
    volume = "2025",
    number = "05",
    pages = "185",
    year = "2025"
}

@article{Song:2025pwy,
    author = "Song, Zhuo-Yang and Yang, Tong-Zhi and Cao, Qing-Hong and Luo, Ming-xing and Zhu, Hua Xing",
    title = "{Explainable AI-assisted optimization for Feynman integral reduction}",
    eprint = "2502.09544",
    archivePrefix = "arXiv",
    primaryClass = "hep-ph",
    reportNumber = "ZU-TH 07/25",
    doi = "10.1007/JHEP06(2026)225",
    journal = "JHEP",
    volume = "2026",
    number = "06",
    pages = "225",
    year = "2026"
}

@article{Butter:2019eyo,
    author = "Butter, Anja and Plehn, Tilman and Winterhalder, Ramon",
    title = "{How to GAN Event Subtraction}",
    eprint = "1912.08824",
    archivePrefix = "arXiv",
    primaryClass = "hep-ph",
    doi = "10.21468/SciPostPhysCore.3.2.009",
    journal = "SciPost Phys. Core",
    volume = "3",
    pages = "009",
    year = "2020"
}

@article{Janssen:2025zke,
    author = "Jan{\ss}en, Timo and Poncelet, Rene and Schumann, Steffen",
    title = "{Sampling NNLO QCD phase space with normalizing flows}",
    eprint = "2505.13608",
    archivePrefix = "arXiv",
    primaryClass = "hep-ph",
    reportNumber = "IFJPAN-IV-2025-11, MCNET-25-11, COMETA-2025-22",
    doi = "10.1007/JHEP09(2025)194",
    journal = "JHEP",
    volume = "2025",
    number = "09",
    pages = "194",
    year = "2025"
}

@misc{Ellenberg:2025umt,
    author = "Ellenberg, Jordan S. and Fraser-Taliente, Cristofero S. and Harvey, Thomas R. and Srivastava, Karan and Sutherland, Andrew V.",
    title = "{Generative Modeling for Mathematical Discovery}",
    eprint = "2503.11061",
    archivePrefix = "arXiv",
    primaryClass = "cs.LG",
    month = "3",
    year = "2025"
}

@article{Cybenko:1989iql,
    author = "Cybenko, G.",
    title = "{Approximation by superpositions of a sigmoidal function}",
    doi = "10.1007/BF02551274",
    journal = "Math. Control Signals Syst.",
    volume = "2",
    number = "4",
    pages = "303--314",
    year = "1989"
}

@article{Gehrmann-DeRidder:2009fgd,
    author = "Gehrmann-De Ridder, A. and Gehrmann, T. and Glover, E. W. N. and Heinrich, G.",
    title = "{NNLO moments of event shapes in e+e- annihilation}",
    eprint = "0903.4658",
    archivePrefix = "arXiv",
    primaryClass = "hep-ph",
    reportNumber = "ZU-TH-04-09, IPPP-09-15",
    doi = "10.1088/1126-6708/2009/05/106",
    journal = "JHEP",
    volume = "2009",
    number = "05",
    pages = "106",
    year = "2009"
}

@article{Kinoshita:1962ur,
    author = "Kinoshita, T.",
    title = "{Mass singularities of Feynman amplitudes}",
    doi = "10.1063/1.1724268",
    journal = "J. Math. Phys.",
    volume = "3",
    pages = "650--677",
    year = "1962"
}

@article{Lee:1964is,
    author = "Lee, T. D. and Nauenberg, M.",
    title = "{Degenerate Systems and Mass Singularities}",
    doi = "10.1103/PhysRev.133.B1549",
    journal = "Phys. Rev.",
    volume = "133",
    pages = "B1549--B1562",
    year = "1964"
}

@article{Sterman:1977wj,
    author = "Sterman, George F. and Weinberg, Steven",
    title = "{Jets from Quantum Chromodynamics}",
    reportNumber = "HUTP-77/A044",
    doi = "10.1103/PhysRevLett.39.1436",
    journal = "Phys. Rev. Lett.",
    volume = "39",
    pages = "1436",
    year = "1977"
}

@article{Bauer:2000ew,
    author = "Bauer, Christian W. and Fleming, Sean and Luke, Michael E.",
    title = "{Summing Sudakov logarithms in $B \to  X_s \gamma $in effective field theory.}",
    eprint = "hep-ph/0005275",
    archivePrefix = "arXiv",
    reportNumber = "UTPT-00-03",
    doi = "10.1103/PhysRevD.63.014006",
    journal = "Phys. Rev. D",
    volume = "63",
    pages = "014006",
    year = "2000"
}

@article{Bauer:2000yr,
    author = "Bauer, Christian W. and Fleming, Sean and Pirjol, Dan and Stewart, Iain W.",
    title = "{An Effective field theory for collinear and soft gluons: Heavy to light decays}",
    eprint = "hep-ph/0011336",
    archivePrefix = "arXiv",
    reportNumber = "UCSD-PTH-00-28",
    doi = "10.1103/PhysRevD.63.114020",
    journal = "Phys. Rev. D",
    volume = "63",
    pages = "114020",
    year = "2001"
}

@article{Bauer:2001ct,
    author = "Bauer, Christian W. and Stewart, Iain W.",
    title = "{Invariant operators in collinear effective theory}",
    eprint = "hep-ph/0107001",
    archivePrefix = "arXiv",
    reportNumber = "UCSD-PTH-01-09",
    doi = "10.1016/S0370-2693(01)00902-9",
    journal = "Phys. Lett. B",
    volume = "516",
    pages = "134--142",
    year = "2001"
}

@article{Bauer:2001yt,
    author = "Bauer, Christian W. and Pirjol, Dan and Stewart, Iain W.",
    title = "{Soft collinear factorization in effective field theory}",
    eprint = "hep-ph/0109045",
    archivePrefix = "arXiv",
    reportNumber = "UCSD-PTH-01-15",
    doi = "10.1103/PhysRevD.65.054022",
    journal = "Phys. Rev. D",
    volume = "65",
    pages = "054022",
    year = "2002"
}

@article{Beneke:2002ph,
    author = "Beneke, M. and Chapovsky, A. P. and Diehl, M. and Feldmann, T.",
    title = "{Soft collinear effective theory and heavy to light currents beyond leading power}",
    eprint = "hep-ph/0206152",
    archivePrefix = "arXiv",
    reportNumber = "PITHA-02-09",
    doi = "10.1016/S0550-3213(02)00687-9",
    journal = "Nucl. Phys. B",
    volume = "643",
    pages = "431--476",
    year = "2002"
}

@article{Jouttenus:2011wh,
    author = "Jouttenus, Teppo T. and Stewart, Iain W. and Tackmann, Frank J. and Waalewijn, Wouter J.",
    title = "{The Soft Function for Exclusive N-Jet Production at Hadron Colliders}",
    eprint = "1102.4344",
    archivePrefix = "arXiv",
    primaryClass = "hep-ph",
    doi = "10.1103/PhysRevD.83.114030",
    journal = "Phys. Rev. D",
    volume = "83",
    pages = "114030",
    year = "2011"
}

@article{Alwall:2014hca,
    author = "Alwall, J. and Frederix, R. and Frixione, S. and Hirschi, V. and Maltoni, F. and Mattelaer, O. and Shao, H. -S. and Stelzer, T. and Torrielli, P. and Zaro, M.",
    title = "{The automated computation of tree-level and next-to-leading order differential cross sections, and their matching to parton shower simulations}",
    eprint = "1405.0301",
    archivePrefix = "arXiv",
    primaryClass = "hep-ph",
    reportNumber = "CERN-PH-TH-2014-064, CP3-14-18, LPN14-066, MCNET-14-09, ZU-TH-14-14",
    doi = "10.1007/JHEP07(2014)079",
    journal = "JHEP",
    volume = "07",
    pages = "079",
    year = "2014"
}

@article{Sherpa:2019gpd,
    author = "Bothmann, Enrico and others",
    collaboration = "Sherpa",
    title = "{Event Generation with Sherpa 2.2}",
    eprint = "1905.09127",
    archivePrefix = "arXiv",
    primaryClass = "hep-ph",
    reportNumber = "FERMILAB-PUB-19-218-T, SLAC-PUB-17433, IPPP/19/42, MCNET-19-11",
    doi = "10.21468/SciPostPhys.7.3.034",
    journal = "SciPost Phys.",
    volume = "7",
    number = "3",
    pages = "034",
    year = "2019"
}

@article{Larkoski:2014uqa,
    author = "Larkoski, Andrew J. and Neill, Duff and Thaler, Jesse",
    title = "{Jet Shapes with the Broadening Axis}",
    eprint = "1401.2158",
    archivePrefix = "arXiv",
    primaryClass = "hep-ph",
    reportNumber = "MIT--CTP-4512",
    doi = "10.1007/JHEP04(2014)017",
    journal = "JHEP",
    volume = "04",
    pages = "017",
    year = "2014"
}

@article{JADE:1986kta,
    author = "Bartel, W. and others",
    collaboration = "JADE",
    title = "{Experimental Studies on Multi-Jet Production in e+ e- Annihilation at PETRA Energies}",
    reportNumber = "DESY-86-086",
    doi = "10.1007/BF01410449",
    journal = "Z. Phys. C",
    volume = "33",
    pages = "23",
    year = "1986"
}

@article{Farhi:1977sg,
    author = "Farhi, Edward",
    title = "{A QCD Test for Jets}",
    reportNumber = "HUTP-77-A059",
    doi = "10.1103/PhysRevLett.39.1587",
    journal = "Phys. Rev. Lett.",
    volume = "39",
    pages = "1587--1588",
    year = "1977"
}

@article{Basham:1978bw,
    author = "Basham, C. Louis and Brown, Lowell S. and Ellis, Stephen D. and Love, Sherwin T.",
    title = "{Energy Correlations in electron - Positron Annihilation: Testing QCD}",
    reportNumber = "RLO-1388-759",
    doi = "10.1103/PhysRevLett.41.1585",
    journal = "Phys. Rev. Lett.",
    volume = "41",
    pages = "1585",
    year = "1978"
}

@article{Catani:1992jc,
    author = "Catani, S. and Turnock, G. and Webber, B. R.",
    title = "{Jet broadening measures in $e^{+} e^{-}$ annihilation}",
    reportNumber = "CERN-TH-6570-92",
    doi = "10.1016/0370-2693(92)91565-Q",
    journal = "Phys. Lett. B",
    volume = "295",
    pages = "269--276",
    year = "1992"
}

@article{Lepage:1977sw,
    author = "Lepage, G. Peter",
    title = "{A New Algorithm for Adaptive Multidimensional Integration}",
    reportNumber = "SLAC-PUB-1839-REV, SLAC-PUB-1839",
    doi = "10.1016/0021-9991(78)90004-9",
    journal = "J. Comput. Phys.",
    volume = "27",
    pages = "192",
    year = "1978"
}

@article{Bell:2018oqa,
    author = "Bell, Guido and Rahn, Rudi and Talbert, Jim",
    title = "{Generic dijet soft functions at two-loop order: correlated emissions}",
    eprint = "1812.08690",
    archivePrefix = "arXiv",
    primaryClass = "hep-ph",
    reportNumber = "DESY-18-209",
    doi = "10.1007/JHEP07(2019)101",
    journal = "JHEP",
    volume = "2019",
    number = "07",
    pages = "101",
    year = "2019"
}

@article{Bell:2022tmi,
    author = "Bell, Guido and Brune, Kevin and Das, Goutam and Wald, Marcel",
    title = "{Automated Calculation of Beam Functions at NNLO}",
    eprint = "2208.04847",
    archivePrefix = "arXiv",
    primaryClass = "hep-ph",
    reportNumber = "SI-HEP-2022-18, P3H-22-085",
    doi = "10.22323/1.416.0026",
    journal = "PoS",
    volume = "LL2022",
    pages = "026",
    year = "2022"
}

@article{Bruser:2018rad,
    author = {Br{\"u}ser, Robin and Liu, Ze Long and Stahlhofen, Maximilian},
    title = "{Three-Loop Quark Jet Function}",
    eprint = "1804.09722",
    archivePrefix = "arXiv",
    primaryClass = "hep-ph",
    reportNumber = "MITP/18-031, MITP-18-031",
    doi = "10.1103/PhysRevLett.121.072003",
    journal = "Phys. Rev. Lett.",
    volume = "121",
    number = "7",
    pages = "072003",
    year = "2018"
}

@article{Banerjee:2018ozf,
    author = "Banerjee, Pulak and Dhani, Prasanna K. and Ravindran, V.",
    title = "{Gluon jet function at three loops in QCD}",
    eprint = "1805.02637",
    archivePrefix = "arXiv",
    primaryClass = "hep-ph",
    reportNumber = "IMSc/2018/05/04",
    doi = "10.1103/PhysRevD.98.094016",
    journal = "Phys. Rev. D",
    volume = "98",
    number = "9",
    pages = "094016",
    year = "2018"
}

@article{Baranowski:2022vcn,
    author = "Baranowski, Daniel and Behring, Arnd and Melnikov, Kirill and Tancredi, Lorenzo and Wever, Christopher",
    title = "{Beam functions for N-jettiness at N$^{3}$LO in perturbative QCD}",
    eprint = "2211.05722",
    archivePrefix = "arXiv",
    primaryClass = "hep-ph",
    reportNumber = "TTP22-067, P3H-22-108, TUM-HEP-1429/22, CERN-TH-2022-178, ZU-TH
  51/22",
    doi = "10.1007/JHEP02(2023)073",
    journal = "JHEP",
    volume = "2023",
    number = "02",
    pages = "073",
    year = "2023"
}

@misc{Bell:2026pou,
    author = "Bell, Guido and Broggio, Alessandro and Edelmann, Sebastian and Lim, Matthew A. and Rahn, Rudi",
    title = "{NNLO soft functions for heavy-quark pair production at hadron colliders}",
    eprint = "2608.09621",
    archivePrefix = "arXiv",
    primaryClass = "hep-ph",
    reportNumber = "UWThPh 2026-7, SI-HEP-2026-17, P3H-26-061",
    month = "8",
    year = "2026"
}

@article{Ebert:2020unb,
    author = "Ebert, Markus A. and Mistlberger, Bernhard and Vita, Gherardo",
    title = "{$N$-jettiness beam functions at N$^{3}$LO}",
    eprint = "2006.03056",
    archivePrefix = "arXiv",
    primaryClass = "hep-ph",
    reportNumber = "MIT-CTP 5208",
    doi = "10.1007/JHEP09(2020)143",
    journal = "JHEP",
    volume = "2020",
    number = "09",
    pages = "143",
    year = "2020"
}

@article{Baranowski:2024ene,
    author = "Baranowski, Daniel and Delto, Maximilian and Melnikov, Kirill and Pikelner, Andrey and Wang, Chen-Yu",
    title = "{One-loop corrections to the double-real emission contribution to the zero-jettiness soft function at N3LO in QCD}",
    eprint = "2401.05245",
    archivePrefix = "arXiv",
    primaryClass = "hep-ph",
    reportNumber = "MPP-2024-2,P3H-24-003,TTP24-001,TUM-HEP-1493/24,ZU-TH 03/24",
    doi = "10.1007/JHEP04(2024)114",
    journal = "JHEP",
    volume = "2024",
    number = "04",
    pages = "114",
    year = "2024"
}

@article{Chen:2022yre,
    author = "Chen, Wen and Feng, Feng and Jia, Yu and Liu, Xiaohui",
    title = "{Double-real-virtual and double-virtual-real corrections to the three-loop thrust soft function}",
    eprint = "2206.12323",
    archivePrefix = "arXiv",
    primaryClass = "hep-ph",
    doi = "10.1007/JHEP12(2022)094",
    journal = "JHEP",
    volume = "2022",
    number = "12",
    pages = "094",
    year = "2022"
}

@article{Bell:2023yso,
    author = "Bell, Guido and Dehnadi, Bahman and Mohrmann, Tobias and Rahn, Rudi",
    title = "{The NNLO soft function for N-jettiness in hadronic collisions}",
    eprint = "2312.11626",
    archivePrefix = "arXiv",
    primaryClass = "hep-ph",
    reportNumber = "SI-HEP-2022-20, P3H-22-065, DESY-23-205",
    doi = "10.1007/JHEP07(2024)077",
    journal = "JHEP",
    volume = "2024",
    number = "07",
    pages = "077",
    year = "2024"
}

@article{Boughezal:2015eha,
    author = "Boughezal, Radja and Liu, Xiaohui and Petriello, Frank",
    title = "{$N$-jettiness soft function at next-to-next-to-leading order}",
    eprint = "1504.02540",
    archivePrefix = "arXiv",
    primaryClass = "hep-ph",
    doi = "10.1103/PhysRevD.91.094035",
    journal = "Phys. Rev. D",
    volume = "91",
    number = "9",
    pages = "094035",
    year = "2015"
}

@article{Luo:2019szz,
    author = "Luo, Ming-xing and Yang, Tong-Zhi and Zhu, Hua Xing and Zhu, Yu Jiao",
    title = "{Quark Transverse Parton Distribution at the Next-to-Next-to-Next-to-Leading Order}",
    eprint = "1912.05778",
    archivePrefix = "arXiv",
    primaryClass = "hep-ph",
    doi = "10.1103/PhysRevLett.124.092001",
    journal = "Phys. Rev. Lett.",
    volume = "124",
    number = "9",
    pages = "092001",
    year = "2020"
}

@article{Ebert:2020yqt,
    author = "Ebert, Markus A. and Mistlberger, Bernhard and Vita, Gherardo",
    title = "{Transverse momentum dependent PDFs at N$^3$LO}",
    eprint = "2006.05329",
    archivePrefix = "arXiv",
    primaryClass = "hep-ph",
    reportNumber = "MIT-CTP/5209, SLAC-PUB-17535",
    doi = "10.1007/JHEP09(2020)146",
    journal = "JHEP",
    volume = "2020",
    number = "09",
    pages = "146",
    year = "2020"
}

@article{Luo:2020epw,
    author = "Luo, Ming-xing and Yang, Tong-Zhi and Zhu, Hua Xing and Zhu, Yu Jiao",
    title = "{Unpolarized quark and gluon TMD PDFs and FFs at N$^{3}$LO}",
    eprint = "2012.03256",
    archivePrefix = "arXiv",
    primaryClass = "hep-ph",
    doi = "10.1007/JHEP06(2021)115",
    journal = "JHEP",
    volume = "2021",
    number = "06",
    pages = "115",
    year = "2021"
}

@article{Ebert:2020qef,
    author = "Ebert, Markus A. and Mistlberger, Bernhard and Vita, Gherardo",
    title = "{TMD fragmentation functions at N$^{3}$LO}",
    eprint = "2012.07853",
    archivePrefix = "arXiv",
    primaryClass = "hep-ph",
    reportNumber = "MIT-CTP 5261, SLAC-PUB-17577, MPP-2020-220",
    doi = "10.1007/JHEP07(2021)121",
    journal = "JHEP",
    volume = "2021",
    number = "07",
    pages = "121",
    year = "2021"
}

@misc{Liu:2025tje,
    author = "Liu, Yuanche and Xu, Yingxuan and Zhang, Yang",
    title = "{Uncovering Singularities in Feynman Integrals via Machine Learning}",
    eprint = "2510.10099",
    archivePrefix = "arXiv",
    primaryClass = "hep-ph",
    reportNumber = "USTC-ICTS/PCFT-25-40, P3H-25-073, TTP25-034",
    month = "10",
    year = "2025"
}

@misc{Gao:2026cpd,
    author = "Gao, Haofei and others",
    title = "{LQCDMaster: Agentic Scientific Computing for Lattice Quantum Chromodynamics Research}",
    eprint = "2607.15001",
    archivePrefix = "arXiv",
    primaryClass = "hep-lat",
    month = "7",
    year = "2026"
}

@misc{Ulam:2026Jacobian,
    title = {A Counterexample to the {Jacobian} Conjecture},
    howpublished = {Technical note},
    url = {https://ulam.ai/research/jacobian.pdf},
    month = jul,
    year = {2026}
}

@misc{OpenAI:2026NavierStokes,
    author = {{OpenAI}},
    title = {Finite Time Blowup for {Navier--Stokes}},
    url = {https://cdn.openai.com/pdf/32d9f210-8b73-45e0-91bc-82a30aef8a9a/navier-stokes.pdf},
    month = sep,
    year = {2026}
}

@article{Aveleira:2025svg,
    author = "Aveleira, Benjamin Campillo and Gehrmann-De Ridder, Aude and Gehrmann, Thomas and Glover, Nigel and Heinrich, Gudrun and Preuss, Christian Tobias",
    title = "{EERAD3 version 2: QCD corrections in hadronic colour-singlet decays}",
    eprint = "2503.20610",
    archivePrefix = "arXiv",
    primaryClass = "hep-ph",
    reportNumber = "IPPP/25/18, P3H-25-023, KA-TP-09-2025, MCNET-25-05, ZU-TH 20/25",
    doi = "10.21468/SciPostPhysCodeb.59",
    journal = "SciPost Phys. Codeb.",
    volume = "59",
    pages = "1",
    year = "2025"
}

@article{Chen:2026jxf,
    author = "Chen, Xuan and Chicherin, Dmitry and Fox, Elliot and Glover, Nigel and Marcoli, Matteo and Sotnikov, Vasily and Sun, Huiting and Zhang, Hantian and Zoia, Simone",
    title = "{Four-Jet Rate in Electron-Positron Annihilation at Order $\alpha_s^4$}",
    eprint = "2602.18185",
    archivePrefix = "arXiv",
    primaryClass = "hep-ph",
    reportNumber = "CERN-TH-2026-017, IPPP/26/12, ZU-TH 04/26, LAPTH-007/26",
    doi = "10.1103/xxg6-yd47",
    journal = "Phys. Rev. Lett.",
    volume = "136",
    number = "25",
    pages = "251901",
    year = "2026"
}

@article{Becher:2006qw,
    author = "Becher, Thomas and Neubert, Matthias",
    title = "{Toward a NNLO calculation of the anti-B ---{\ensuremath{>}} X(s) gamma decay rate with a cut on photon energy. II. Two-loop result for the jet function}",
    eprint = "hep-ph/0603140",
    archivePrefix = "arXiv",
    reportNumber = "CLNS-06-1954, FERMILAB-PUB-06-057-T",
    doi = "10.1016/j.physletb.2006.04.046",
    journal = "Phys. Lett. B",
    volume = "637",
    pages = "251--259",
    year = "2006"
}

@article{Becher:2010pd,
    author = "Becher, Thomas and Bell, Guido",
    title = "{The gluon jet function at two-loop order}",
    eprint = "1008.1936",
    archivePrefix = "arXiv",
    primaryClass = "hep-ph",
    doi = "10.1016/j.physletb.2010.11.036",
    journal = "Phys. Lett. B",
    volume = "695",
    pages = "252--258",
    year = "2011"
}

@article{Becher:2008cf,
    author = "Becher, Thomas and Schwartz, Matthew D.",
    title = "{A precise determination of $\alpha_s$ from LEP thrust data using effective field theory}",
    eprint = "0803.0342",
    archivePrefix = "arXiv",
    primaryClass = "hep-ph",
    reportNumber = "FERMILAB-PUB-08-048-T",
    doi = "10.1088/1126-6708/2008/07/034",
    journal = "JHEP",
    volume = "07",
    pages = "034",
    year = "2008"
}

@article{Kelley:2011ng,
    author = "Kelley, Randall and Schwartz, Matthew D. and Schabinger, Robert M. and Zhu, Hua Xing",
    title = "{The two-loop hemisphere soft function}",
    eprint = "1105.3676",
    archivePrefix = "arXiv",
    primaryClass = "hep-ph",
    doi = "10.1103/PhysRevD.84.045022",
    journal = "Phys. Rev. D",
    volume = "84",
    pages = "045022",
    year = "2011"
}

@misc{Claude:2026NineLoops,
    author = {{Claude (Anthropic)}},
    title = {Nine-loop six-gluon {MHV} amplitude in planar {$\mathcal{N}=4$} super-{Yang--Mills} theory: large data files},
    howpublished = {Zenodo dataset},
    doi = {10.5281/zenodo.22949278},
    url = {https://smsharma.io/cosmic-nine-loops/},
    note = {\href{https://smsharma.io/cosmic-nine-loops/}{Results and validation}},
    month = sep,
    year = {2026}
}

@misc{Aether:Project,
       title = {{Aether: Autonomous Engine for Theoretical \& Hands-on Exploration in Research}},
       howpublished = {Official project repository},
       url = {https://github.com/Science-Discovery/Aether}
   }

@misc{Qiu:2026iby,
    author = "Qiu, Shi and Cai, Zeyu and Wei, Jiashen and Li, Zeyu and Yin, Yixuan and Cao, Qing-Hong and Liu, Chang and Luo, Ming-xing and Yuan, Xing-Bo and Zhu, Hua Xing",
    title = "{An End-to-end Architecture for Collider Physics and Beyond}",
    eprint = "2603.14553",
    archivePrefix = "arXiv",
    primaryClass = "hep-ph",
    reportNumber = "CPTNP-2026-012",
    month = "3",
    year = "2026"
}

@misc{Li:2026yrp,
    author = "Li, Wanchen and Shao, Ding Yu and Shi, Hao-Zhe and Sun, Yu-Xuan",
    title = "{Nested-GPT for variable-multiplicity parton showers: A case study in the resummation of non-global logarithms}",
    eprint = "2605.18360",
    archivePrefix = "arXiv",
    primaryClass = "hep-ph",
    month = "5",
    year = "2026"
}

@article{Gao:2025dkn,
    author = "Gao, Meisen and Kang, Zhong-Bo and Penttala, Jani and Shao, Ding Yu",
    title = "{Determination of the initial condition for the Balitsky-Kovchegov equation with transformers}",
    eprint = "2510.26779",
    archivePrefix = "arXiv",
    primaryClass = "hep-ph",
    doi = "10.1007/JHEP03(2026)034",
    journal = "JHEP",
    volume = "03",
    pages = "034",
    year = "2026"
}

@article{Fu:2024fgj,
    author = "Fu, Rong-Jun and Rahn, Rudi and Shao, Ding Yu and Waalewijn, Wouter J. and Wu, Bin",
    title = "{qT Slicing with Multiple Jets}",
    eprint = "2412.05358",
    archivePrefix = "arXiv",
    primaryClass = "hep-ph",
    reportNumber = "UWThPh 2025-3",
    doi = "10.1103/htvz-wz1p",
    journal = "Phys. Rev. Lett.",
    volume = "135",
    number = "17",
    pages = "171903",
    year = "2025"
}

@article{vanBeekveld:2025lpz,
    author = "van Beekveld, Melissa and Ferrario Ravasio, Silvia and Helliwell, Jack and Karlberg, Alexander and Salam, Gavin P. and Scyboz, Ludovic and Soto-Ontoso, Alba and Soyez, Gregory and Zanoli, Silvia",
    title = "{Logarithmically-accurate and positive-definite NLO shower matching}",
    eprint = "2504.05377",
    archivePrefix = "arXiv",
    primaryClass = "hep-ph",
    reportNumber = "CERN-TH-2025-004, OUTP-25-01P, Nikhef 2025-003",
    doi = "10.1007/JHEP10(2025)038",
    journal = "JHEP",
    volume = "2025",
    number = "10",
    pages = "038",
    year = "2025"
}

@article{Celmaster:1979xr,
    author = "Celmaster, William and Gonsalves, Richard J.",
    title = "{An Analytic Calculation of Higher Order Quantum Chromodynamic Corrections in e+ e- Annihilation}",
    reportNumber = "UCSD-10P10-206",
    doi = "10.1103/PhysRevLett.44.560",
    journal = "Phys. Rev. Lett.",
    volume = "44",
    pages = "560",
    year = "1980"
}

\appendix
\section{An analytic NLO projection}
\label{app:fixed-born}

This appendix gives the analytic projection and its inverse used in the local NLO calculation. The projection maps a real configuration to Born kinematics; the inverse generates radiation while keeping all Born momenta fixed. We denote the projection by $\widehat{M}:\Phi_{N+1}\to\Phi_N$. It must preserve on-shell kinematics and four-momentum, and approach the underlying Born state in every single-unresolved limit. The tree-level real coefficient $R_{N+1}$ must give an integrable product with $\Delta J=J_{N+1}-J_N\circ\widehat{M}$, with physical cuts evaluated on each state separately.

For massless final states with $P=(Q,\mathbf0)$, choose the pair $h=(i,j)$ minimizing $d_{ij}$ and define
\begin{equation}
\begin{aligned}
 K_{ij}&=p_i+p_j,\qquad d_{ij}=E_i+E_j-|\mathbf K_{ij}|,\\
 \lambda&=1-d_{ij}/Q,\\
 \bar p_1&=\frac{(|\mathbf K_{ij}|,\mathbf K_{ij})}{\lambda},\qquad
 \bar p_a=\frac{p_{s_a}}{\lambda}\quad(a=2,\ldots,N),
\end{aligned}
\label{eq:app-map}
\end{equation}
where $s_a$ labels the spectators in increasing parton-label order; the merged pair is assigned Born label 1. We denote this projection for a specified pair $h=(i,j)$ by $\widehat{M}_h$; the global map $\widehat{M}$ selects the pair minimizing $d_{ij}$, with a fixed ordering used to break ties. The projected momenta obey $\bar p_a^2=0$ and $\sum_a\bar p_a=P$. In a soft or collinear limit, the selected pair contains the unresolved radiation and $\lambda\to1$, recovering the Born momenta.

For a measurement Lipschitz-continuous in energy flow, $|J(X)-J(Y)|\le L_Jd_E(X,Y)$ for a finite constant $L_J$, where $d_E(X,Y)=Q^{-1}\min_{\{f_{ab}\}}\sum_{ab}f_{ab}|\mathbf n_a-\mathbf n'_b|$ is the minimum energy-transport cost between states of total energy $Q$. Here $f_{ab}\ge0$ transfers energy from direction $\mathbf n_a$ in $X$ to $\mathbf n'_b$ in $Y$, with $\sum_bf_{ab}=E_a$ and $\sum_af_{ab}=E'_b$, the corresponding particle energies. For thrust and the $C$-parameter, this gives $|\Delta J|=O(\sqrt{d_{ij}/Q})$. In a soft-collinear limit, $d_{ij}/Q=O(y\theta_{ij}^2)$, suppressing $\dd y/y\,\dd\theta_{ij}^2/\theta_{ij}^2$; here $y$ is the soft energy divided by $Q$ and $\theta_{ij}$ is the pair's opening angle.

Conversely, fix $b=(\bar p_1,\ldots,\bar p_N)$ with $e_1=\bar p_1^0$ and choose $0<\lambda<1$. Only the $N-1$ spectator momenta are rescaled. The Born leg $\bar p_1$ is replaced by two massless particles with total momentum $K_{ij}$, fixed by four-momentum conservation,
\begin{subequations}
\label{eq:app-inverse}
\begin{equation}
\begin{aligned}
 p_{s_a}&=\lambda\bar p_a,\qquad
 K_{ij}=(1-\lambda)P+\lambda\bar p_1,\\
 K_{ij}^2&=(1-\lambda)\big[Q^2(1-\lambda)+2\lambda Qe_1\big]>0.
\end{aligned}
\label{eq:app-inverse-total}
\end{equation}
Write $\bar p_1=e_1(1,\mathbf n)$, and $k=\lambda e_1$. Choose orthonormal vectors $\mathbf e_x,\mathbf e_y$ transverse to $\mathbf n$ and set $\mathbf e_\perp(\phi)=\cos\phi\,\mathbf e_x+\sin\phi\,\mathbf e_y$. The individual momenta are
\begin{equation}
\begin{aligned}
 p_i^0&=\frac{K_{ij}^0+k\cos\theta}{2},\qquad p_j=K_{ij}-p_i,\\
 \mathbf p_i&=\frac12\big[(k+K_{ij}^0\cos\theta)\mathbf n
             +\sqrt{K_{ij}^2}\sin\theta\,\mathbf e_\perp(\phi)\big].
\end{aligned}
\label{eq:app-inverse-daughters}
\end{equation}
\end{subequations}
Here $\theta,\phi$ specify the direction of $p_i$ in the $K_{ij}$ rest frame, with polar axis $\mathbf n$, $-1<\cos\theta<1$ and $0\le\phi<2\pi$. Equation~\eqref{eq:app-inverse-daughters} boosts the two back-to-back massless daughters to the total centre-of-mass frame, ensuring $p_i^2=p_j^2=0$. Define $G_h(b,\xi)$ as the complete event $\{p_i,p_j,p_{s_2},\ldots,p_{s_N}\}$, with $\xi=(\lambda,\cos\theta,\phi)$. Reapplying the specified-pair projection gives $\widehat{M}_h(G_h(b,\xi))=b$. The selector $\chi_h$ is one when $h$ is the pair chosen by $\widehat{M}$ and zero otherwise. Selecting charts with $\chi_h=1$ therefore ensures $\widehat{M}(G_h(b,\xi))=b$. Before this selection, the labelled Lorentz-invariant measure on each chart is
\begin{equation}
 \dd\Phi_{N+1}=\dd\Phi_N(b)\,
 \underbrace{\frac{Qe_1\lambda^{2N-3}}{32\pi^3}}_{\mathcal J_h(b,\xi)}
 \dd\lambda\,\dd\cos\theta\,\dd\phi.
\label{eq:app-jacobian}
\end{equation}
Summing branches with $\chi_h$ covers phase space without double counting. Adding and subtracting the projected real measurement gives the NLO correction
\begin{equation}
 \Sigma_N^{(1)}[J]=\int\dd\Phi_{N+1}R_{N+1}\Delta J
 +\int\dd\Phi_N(b)c_{\widehat{M}}(b)J_N(b),
\label{eq:app-subtraction}
\end{equation}
where $c_{\widehat{M}}$ combines the virtual and projected real contributions into a finite density relative to $\dd\Phi_N$. For the pointwise matching below, $\widehat{M}$ in Eq.~\eqref{eq:app-subtraction} must be the projection in Eq.~\eqref{eq:app-map}, with inverse and measure given by Eqs.~\eqref{eq:app-inverse}--\eqref{eq:app-jacobian}. Direct sampling in $\Phi_{N+1}$ needs no additional Jacobian, but does not permit changing this projection while reusing the contact obtained from Eq.~\eqref{eq:app-pointwise-contact}.

The inverse map allows $c_{\widehat{M}}(b)$ to be determined separately at every resolved $N$-body Born phase-space point $b$. The required radiation input is the tree-level jettiness cumulant above $t$, evaluated at this fixed Born point,
\begin{equation}
\begin{aligned}
 c_{\widehat{M}}(b)&=\lim_{t\to0}\big[\Sigma_{N,\mathrm{sing}}^{(1)}(b,t)
                  +\mathcal R_{>t}(b)\big],\\
 \mathcal R_{>t}(b)&=\left.\int\dd\Phi_{N+1}\,R_{N+1}
                   \Theta(\tau_N-t)\right|_b.
\end{aligned}
\label{eq:app-pointwise-contact}
\end{equation}
The notation $|_b$ denotes a density relative to $\dd\Phi_N(b)$. Generate the real momenta with $G_h(b,\xi)$, select the branch with $\chi_h=1$, and integrate only the radiation variables with the Jacobian in Eq.~\eqref{eq:app-jacobian}. One then evaluates $\tau_N$ on each generated state and includes its weight when $\tau_N>t$. For the geometric measure in Eq.~\eqref{eq:tau3}, this is simply $\lambda<1-t$ because $\tau_N=1-\lambda$. The Born point $b$ is held fixed throughout. Here $\Sigma_{N,\mathrm{sing}}^{(1)}(b,t)$ is the NLO correction to the EFT cumulant for $\tau_N\le t$, differential in the same Born variables at leading power and including the coefficient of $\delta(\tau_N)$~\cite{Gaunt:2015pea}. Its addition to $\mathcal R_{>t}$ compensates the negative radiation contribution to the below-$t$ cumulant. Equation~\eqref{eq:app-pointwise-contact} then determines $c_{\widehat{M}}(b)$ point by point. This limiting relation fixes the contact convention. Appendix~\ref{app:local-contact} evaluates the same contact by finite four-dimensional integrals, so neither its extraction nor Eq.~\eqref{eq:app-subtraction} requires a small slicing parameter.

For $N=3$, Table~\ref{tab:fixed-born} confirms the kinematic identities and phase-space normalization using momenta and phase-space weights alone.

\smallskip
\noindent\begin{minipage}{\columnwidth}
\makeatletter\def\@captype{table}\makeatother
\small
\caption{Phase-space checks of the analytic $4\to3$ map. Residuals are maximum componentwise errors, with momenta normalized to $Q$ and squared masses to $Q^2$. The scans use 4096 points per branch and 4096 independent four-body points. The volume test uses $2^{28}$ points in 32 independent batches; its uncertainty is one standard error.}
\label{tab:fixed-born}
\centering
\begin{tabular}{@{}lr@{}}
\toprule
Check & Result \\
\midrule
On-shell and momentum conservation & $<4\times10^{-14}$ \\
Complete four-body round trip & $<5\times10^{-13}$ \\
Fixed-Born reconstruction & $<3\times10^{-16}$ \\
Volume / $[Q^4/(24576\pi^5)]$ & $1.00006\pm0.00008$ \\
\bottomrule
\end{tabular}
\end{minipage}

\par\smallskip

We now test local cancellation with the four-parton tree matrix elements from \texttt{EERAD3}~\cite{Gehrmann-DeRidder:2014hxk}. Fix a resolved Born point with $2E_i/Q=(0.8,0.7,0.5)$, for which $1-T=0.2$ and $C=9/14$. Choose $\bar{\mathbf p}_1$ along $+z$ and $\bar{\mathbf p}_2$ in the $xz$ plane with positive $x$, fixing the overall orientation. For either measurement $J=1-T,C$, the radiation integral normalized to the three-parton Born density $B_3(b)$ is
\begin{equation}
\begin{aligned}
 \mathcal F_J(b;\eta)
 &=\frac{1}{B_3(b)}\sum_h\int\dd\xi\,\mathcal J_h\chi_h R_4(G_h)\\
 &\quad\times[J_4(G_h)-J_3(b)]\,
       \Theta(y_{\min}-\eta),
\end{aligned}
\label{eq:app-fixed-born-real-test}
\end{equation}
where $y_{\min}=\min_{i<j}s_{ij}/Q^2$. Replacing the measurement difference by $J_4(G_h)$ or $J_3(b)$ defines the separate real and projected integrals, $\mathcal F_J^R$ and $\mathcal F_J^P$, so that $\mathcal F_J=\mathcal F_J^R-\mathcal F_J^P$. The powers of $\alpha_s/(2\pi)$ are stripped from the tree coefficients $R_4$ and $B_3$. All radiation variables are integrated using Eqs.~\eqref{eq:app-inverse}--\eqref{eq:app-jacobian}, while $b$ remains fixed.

\begin{figure*}[t]
\centering
\includegraphics[width=\textwidth]{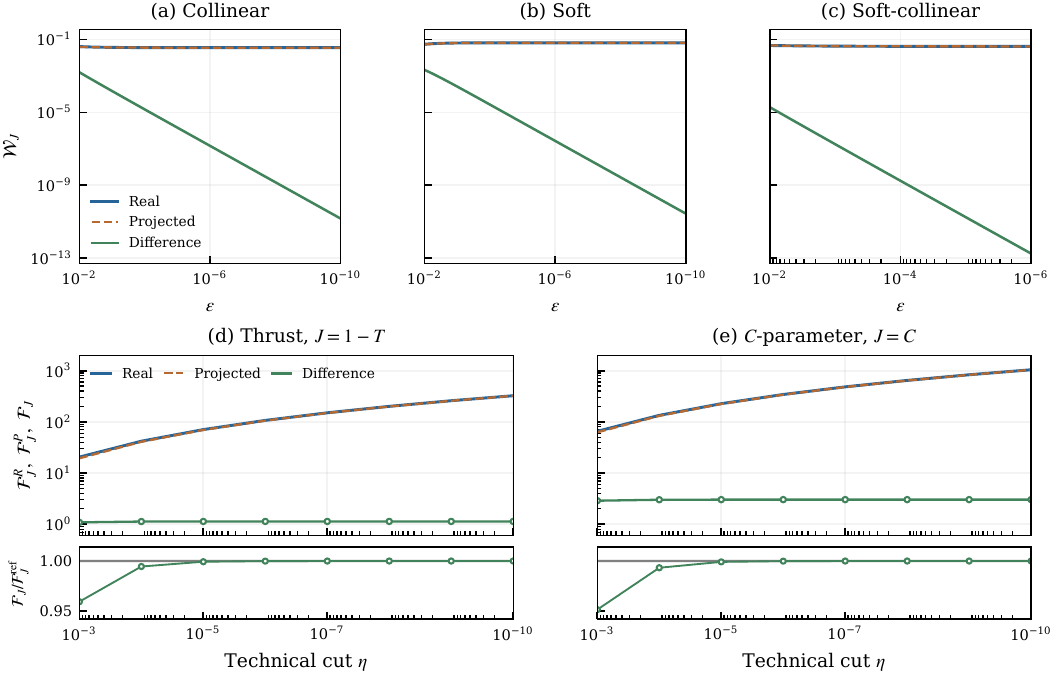}
\caption{Local P2B cancellation at a fixed three-parton Born point. The upper panels show the $q\bar qgg$ channel with the $qg$ pair $(1,3)$ generated by Eq.~\eqref{eq:app-inverse}, at $\phi=0.7$. Write $\delta=1-\lambda$ and $w=1-\cos\theta$. The collinear, soft and soft-collinear paths use $(\delta,w)=(\epsilon,0.8),(\epsilon,20\epsilon),(\epsilon^2,2\epsilon)$, respectively. The ordinate $\mathcal W_J$ denotes the logarithmic radiation density for $J=1-T$. Blue and orange show $\omega\mathcal J_hR_4J_4/B_3$ and $\omega\mathcal J_hR_4J_3(b)/B_3$; green shows their difference. Here $\omega=\delta$ in (a) and $\omega=\delta w$ in (b,c), corresponding to logarithmic radiation coordinates. The lower panels integrate all six branches and both partonic channels. Their lower strips show $\mathcal F_J(\eta)/\mathcal F_J^{\mathrm{ref}}$, with $\mathcal F_J^{\mathrm{ref}}=\mathcal F_J(10^{-10})$ and correlated statistical errors. Grey bands give the relative one-standard-error uncertainty of the reference integral. The technical cut $\eta$ on pair invariants probes the numerical stability of the locally subtracted integral.}
\label{fig:fixed-born-real}
\end{figure*}

Figure~\ref{fig:fixed-born-real} shows the local cancellation and the convergence of this integral. The separate real and projected logarithmic radiation densities approach nonzero limits, while their difference vanishes in soft, collinear and soft-collinear scans. The integral includes both $q\bar qgg$ and four-quark channels with $n_f=5$, using $32$ independent batches of $6\times2^{16}$ radiation points. Every accepted configuration projects back to the same $b$, with maximum componentwise error below $2\times10^{-16}Q$. At $\eta=10^{-10}$ we obtain $\mathcal F_{1-T}=1.1372\pm0.0013$ and $\mathcal F_C=3.0338\pm0.0035$. Reducing $\eta$ from $10^{-6}$ to $10^{-10}$ changes either result by less than $0.011\%$, while the separate real and projected integrals continue to grow. These tests demonstrate local infrared cancellation with the Born phase-space point held fixed throughout.

\section{A proposal for local three-jet NNLO subtraction at fixed Born kinematics}
\label{app:nnlo-fixed-born}

This appendix develops the radiation part of the proposed extension of the local subtraction to three-jet NNLO, using a four-jet NLO calculation as input~\cite{Cacciari:2015jma}. To preserve local cancellation at a specified three-parton Born point, the real contribution and its counterterm must be generated on a common phase space with those Born momenta fixed. The LLM constructs it by composing the inverse in Appendix~\ref{app:fixed-born} with the inverse of the NLO subtraction map. We first give the phase space and paired radiation weights, then discuss their integrability and numerical checks. Appendix~\ref{app:local-contact} gives the complementary proposal for direct contact calculation.

Let $R_5$ denote the five-parton tree contribution, $S_a$ the counterterms grouped by their maps $\widehat{m}_a:\Phi_5\to\Phi_4$, and $K_4$ the sum of the renormalized four-parton one-loop contribution and integrated counterterms, finite at resolved four-parton kinematics. For an infrared-safe four-jet measurement $f_m$ on an $m$-parton state, the NLO correction is
\begin{equation}
\begin{aligned}
 \mathcal I_4^{(1)}[f]
 &=\int\dd\Phi_5\left[R_5f_5-\sum_aS_a f_4(\widehat{m}_a\Phi_5)\right]\\
 &\quad+\int\dd\Phi_4\,K_4f_4.
\end{aligned}
\label{eq:nnlo-nlo-input}
\end{equation}
Only this correction, at the order of the three-jet NNLO coefficient, is needed. The matrix elements, counterterms, maps and finite $K_4$ are reused without modification, while their four- and five-parton integrations are replaced by the common phase space below. No new integrated counterterms or $d$-dimensional calculation are required. Flavour sums, symmetry factors and normalization are inherited from the NLO input, whose generation cuts must cover the new measurement.

\subsection{A common phase space}

We illustrate the pairing with antenna subtraction~\cite{Gehrmann-DeRidder:2005btv}; other NLO schemes can follow the same strategy with compatible inverse maps, measures and infrared bounds.

Use the P2B projection $\widehat{M}_4:\Phi_4\to\Phi_3$ of Eq.~\eqref{eq:app-map}, ordering its Born output by decreasing energy. The inverse sums six pair choices and six assignments of the Born momenta to the merged leg and spectators. Here $h$ includes both choices. Write $G_{4,h}$ for the $N=3$ inverse in Eqs.~\eqref{eq:app-inverse}--\eqref{eq:app-jacobian}, generating $x=G_{4,h}(b,\xi)$. At fixed $b$, its radiation measure is $\dd\mu_4^b=\sum_h\chi_h\mathcal J_h\dd\xi$. The Born measure $\dd\Phi_3(b)$ is restricted to this energy ordering. Denote the two angles in $\xi$ by $\theta_4,\phi_4$, to distinguish them from the antenna azimuth $\phi$. For each antenna channel, write $x=(I,K,q_1,q_2)$, where $I,K$ are the massless parent four-momenta replacing partons $i,j,k$ under $\widehat{m}_a$, and $q_1,q_2$ are unchanged spectators.

In the $I+K$ rest frame, divide momenta by $\sqrt{s_{IK}}$, where $s_{IK}=(I+K)^2$, giving a dimensionless total $(1,\mathbf0)$. For $r,s>0$ and $r+s<1$, choose the massless daughters
\begin{equation}
\begin{aligned}
 \hat p_i&=\tfrac12(1-s)(1,0,0,1),\\
 \hat p_j&=\left(\frac{r+s}{2},\frac{\sqrt{rs(1-r-s)}}{1-s},0,
             \frac{r+s}{2}-\frac{r}{1-s}\right),\\
 \hat p_k&=(1,\mathbf0)-\hat p_i-\hat p_j.
\end{aligned}
\label{eq:nnlo-inverse}
\end{equation}
The hatted momenta satisfy $2\hat p_i\cdot\hat p_j=r$ and $2\hat p_j\cdot\hat p_k=s$. The antenna subtraction map gives the back-to-back parents
\begin{equation}
\begin{aligned}
 c&=1-r-s,\qquad y=\frac{s}{r+s},\\
 \rho_{\mathrm{ant}}&=\sqrt{1+\frac{4y(1-y)rs}{c}},\\
 a_i&=\frac{1+\rho_{\mathrm{ant}}-2ys}{2(1-s)},\qquad
 a_k=\frac{1-\rho_{\mathrm{ant}}-2yr}{2(1-r)},\\
 \hat I&=a_i\hat p_i+y\hat p_j+a_k\hat p_k,\\
 \hat K&=(1,\mathbf0)-\hat I.
\end{aligned}
\label{eq:nnlo-native-parents}
\end{equation}
Rotate $\hat I$ onto the direction of $I$ in the $I+K$ rest frame, rotate the decay plane about that direction by $\phi$, multiply all daughters by $\sqrt{s_{IK}}$, and boost from the $I+K$ rest frame to the total centre-of-mass frame. The two spectators are unchanged. This defines the inverse $p=G_a^{\mathrm{ant}}(x;r,s,\phi)$ explicitly. For $0\le\phi<2\pi$, define the azimuthal origin using the spectator with the larger component transverse to $I$ in the antenna rest frame. If both components vanish, use any nonparallel Cartesian reference axis.

The resulting phase space obeys
\begin{equation}
\begin{aligned}
 \widehat{m}_aG_a^{\mathrm{ant}}(x;r,s,\phi)&=x,\qquad \widehat{M}_{5,a}=\widehat{M}_4\circ\widehat{m}_a,\\
 \widehat{M}_{5,a}G_a^{\mathrm{ant}}(G_{4,h}(b,\xi);r,s,\phi)&=b,\\
 \dd\Phi_5&=\dd\Phi_4(x)\,
 \underbrace{\frac{s_{IK}}{32\pi^3}\dd r\,\dd s\,\dd\phi}
             _{\dd\Phi_{\mathrm{ant},a}}.
\end{aligned}
\label{eq:nnlo-five-measures}
\end{equation}
Thus all three Born momenta are fixed exactly in both members of a subtraction pair. The phase-space Jacobian is the product of the measure in Appendix~\ref{app:fixed-born} and the antenna measure. Table~\ref{tab:nnlo-fixed-born} checks this construction against the NLO subtraction maps.

\begin{table}[t]
\caption{Kinematic checks of the common fixed-Born phase space at $2E_i/Q=(0.8,0.7,0.5)$. Of 600 attempted four-parton configurations, 308 satisfy the selected-chart condition. Each is tested at azimuths $\phi$ and $\phi+\pi/2$. Entries are maximum componentwise residuals normalized to $Q$, or $Q^2$ for the mass shell, using extended-precision geometry.}
\label{tab:nnlo-fixed-born}
\centering
\begin{tabular}{@{}lc@{}}
\toprule
Check & Maximum residual \\
\midrule
Mass shell & $<7\times10^{-17}$ \\
Momentum conservation & $<2\times10^{-16}$ \\
Subtraction inverse, $\widehat{m}_aG_a^{\mathrm{ant}}=x$ & $<3\times10^{-13}$ \\
Fixed Born point, $\widehat{M}_{5,a}G_a^{\mathrm{ant}}G_{4,h}=b$ & $<3\times10^{-16}$ \\
\bottomrule
\end{tabular}
\end{table}

\subsection{Local subtraction on the common phase space}

Different counterterms have different maps. We therefore partition the real kernel into contributions $R_a$ and assign each the projection $\widehat{M}_{5,a}=\widehat{M}_4\circ\widehat{m}_a$. An additional smooth partition suppresses counterterms whose reduced four-parton state becomes singular while the real state remains resolved.

Let $v_a=\tau_3(\widehat{m}_a p)$, with the minimized geometric measure of Eq.~\eqref{eq:tau3}, and define the dimensionless transport distance between the three daughters and their two parents by
\begin{equation}
 \ell_a=\frac1Q\min_{\substack{0\le z_l\le E_l\\\sum_lz_l=E_I}}
 \sum_{l=i,j,k}\left[z_l|\mathbf n_l-\mathbf n_I|
       +(E_l-z_l)|\mathbf n_l-\mathbf n_K|\right].
\label{eq:nnlo-transport}
\end{equation}
All energies and unit directions in Eq.~\eqref{eq:nnlo-transport} refer to the total centre-of-mass frame. The variable $z_l$ is the energy transferred from daughter $l$ to parent $I$; the remainder goes to $K$. Sorting $|\mathbf n_l-\mathbf n_I|-|\mathbf n_l-\mathbf n_K|$ and assigning energy to $I$ in that order until $E_I=I^0$ is reached solves the minimization. It bounds the change of an energy-flow Lipschitz measurement, $|J_5-J_4|\le L_J\ell_a$, and also $|\tau_3(p)-v_a|\le\ell_a$. Choose
\begin{equation}
\begin{aligned}
 h_a&=h\!\left(\frac{v_a}{v_a+\ell_a}\right),\\
 h(z)&=\begin{cases}
 0,&z\le\tfrac12,\\
 w^2(3-2w),\quad w=4z-2,&\tfrac12<z<\tfrac34,\\
 1,&z\ge\tfrac34.
 \end{cases}
\end{aligned}
\label{eq:nnlo-screen}
\end{equation}
Near a single-unresolved limit of channel $a$ with resolved $x$, $h_a=1$. It vanishes when $v_a\to0$ while $\tau_3(p)$ stays nonzero, suppressing a counterterm singularity absent from the real configuration. The complementary contribution is retained, so this partition removes no phase-space region.

A suitable signed partition of the real contribution is
\begin{equation}
\begin{aligned}
 \hat d_{ij}&=\frac{s_{ij}}{Q(E_i+E_j+|\mathbf p_i+\mathbf p_j|)},\qquad
 u_a=\min(\hat d_{ij},\hat d_{jk}),\\
 \omega_a&=\frac{u_a^{-2}}{\sum_cu_c^{-2}},\qquad
 D=R_5-\sum_c h_cS_c,\\
 R_a&=h_aS_a+\omega_aD,\qquad \sum_aR_a=R_5.
\end{aligned}
\label{eq:nnlo-real-partition}
\end{equation}
Here $\hat d_{ij}=d_{ij}/Q$ is the dimensionless deficit corresponding to Eq.~\eqref{eq:app-map}. The $R_a$ need not be positive. The 30 channels have three distinct labels $(i,j,k)$ with $i<k$; $j$ is the parton removed by the subtraction map. We average $R_5$ and $S_a$ over the same $5!$ label permutations, grouping counterterms with identical maps, and average $K_4$ over $4!$ permutations. These averages relabel integration variables and introduce no extra multiplicity.

On an accepted chart, $x=G_{4,h}(b,\xi)$ and $p=G_a^{\mathrm{ant}}(x;r,s,\phi)$, set $J_5=J(p)$, $J_4=J(x)$ and $\Delta J_m=J_m-J_3(b)$. The common local weight is
\begin{equation}
\begin{aligned}
 W_a^J&=R_a\Delta J_5-S_a\Delta J_4\\
 &=\omega_aD\Delta J_5+h_aS_a(J_5-J_4)
                      -(1-h_a)S_a\Delta J_4.
\end{aligned}
\label{eq:nnlo-paired-weight}
\end{equation}
Every difference is formed before integration, at the same five-parton point and its mapped four-parton state. Scalar antennae require an additional treatment of collinear spin correlations~\cite{Gehrmann-DeRidder:2005btv}. At fixed $x,r,s$, define $\langle W_a^J\rangle_\phi=[W_a^J(\phi)+W_a^J(\phi+\pi/2)]/2$. This average preserves the full azimuthal integral. The leading $\cos2\phi$ and $\sin2\phi$ harmonics cancel within this pair, which is also the unit for estimating statistical errors.

At fixed $b$, the radiation contribution and its combination with the required NNLO contact take the form
\begin{equation}
\begin{aligned}
 F_J^{\mathcal P}(b)
 &=\sum_a\int\dd\mu_4^b\,\dd\Phi_{\mathrm{ant},a}\,\langle W_a^J\rangle_\phi
       +\int\dd\mu_4^b\,K_4\Delta J_4,\\
 \Sigma_3^{(2)}[J]
 &=\int\dd\Phi_3(b)\left[F_J^{\mathcal P}(b)
                     +c_{\mathcal P}^{(2)}(b)J_3(b)\right].
\end{aligned}
\label{eq:nnlo-p2b-scheme}
\end{equation}
Here $\mathcal P$ denotes the combined choice of $\widehat{M}_4$, $\widehat{m}_a$ and $R_a$. Integrating over $b$ and using $\sum_aR_a=R_5$ recovers the real and four-parton terms of Eq.~\eqref{eq:nnlo-nlo-input}; the remaining projected contribution is absorbed into $c_{\mathcal P}^{(2)}$. For the constant measurement $J=1$, $F_1^{\mathcal P}=0$ sample by sample. The four-parton weight $K_4\Delta J_4$ is separately integrable.

\subsection{Integrability and pointwise contact matching}

For a fixed Born point $b$ away from soft and collinear boundaries, the maps give the bounds $d_E(x,b)\le \mathcal{C}\sqrt{v_a}$ and $\ell_a\le \mathcal{C}\sqrt{\min(r,s)}$, with $d_E$ defined in Appendix~\ref{app:fixed-born}. The positive constants $\mathcal{C}$, $\mathcal{C}'$, $\mathcal{C}_J$ and $c_0$ below remain finite on compact resolved Born regions. The first bound follows by transporting the merged pair onto its total direction and rescaling; the second follows from the explicit antenna inverse. Combining the latter bound with $\ell_a\le \mathcal{C}\sqrt{s_{IK}}/Q$ and $\min(s_{ij},s_{jk})/Q^2\le2u_a$ gives $\ell_a\le \mathcal{C}u_a^{1/4}$. With $u_{\min}=\min_a u_a\le v=\tau_3(p)$, these imply
\begin{equation}
\begin{aligned}
 \omega_a|\Delta J_5|&\le \mathcal{C}_Jv^{1/8},\\
 h_a|J_5-J_4|&\le \mathcal{C}_J\min(v_a,\sqrt r,\sqrt s).
\end{aligned}
\label{eq:nnlo-power-bounds}
\end{equation}
Each antenna counterterm has a logarithmic unresolved measure, bounded by $\dd r\,\dd s/(rs)$ times a reduced tree density. Integrating the second line therefore yields $v_a$ times at most two logarithms. On the support of $1-h_a$, the transport bound instead implies $r,s>c_0v_a^2$, so the last term of Eq.~\eqref{eq:nnlo-paired-weight} is bounded after antenna integration by a reduced tree density times $\sqrt{v_a}(1+|\log v_a|^2)$. Both terms are integrable at the remaining four-parton boundary.

For the first term, NLO factorization gives cancellation of the leading single-unresolved residues in $D$ at fixed resolved four-parton kinematics, after the correlated azimuthal average. A further bound is needed when the reduced state also becomes unresolved. For two ordered scales $\kappa$ and $\kappa z$, the following inequality gives a useful estimate of the integral over their ratio $z$.
\begin{equation}
 \int_0^1\frac{\dd z}{z}
     \min(1,\mathcal{C}\kappa^{-A}z^\gamma)
 \le \mathcal{C}'(1+|\log\kappa|).
\label{eq:nnlo-ordered-bound}
\end{equation}
Here $A\ge0$ and $\gamma>0$. 
If the remaining integral over $z$ is bounded by Eq.~\eqref{eq:nnlo-ordered-bound}, the positive power in Eq.~\eqref{eq:nnlo-power-bounds} makes the outer logarithms integrable. Establishing this bound for the complete $D$ contribution in all overlapping soft and collinear limits remains necessary. If $p$ approaches two jets while $\widehat{M}_{5,a}p=b$ remains resolved, $u_a$ stays nonzero and $\omega_a=O(u_{\min}^2)$ ensures the required suppression. Finally, $K_4$ has only logarithmic enhancements of its single-unresolved tree behaviour, rendered integrable by $\Delta J_4=O(\sqrt{v_a})$. These statements use the established factorization properties of the NLO input and apply after this azimuthal average. The Monte Carlo variance still depends on the sampling density.

Contact matching uses the same radiation integral at fixed $b$, with jettiness as the measurement. For $t>0$, let
\begin{equation}
\begin{aligned}
 \mathcal R_{>t}^{(2),\mathcal P}(b)
    &=F_{\Theta(\tau_3-t)}^{\mathcal P}(b),\\
 c_{\mathcal P}^{(2)}(b)
    &=\lim_{t\to0}\left[\Sigma_{3,\mathrm{sing}}^{(2)}(b,t)
                      +\mathcal R_{>t}^{(2),\mathcal P}(b)\right].
\end{aligned}
\label{eq:nnlo-pointwise-contact}
\end{equation}
This relation defines the contact separately at each resolved three-parton point, integrating only radiation variables. Here $\Sigma_{3,\mathrm{sing}}^{(2)}$ is the NNLO counterpart of the below-$t$ EFT cumulant in Eq.~\eqref{eq:app-pointwise-contact}, including the coefficient of $\delta(\tau_3)$. Both terms use the same jettiness definition and leading-power Born variables. The positive $t$ defines this limiting characterization; Appendix~\ref{app:local-contact} develops a finite-difference representation intended to evaluate it without numerical extrapolation. A different projection prescription changes the finite contact, which must therefore be determined with the same $\mathcal P$ used in Eq.~\eqref{eq:nnlo-p2b-scheme}.

On the mapped four-parton state $x$, $\tau_3(x)=1-\lambda$. This explicit coordinate dependence also determines the powers of the vanishing radiation coordinates used in the finite contact construction. The real-state measurement is evaluated separately, preserving the NLO input's local cancellation without requiring the map to preserve jettiness.

Equations~\eqref{eq:nnlo-p2b-scheme} and~\eqref{eq:nnlo-pointwise-contact} specify the ingredients of the proposed local three-jet NNLO subtraction. Appendix~\ref{app:local-contact} develops the finite contact representation on the same phase space at fixed $b$. The numerical checks below test the radiation contribution.

\subsection{Numerical checks of the fixed-Born radiation contribution}

\begin{figure*}[tp]
\centering
\includegraphics[width=\textwidth]{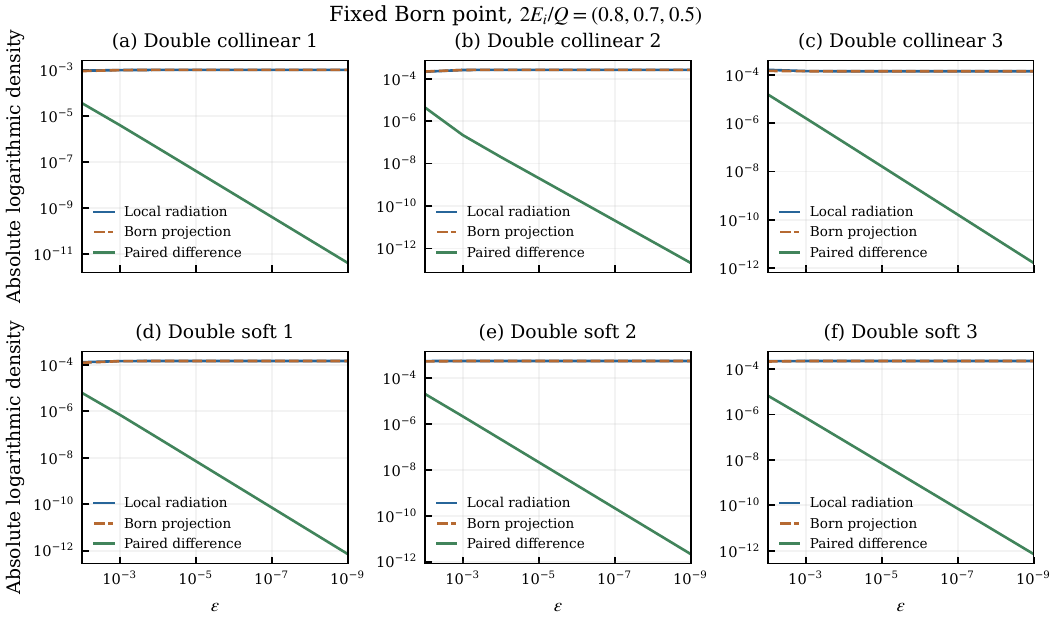}
\caption{Local cancellation in the three-jet NNLO radiation contribution at fixed Born kinematics $2E_i/Q=(0.8,0.7,0.5)$, for thrust $J=1-T$. The upper and lower rows approach three double-collinear and three double-soft limits, respectively. The curves show the magnitudes of the locally subtracted radiation density (blue), its Born-projected counterpart (orange), and their paired difference (green), including the four-jet NLO counterterms. All densities include the logarithmic phase-space Jacobian and are normalized to the tree-level Born density. The paired difference is evaluated before integration and vanishes along all six paths. No technical cut is applied. The paths, normalization and correlated azimuthal average are specified in the text. These checks concern the radiation contribution and do not include the NNLO contact.}
\label{fig:fixed-born-nnlo-local}
\end{figure*}

\begin{figure*}[tp]
\centering
\includegraphics[width=\textwidth]{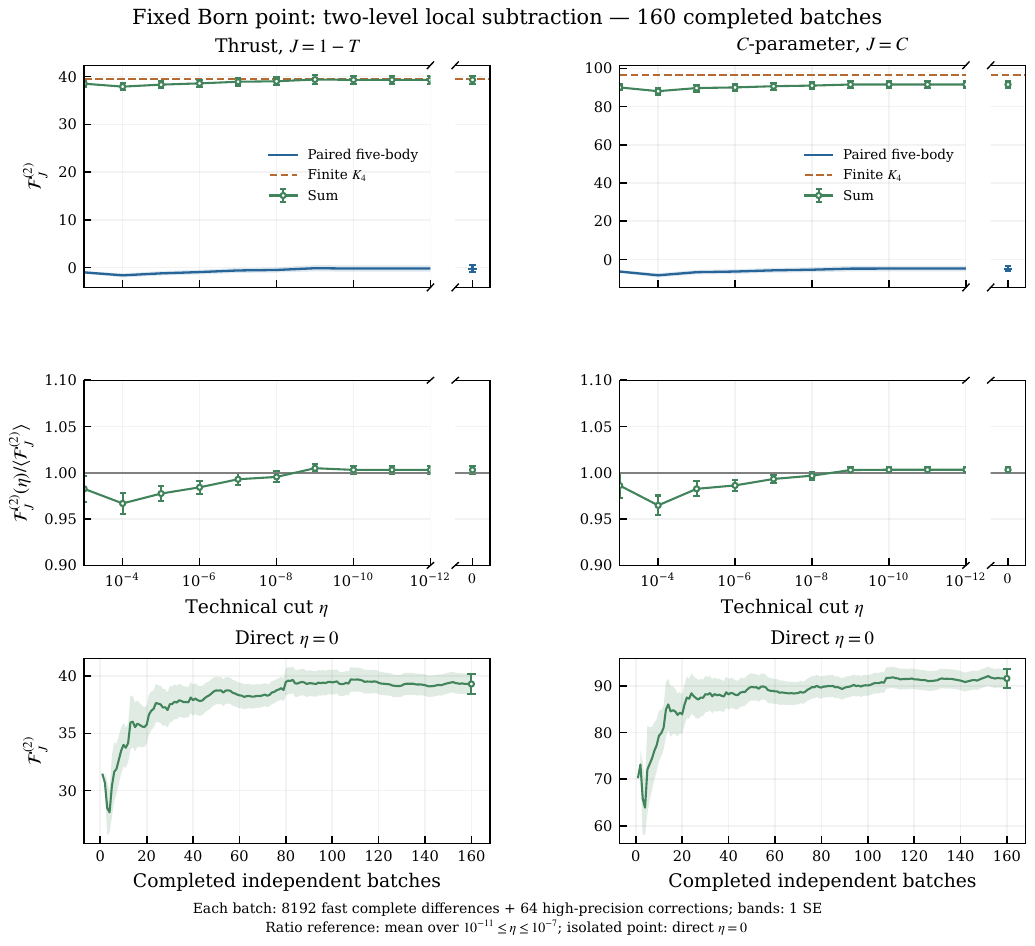}
\caption{NNLO radiation integrals at the fixed three-parton Born point of Fig.~\ref{fig:fixed-born-nnlo-local}. The ordinate $\mathcal F_J^{(2)}$ is the radiation integral normalized to the tree-level Born density, with the coupling stripped as in Eq.~\eqref{eq:nnlo-fixed-born-test}. The upper panels show the paired five-parton integral (blue), the four-parton term containing the finite one-loop and integrated-counterterm weight $K_4$ (orange), and their sum (green). In each sector $a$, the paired integrand is $R_a[J_5(p)-J_3(b)]-S_a[J_4(\widehat m_a p)-J_3(b)]$, where $R_a$ is the real sector weight and $S_a$ its NLO counterterm, evaluated on the same five-parton state $p$; $b$ is the common Born point. The middle panels divide the sum by its mean over $10^{-11}\le\eta\le10^{-7}$, accounting for correlations. The first two rows share aligned axes, with the directly evaluated $\eta=0$ point shown separately. The lower panels show the cumulative mean at $\eta=0$ over 160 independent batches. Bands and error bars denote one standard error, including the independent four-parton uncertainty in the sum. The parameter $\eta$ is a technical cut for numerical-stability checks. These results evaluate the locally subtracted radiation contribution without a small slicing parameter; the NNLO contact is a separate contribution.}
\label{fig:fixed-born-nnlo}
\end{figure*}

Figures~\ref{fig:fixed-born-nnlo-local} and~\ref{fig:fixed-born-nnlo} use the complete four-jet NLO input from \texttt{EERAD3}~\cite{Aveleira:2025svg}, with $n_f=5$, $\mu_R=Q$ and $2E_i/Q=(0.8,0.7,0.5)$. All three Born momenta remain fixed throughout. For $J=1-T,C$, define
\begin{equation}
 \mathcal F_J^{(2)}(b;\eta)
 =\frac{F_J^{\mathcal P}(b;\eta)}{\overline B_3(b)},
\label{eq:nnlo-fixed-born-test}
\end{equation}
Here $\overline B_3(b)=\sum_{\pi\in S_3}B_3(b_\pi)$ is the density on the energy-ordered Born region, with $b_\pi$ running over all six permutations of the Born momenta; powers of $\alpha_s/(2\pi)$ are stripped. The five-parton weights include both correlated azimuths before statistical accumulation. For the cut scan, the minimum pair invariant $y_{\min}=\min_{i<j}s_{ij}/Q^2$ must exceed $\eta$ at both azimuths; this common condition multiplies the entire paired weight. We also evaluate $\eta=0$ directly. The four-parton $K_4\Delta J_4$ contribution is included for every cut, and no auxiliary jettiness threshold is imposed.

For Fig.~\ref{fig:fixed-born-nnlo-local}, the radiation, Born-projected and paired densities are $\mathcal J_{\log}\langle R_aJ_5-S_aJ_4\rangle_\phi/\overline B_3$, $\mathcal J_{\log}\langle R_a-S_a\rangle_\phi J_3(b)/\overline B_3$ and $\mathcal J_{\log}\langle W_a^J\rangle_\phi/\overline B_3$, respectively. The brackets denote the two-azimuth average defined above Eq.~\eqref{eq:nnlo-p2b-scheme}; the paired density is evaluated directly from the complete weight.

The paths are generated by the common inverse in Eq.~\eqref{eq:nnlo-five-measures}. The upper row takes $\delta_4=1-\lambda=r=\epsilon$ at fixed $s$ and four-parton decay angle $\theta_4$. The lower row takes $\delta_4=\epsilon$, $1-\cos\theta_4=A\epsilon$ and $r+s=B\epsilon$ at fixed $r/(r+s)$. The columns use cyclic Born assignments $(1,2,3)$, $(2,3,1)$ and $(3,1,2)$, with outer pair $(3,4)$ and antenna labels $(i,j,k)=(2,1,3)$. From left to right, $(\cos\theta_4,s)=(-0.35,0.22),(0.20,0.35),(0.55,0.47)$ above, and $(A,B,r/(r+s))=(8,1,0.3),(12,1.3,0.4),(16,1.6,0.5)$ below. Both rows use $(\phi_4,\phi)=(0.73,0.419),(1.23,1.119),(1.73,1.819)$. The logarithmic Jacobian $\mathcal J_{\log}$ is the product of the four-parton and antenna Jacobians, multiplied by $\delta_4r$ in the upper row and $\delta_4(1-\cos\theta_4)(r+s)^2$ in the lower row. No technical cut is applied. The $C$-parameter gives the same limiting behaviour, and separate single-unresolved scans at resolved four-parton kinematics test the NLO input's real--counterterm cancellation.

Figure~\ref{fig:fixed-born-nnlo-local} shows three double-collinear and three double-soft paths. The locally subtracted radiation density and its Born-projected counterpart approach the same limit, while their paired difference decreases linearly with $\epsilon$ down to $10^{-9}$. These tests support local cancellation along the sampled paths at this nontrivial fixed Born point.

Figure~\ref{fig:fixed-born-nnlo} integrates the same paired construction for thrust and the $C$-parameter. The upper panels separate the paired five-parton term from the finite four-parton contribution and show their sum. The small-cut results agree with the direct $\eta=0$ calculation, and the lower panels show the cumulative estimate settling as independent batches are added. These results demonstrate numerical integration of the NNLO radiation contribution at fixed Born momenta without a small slicing parameter for the measurements tested here.

Figure~\ref{fig:fixed-born-nnlo} integrates the complete paired weight with importance sampling. To reduce the cost of high-precision evaluation, we use $\langle w_{\mathrm f}\rangle+\langle w_{\mathrm h}-w_{\mathrm f}\rangle$, where $w_{\mathrm f}$ and $w_{\mathrm h}$ are the importance-weighted local differences evaluated with double- and higher-precision arithmetic. The two averages use independent samples, while each correction evaluates both weights at the same coordinates. In numerically delicate regions the auxiliary $w_{\mathrm f}$ is set to zero and the correction includes the full $w_{\mathrm h}$. This rearrangement preserves the locally subtracted radiation integral.

A four-hour run on the same M1 laptop produced 160 independently randomized batches, each containing 8192 fast weights and 64 high-precision corrections. Zero-weight unselected charts remain in the sample count. The finite four-parton contribution uses a separate sample of 320 batches of 64 points. The direct zero-cut results are $\mathcal F_{1-T}^{(2)}=39.32\pm0.89$ and $\mathcal F_C^{(2)}=91.61\pm2.01$, with relative uncertainties of $2.25\%$ and $2.19\%$. The paired five-parton contributions are $-0.19\pm0.66$ and $-4.78\pm1.47$, respectively; the corresponding four-parton contributions are $39.51\pm0.59$ and $96.39\pm1.37$. The middle panels use the arithmetic mean over the five cuts $10^{-11}\le\eta\le10^{-7}$ as their reference, accounting for all common-sample correlations. The cumulative estimates include the same independent four-parton contribution at every batch count.

These integrals contain the measurement differences $J_m-J_3(b)$ and test the locally subtracted radiation contribution. The proposed contact extraction uses the finite-difference weights developed in Appendix~\ref{app:local-contact} on the same fixed-Born phase space.

\par
Algorithm~\ref{alg:nnlo-fixed-born} specifies the estimator used for Fig.~\ref{fig:fixed-born-nnlo}, including the frozen importance density, the independent fast and correction samples, and the separately sampled four-parton contribution. In the code, \texttt{M4}, \texttt{m[a]}, \texttt{g.h} and \texttt{g.omega} denote $\widehat{M}_4$, $\widehat{m}_a$, $h_a$ and $\omega_a$, respectively. Both precision levels form the complete local difference before integration. The guard acts only on the auxiliary fast weight, whose omission is restored by the high-minus-fast correction. Rejected charts remain in the sample count, and both azimuths form one observation. The ancillary reproduction instructions specify the sampling transforms, normalization conventions and numerical treatment for an independent reproduction of Figs.~\ref{fig:fixed-born-nnlo-local} and~\ref{fig:fixed-born-nnlo}. Figure~\ref{fig:fixed-born-nnlo} uses $J=1-T,C$.

\Needspace{8\baselineskip}
\begingroup
\renewcommand{\lstlistingname}{Algorithm}
\begin{lstlisting}[
  language=Python,
  basicstyle=\fontsize{7.4}{8.7}\selectfont\ttfamily,
  keywordstyle=\bfseries,
  commentstyle=\normalfont\itshape\fontsize{7.4}{8.7}\selectfont,
  numbers=left,
  numberstyle=\tiny,
  numbersep=4pt,
  xleftmargin=12pt,
  frame=tb,
  breaklines=true,
  breakatwhitespace=true,
  columns=fullflexible,
  keepspaces=true,
  showstringspaces=false,
  caption={Integration of locally subtracted NNLO radiation at fixed Born momenta, with the sampling and precision treatment used in Fig.~\ref{fig:fixed-born-nnlo}.},
  label={alg:nnlo-fixed-born}
]
# Fixed-Born NNLO radiation. Q=1; nf=5; mu_R=Q; J=(1-T,C).
# b=((.4,0,0,.4),(.35,sqrt(3)/8,0,-.275),
#    (.25,-sqrt(3)/8,0,-.125)); JB=J(b).
# R,S: average all 120 labels, then multiply by
# 8192*pi**7/Bbar3; K4: all 24, 2048*pi**5/Bbar3.
# Bbar3 is the SUM over six Born labels, not a mean.
eta = [10.**(-k) for k in range(3,13)] + [0.]
proposal = load_frozen("production-proposal.json")
# q(u)=.2+.8*q_tree(u); full support, frozen 32 boxes.

def draw_four(v):  # Seven coordinates; Appendix A.
    h,perm,delta,c4,phi4,pdelta,pc = outer_sample(v)
    # Six pairs and six Born assignments, uniform.
    # delta=1-lambda; densities include both mixture
    # branches; explicit transforms in companion file.
    x = G4(b,h,perm,1-delta,c4,phi4)
    if not selected_minimum_pair(x,h):
        return x,0,delta  # Keep zero in sample count.
    w4 = 36*b[perm[0]].energy*(1-delta)**3
    w4 /= 16*pi**2*pdelta*pc
    return x,w4,delta

def packet(u,precision):  # No outer q(u) division yet.
    a = int(30*u[0])  # All (i,j,k), i<k, distinct.
    x,w4,delta = draw_four(u[1:8])
    if w4 == 0: return zeros((11,2))
    r,s,phi,qant = balanced_antenna(u[8:13],delta**2)
    # qant: full density in dr ds dphi. 20% uniform,
    # 80% three-corner mixture; no infrared floor.
    blocks = []; ymin = infinity
    for angle in (phi,phi+pi/2):
        p,jac = inverse_antenna(x,a,r,s,angle)
        # 60-digit kinematics; jac=(I+K)^2/(32*pi^3).
        ymin = min(ymin,min_invariant(p))
        if precision == "fast" and ymin<1e-8:
            return zeros((11,2))  # Check before fast maps.
        g = geometry(p,precision)  # Cache all 30 maps.
        active = [c for c in CHANNELS if g.h[c]>0]
        needed = union(active,[a])
        if precision == "fast" and min(
            g.reduced_min[c] for c in needed)<1e-8:
            return zeros((11,2))  # L only; H is included in the correction.
        if precision == "high": assert_close(M4(m[a](p)),b)
        blocks.append((p,g,jac,active,needed))
    values = []
    for p,g,jac,active,needed in blocks:
        R,S = native_kernels(p,needed,precision)
        # Persistent processes; full label averages;
        # exact groups sharing a map, sign R-S.
        D = accurate_sum([R]+[-g.h[c]*S[c] for c in active])
        W = accurate_sum([g.omega[a]*D*(J(p)-JB),
             g.h[a]*S[a]*(J(p)-J(x)),
             -(1-g.h[a])*S[a]*(J(x)-JB)])
        # High: quad kernels and 50-digit sums; if the
        # packet error bound exceeds 1e-18*abs(W),
        # recompute the unaggregated packet at SAME p.
        values.append(w4*30*jac*W/(2*qant))
    value = accurate_sum(values)  # Average both azimuths.
    return (ymin>array(eta))[:,None]*value[None,:]

def design(batch,seed,power):
    sd = SeedSequence([seed,batch]).generate_state(1)[0]
    raw = scrambled_Sobol(14,sd,2**power,float64)
    raw = clip_to_open_unit_cube(raw)
    return proposal.sample_many(raw[:,:13],raw[:,13])
    # Returns (u,q), where q is the FULL mixture density.

L=[]; C=[]  # Independent fast and correction batch means.
for batch in range(160):  # Complete batches 0,...,159.
    ul,ql = design(batch,2609179501,13)  # 8192 points.
    uc,qc = design(batch,2609179502,6)   # 64 points.
    low=[]; correction=[]
    for u,q in zip(ul,ql):
        low.append(packet(u,"fast")/q)
    for u,q in zip(uc,qc):
        l = packet(u,"fast")/q
        h = packet(u,"high")/q  # SAME u as its l.
        correction.append(h-l)  # Includes every guard-zero.
    # Retry failures at same coordinates; never drop them.
    L.append(mean(low,axis=0)); C.append(mean(correction,axis=0))
    save_complete_batch(L[-1],C[-1])
# The 4h run completed 128 primary + 32 extension batches.
# Its incomplete final batch is saved, but not plotted.
# Independent finite-four data, already importance weighted.
# Each point integrates w4*K4(x)*(J(x)-JB), with its own
# proposal weight. Reuse the 320 saved means (64 points each).
K = load_independent_K4_means()
km = mean(K,axis=0); VK = cov_rows(K)/320
paired = mean(L,axis=0)+mean(C,axis=0)  # Two means.
F = paired+km  # Same K4 at EVERY eta, including eta=0.
# cov_rows uses rowvar=False, ddof=1; flatten keeps batches.
VP = cov_rows(flatten_batches(L))/160
VP += cov_rows(flatten_batches(C))/160
VF = VP+tile(VK,(11,11))  # Full 22 x 22 covariance.
SE = sqrt(diag(VF)).reshape(11,2)
# Ratio: A=mean(F[1e-11<=eta<=1e-7]), independently for
# each J; use grad(F_i/A) @ VF @ grad(F_i/A) for variance.
# Running eta=0: replace 160 by each complete prefix size;
# use the SAME km,VK. Last point equals F[-1],SE[-1].
\end{lstlisting}
\endgroup

\section{Fixed-Born NNLO radiation for arbitrary jet multiplicity}
\label{app:njet-fixed-born}

The radiation construction in Appendix~\ref{app:nnlo-fixed-born} extends to any fixed finite $N\ge2$ for massless final-state jets. The outer inverse already holds for general $N$, while the antenna inverse acts on three partons and leaves all spectators unchanged. We give the phase-space measure at fixed Born momenta and combine each real contribution with its counterterm before integration, assuming an $(N+1)$-jet NLO antenna input with the map in Eq.~\eqref{eq:nnlo-native-parents}. The integrability argument applies on compact resolved Born regions and to the energy-flow Lipschitz measurements specified in Appendix~\ref{app:fixed-born}. Together with the required Born contact, these ingredients specify a proposal for extending the local subtraction to higher jet multiplicities.

Let $R_{N+2}$ be the tree contribution, $S_a$ the NLO counterterms grouped by their maps $\widehat m_a:\Phi_{N+2}\to\Phi_{N+1}$, and $K_{N+1}$ the finite one-loop plus integrated-counterterm contribution. For an infrared-safe $(N+1)$-jet measurement $f_m$ on $m$ partons, the NLO correction is
\begin{equation}
\begin{aligned}
 \mathcal I_{N+1}^{(1)}[f]
 &=\int\dd\Phi_{N+2}\Big[R_{N+2}f_{N+2}\\
 &\hspace{3em}-\sum_aS_a f_{N+1}(\widehat m_a\Phi_{N+2})\Big]\\
 &\quad+\int\dd\Phi_{N+1}\,K_{N+1}f_{N+1}.
\end{aligned}
\label{eq:njet-nlo-input}
\end{equation}
For $e^+e^-\to N$ jets, these terms are of order $\alpha_s^N$. The NLO kernels, maps and integrated counterterms are recycled~\cite{Gehrmann-DeRidder:2005btv}; their outer phase-space integrations are replaced below. This requires access to the differential NLO contributions rather than previously integrated histograms.

\subsection{Radiation phase space at fixed Born momenta}

Fix $b=(\bar p_1,\ldots,\bar p_N)$, with $\bar p_A^2=0$ and $\sum_A\bar p_A=P=(Q,\mathbf0)$, away from soft and collinear boundaries. Order the Born momenta by decreasing energy, breaking ties by a fixed ordering of direction components. This also removes the label ambiguity for the equal-energy two-parton Born state. A chart $h$ chooses a pair $(i,j)$ among $N+1$ labels and a permutation $\sigma$ assigning the Born momenta to the merged leg and spectators $s_A$. Define $e=\bar p_{\sigma(1)}^0$, $\mathbf n=\bar{\mathbf p}_{\sigma(1)}/e$, $k=\lambda e$ and $\xi=(\lambda,\cos\theta,\phi_0)$. The point $x=G_{N+1,h}(b,\xi)$ is
\begin{equation}
\begin{aligned}
 x_{s_A}&=\lambda\bar p_{\sigma(A)},\qquad A=2,\ldots,N,\\
 K_{ij}&=(1-\lambda)P+\lambda\bar p_{\sigma(1)},\\
 K_{ij}^2&=(1-\lambda)\big[Q^2(1-\lambda)+2\lambda Qe\big],\\
 x_i^0&=\tfrac12(K_{ij}^0+k\cos\theta),\qquad x_j=K_{ij}-x_i,\\
 \mathbf x_i&=\tfrac12\big[(k+K_{ij}^0\cos\theta)\mathbf n
       +\sqrt{K_{ij}^2}\sin\theta\,\mathbf e_\perp(\phi_0)\big].
\end{aligned}
\label{eq:njet-first-inverse}
\end{equation}
Here $0<\lambda<1$, and the transverse basis and angular ranges are those of Eq.~\eqref{eq:app-inverse-daughters}. These momenta are massless and sum to $P$ for every $N$. The projection $\widehat M_{N+1}$ merges the minimum-deficit pair as in Eq.~\eqref{eq:app-map} and orders the result. The selector $\chi_h$ selects the chart when this pair is $(i,j)$, giving $\widehat M_{N+1}x=b$. The $\binom{N+1}{2}N!$ charts yield
\begin{equation}
\begin{aligned}
 \dd\mu_{N+1}^b&=\sum_h\chi_h\frac{Qe\lambda^{2N-3}}{32\pi^3}
                   \dd\lambda\,\dd\cos\theta\,\dd\phi_0,\\
 \dd\Phi_{N+1}&=\dd\Phi_N(b)\,\dd\mu_{N+1}^b.
\end{aligned}
\label{eq:njet-first-measure}
\end{equation}
The Born measure is restricted to the stated ordering. Summing assignments restores labelled phase space without adding a physical symmetry factor. The minimum deficit is at most the energy of the two least energetic partons, so $\lambda\ge(N-1)/(N+1)>0$ on selected charts. Thus the rescaling introduces no zero denominator for any $N\ge2$.

For channel $a$, write $x=(I,K,q_1,\ldots,q_{N-1})$, where $I,K$ are the massless antenna parents and the $q_A$ are unchanged spectators. Generate their three daughters using Eqs.~\eqref{eq:nnlo-inverse}--\eqref{eq:nnlo-native-parents}, with the rotations and boost specified there. This defines $p=G_a^{\mathrm{ant}}(x;r,s,\phi)\in\Phi_{N+2}$. The azimuthal reference uses the spectator with the largest transverse component, with a nonparallel Cartesian axis when all such components vanish. The inverse and factorized measure obey
\begin{equation}
\begin{aligned}
 \widehat m_aG_a^{\mathrm{ant}}(x;r,s,\phi)&=x,\\
 \widehat M_{N+2,a}&=\widehat M_{N+1}\circ\widehat m_a,\\
 \widehat M_{N+2,a}p&=\widehat M_{N+1}x=b,\\
 \dd\Phi_{N+2}&=\dd\Phi_N(b)\,\dd\mu_{N+1}^b\,
                \frac{s_{IK}}{32\pi^3}\dd r\,\dd s\,\dd\phi.
\end{aligned}
\label{eq:njet-second-inverse}
\end{equation}
The antenna factorization is independent of the spectator multiplicity~\cite{Gehrmann-DeRidder:2005btv}. On each selected chart, the combined Jacobian is $Qe\lambda^{2N-3}s_{IK}/(1024\pi^6)$. There are six radiation coordinates at fixed $b$, irrespective of $N$. Reapplying the NLO subtraction map recovers exactly the same $x$ used in its counterterm, and both states project to the same $b$.

\subsection{Local pairing and integrability}

Use minimized geometric $N$-jettiness as defined in Eq.~\eqref{eq:tau3}. For $N+1$ massless partons its minimizing partition can be chosen as one pair and $N-1$ singletons. Hence $v_a=\tau_N(x)=1-\lambda$ on a selected chart. Define $\ell_a$, $h_a$, $u_a$ and $\omega_a$ by Eqs.~\eqref{eq:nnlo-transport}--\eqref{eq:nnlo-real-partition}, using this $v_a$ and the $N+2$ real labels. Include all $(N+2)\binom{N+1}{2}$ geometric channels $(i,j,k)$ with distinct labels and $i<k$, setting $S_a=0$ for absent counterterms. As in Appendix~\ref{app:nnlo-fixed-born}, use the same full $(N+2)!$ label average for $R_{N+2},S_a$ and an $(N+1)!$ average for $K_{N+1}$, preserving the input's flavour sums and symmetry factors. Set
\begin{equation}
\begin{aligned}
 D&=R_{N+2}-\sum_c h_cS_c,\qquad R_a=h_aS_a+\omega_aD,\\
 \Delta J_{N+2}&=J(p)-J_N(b),\qquad
 \Delta J_{N+1}=J(x)-J_N(b),\\
 W_a^J&=R_a\Delta J_{N+2}-S_a\Delta J_{N+1}\\
      &=\omega_aD\Delta J_{N+2}+h_aS_a[J(p)-J(x)]\\
      &\quad-(1-h_a)S_a\Delta J_{N+1}.
\end{aligned}
\label{eq:njet-local-weight}
\end{equation}
Form this complete weight on the common phase space before integration. For scalar antennae, average it at $\phi$ and $\phi+\pi/2$ to cancel the leading collinear spin harmonics as in Appendix~\ref{app:nnlo-fixed-born}. The two states share Born coordinates but need not share jettiness. With $\mathcal P_N$ denoting this projection and partition prescription, the radiation integral at fixed $b$ and the NNLO correction are
\begin{equation}
\begin{aligned}
 F_J^{\mathcal P_N}(b)
 &=\sum_a\int\dd\mu_{N+1}^b\,\dd\Phi_{\mathrm{ant},a}
                      \langle W_a^J\rangle_\phi\\
 &\quad+\int\dd\mu_{N+1}^b\,K_{N+1}\Delta J_{N+1},\\
 \Sigma_N^{(2)}[J]
 &=\int\dd\Phi_N(b)\left[F_J^{\mathcal P_N}(b)
                    +c_{\mathcal P_N}^{(2)}(b)J_N(b)\right].
\end{aligned}
\label{eq:njet-p2b}
\end{equation}
The identity $\sum_aR_a=R_{N+2}$ restores the physical real and one-loop terms of Eq.~\eqref{eq:njet-nlo-input} after integration over $b$. The remaining projected term defines the Born contact. For $J=1$, the radiation weight vanishes before integration.

To establish integrability for general $N$, first note that the unchanged spectators contribute nothing to the transport distance $\ell_a$. The bounds $d_E(x,b)\le\mathcal C\sqrt{v_a}$, $\ell_a\le\mathcal C\sqrt{\min(r,s)}$ and $\ell_a\le\mathcal C u_a^{1/4}$ therefore hold as in Appendix~\ref{app:nnlo-fixed-born}. Set $v=\tau_N(p)$. Every partition of $N+2$ momenta into $N$ clusters contains an internal pair. The triangle inequality bounds that pair's deficit by its cluster's deficit, giving $u_{\min}\le v$ for every $N$. Since $v_a\le v+\ell_a$, these facts imply
\begin{equation}
\begin{aligned}
 d_E(p,b)&\le\mathcal C(\sqrt v+u_a^{1/8}),\\
 \omega_a|\Delta J_{N+2}|&\le\mathcal C_Jv^{1/8},\\
 h_a|J(p)-J(x)|&\le\mathcal C_J\min(v_a,\sqrt r,\sqrt s).
\end{aligned}
\label{eq:njet-bounds}
\end{equation}
For the second line, use $\omega_a\le\min[1,(v/u_a)^2]$ and split at $u_a=v$. The constants may depend on $N$ and the compact Born region, but the positive exponents do not.

The last line makes the explicit antenna difference integrable with its logarithmic measure. On the support of $1-h_a$, $r,s>c_0v_a^2$, leaving a reduced tree density times $\sqrt{v_a}(1+|\log v_a|^2)$ after antenna integration. The remaining $(N+1)$-parton single-unresolved integration is finite. For $D$, the NLO input cancels single-unresolved residues, including collinear spin correlations after averaging over $\phi$ and $\phi+\pi/2$. When the reduced state also becomes unresolved, controlling $D$ requires the uniform bound discussed in Appendix~\ref{app:nnlo-fixed-born}.

If $p$ approaches fewer than $N$ jets while its projection remains at the resolved $b$, $u_a$ cannot vanish. Otherwise $\ell_a\to0$ and $v_a\to0$ would force $b$ to the same lower-multiplicity boundary. Hence $\omega_a=O(u_{\min}^2)$ suppresses such limits. Finally, $\Delta J_{N+1}=O(\sqrt{v_a})$ renders the logarithmically enhanced single-unresolved behaviour of $K_{N+1}$ integrable.

The required contact is defined at each Born point by
\begin{equation}
 c_{\mathcal P_N}^{(2)}(b)=\lim_{t\to0}\left[
       \Sigma_{N,\mathrm{sing}}^{(2)}(b,t)
       +F_{\Theta(\tau_N-t)}^{\mathcal P_N}(b)\right].
\label{eq:njet-contact}
\end{equation}
The EFT cumulant uses the same leading-power Born variables and includes the coefficient of $\delta(\tau_N)$, with its process-dependent two-loop hard contribution. Only radiation coordinates are integrated in $F$. The availability of this EFT input and of the $(N+1)$-jet NLO calculation is separate from the explicit kinematic construction above.

\subsection{Recursion and the four-jet example}

The inverse maps admit a recursive form. Starting with $x_1=G_{N+1,h}(b,\xi)$ and $\widehat{\mathcal M}_1=\widehat M_{N+1}$, let $a_k$ choose an antenna insertion into $x_k\in\Phi_{N+k}$, with radiation variables $\xi_k=(r_k,s_k,\phi_k)$. On each chosen sequence of channels,
\begin{equation}
\begin{aligned}
 x_{k+1}&=G_{a_k}^{\mathrm{ant}}(x_k;\xi_k),\\
 \widehat{\mathcal M}_{k+1}
   &=\widehat{\mathcal M}_k\circ\widehat m_{a_k},\qquad
     \widehat{\mathcal M}_kx_k=b,\\
 \dd\mu_{N+k}^b
   &=\dd\mu_{N+1}^b\prod_{j=1}^{k-1}\dd\Phi_{\mathrm{ant},a_j}.
\end{aligned}
\label{eq:njet-recursion}
\end{equation}
The inverse identity $\widehat m_{a_k}G_{a_k}^{\mathrm{ant}}=\mathrm{id}$ proves preservation of $b$ by induction; phase-space factorization proves the product measure. Each insertion adds three coordinates. At NNLO only $x_1,x_2$ are required, and Eq.~\eqref{eq:njet-local-weight} gives their locally subtracted weight. Further iteration generates additional radiation while keeping $b$ fixed; higher perturbative orders also require the corresponding radiation subtraction and EFT singular distribution.

For $N=3$, these equations reproduce Appendix~\ref{app:nnlo-fixed-born}. For $N=4$, they give 240 outer charts, 60 antenna channels and the Jacobian $Qe\lambda^5s_{IK}/(1024\pi^6)$, with no planarity restriction on the four-parton Born state. A five-jet NLO input has been implemented and checked in the recent \texttt{NNLOJET} four-jet NNLO calculation~\cite{Chen:2026jxf}, using \texttt{OpenLoops2} for the six-parton tree and five-parton one-loop amplitudes. Such an input could support a future application beyond three jets. The radiation subtraction at fixed Born momenta and contact matching at these higher multiplicities have not been evaluated numerically here.

\section{Local contact extraction at NLO and NNLO}
\label{app:local-contact}

This appendix derives the finite radiation integrals that, together with the EFT $\delta(\tau_N)$ coefficient, directly determine the Born contact. The integrals use radiation configurations generated at fixed Born momenta and cover the full phase space without a small slicing parameter. We first illustrate the finite-rescaling construction in one variable, then give the explicit NLO result used in Sec.~\ref{sec:nlo-local-results}. We next describe the recursive two-jet NNLO contact used in Sec.~\ref{sec:twojet-results} and a proposed three-jet NNLO extension using the radiation input of Appendix~\ref{app:nnlo-fixed-born}.

\subsection{A one-dimensional example}

Consider a radiation density $g(x)/x$ on $0<x<1$, where $x$ measures the distance from the singular boundary $x=0$. Let $g$ be regular away from $x=0$, with $g(x)=g_0+O(x^\alpha)$, $\alpha>0$ and $g_0=g(0)$. The integral above a small cut $t$ contains $-g_0\ln t$. Its singular counterpart below the cut is $H+g_0\ln t$, where $H$ is a known constant in the singular cumulant. Contact matching therefore gives
\begin{equation}
\begin{aligned}
 c&=\lim_{t\to0}\left[H+g_0\ln t+\int_t^1\frac{\dd x}{x}g(x)\right]
       =H+A,\\
 A&=\int_0^1\frac{\dd x}{x}[g(x)-g_0].
\end{aligned}
\label{eq:contact-toy-matching}
\end{equation}
The logarithms cancel, leaving the finite radiation contribution $A$. For example, $g(x)=1+x$ gives $A=1$ and $c=H+1$.

We can obtain the same $A$ using only interior values of $g$. Choose the fixed rescaling $x\mapsto x/2$ and define $\Delta g(x)=g(x)-g(x/2)$ and $\ell=\ln2$. Since $\Delta g=O(x^\alpha)$, its integrals with $1/x$ and $\ln x/x$ are finite. Changing variables from $x$ to $x/2$ in the shifted terms gives
\begin{equation}
\begin{aligned}
 \int_0^1\frac{\dd x}{x}\Delta g(x)
 &=\int_{1/2}^1\frac{\dd x}{x}g(x)-g_0\ell,\\
 \int_0^1\frac{\dd x}{x}\ln x\,\Delta g(x)
 &=-\ell A+\int_{1/2}^1\frac{\dd x}{x}\ln(2x)g(x)
          -\frac{g_0\ell^2}{2}.
\end{aligned}
\label{eq:contact-toy-shift}
\end{equation}
Eliminating $g_0$ between these two identities yields
\begin{equation}
\begin{aligned}
 A={}&\int_0^1\frac{\dd x}{x}
       \left(\frac12-\frac{\ln x}{\ell}\right)[g(x)-g(x/2)]\\
 &+\int_{1/2}^1\frac{\dd x}{x}
       \left(\frac12+\frac{\ln x}{\ell}\right)g(x).
\end{aligned}
\label{eq:contact-toy-local}
\end{equation}
Both integrals converge separately. In the first, the common value $g_0$ cancels within each sampled difference. The second restores the finite interval left by the change of variables. It is part of the answer and cannot be dropped. Substituting $g(x)=1+x$ again gives $A=1$. The integration covers the entire interval $0<x<1$; $1/2$ is a rescaling factor, not an infrared cut. Equation~\eqref{eq:contact-toy-local} thus determines the contact from finite weights without evaluating $g$ at $x=0$ or taking a small numerical limit.

Alternatively, identify $A$ as the constant term of an auxiliary radiation integral. This also provides a direct derivation of the finite weights in Eq.~\eqref{eq:contact-toy-local}, which is the route we extend to NLO and NNLO below. For $\operatorname{Re}\zeta>0$, define
\begin{subequations}
\label{eq:contact-toy-constant-to-local}
\begin{equation}
 \mathcal Z(\zeta)=\int_0^1\frac{\dd x}{x}g(x)x^\zeta
    =\frac{g_0}{\zeta}+A+O(\zeta).
\label{eq:contact-toy-laurent}
\end{equation}
Thus $A=[\zeta^0]\mathcal Z$. To evaluate this constant using only interior values of $g$, change variables in the term containing $g(x/2)$, including the interval $[1/2,1]$. This gives
\begin{equation}
\begin{aligned}
 \mathcal Z(\zeta)=\frac{1}{1-2^\zeta}\bigg[
 &\int_0^1\frac{\dd x}{x}x^\zeta\Delta g(x)\\
 &-2^\zeta\int_{1/2}^1\frac{\dd x}{x}x^\zeta g(x)\bigg].
\end{aligned}
\label{eq:contact-toy-rescaled}
\end{equation}
Since $(1-2^\zeta)^{-1}=-1/(\ell\zeta)+1/2+O(\zeta)$, the constant terms of the two weights are
\begin{equation}
\begin{aligned}
 \relax[\zeta^0]\frac{x^\zeta}{1-2^\zeta}
     &=\frac12-\frac{\ln x}{\ell},\\
 [\zeta^0]\frac{-2^\zeta x^\zeta}{1-2^\zeta}
     &=\frac12+\frac{\ln x}{\ell}.
\end{aligned}
\label{eq:contact-toy-constant-weights}
\end{equation}
\end{subequations}
Substituting these weights into Eq.~\eqref{eq:contact-toy-rescaled} gives exactly Eq.~\eqref{eq:contact-toy-local}. The logarithmically weighted difference is integrable, so this constant term can be taken inside the integrals. The strategy is therefore to identify the finite contribution through the auxiliary integral and then convert it into convergent integrals by finite rescaling. The numerical calculation uses these finite weights, not a small numerical $\zeta$. The auxiliary exponent is not a dimensional regulator. At higher orders, repeated differences remove the additional logarithms of the unresolved radiation coordinates.

\subsection{Singular-distribution convention and contact}

For the physical calculation, fix a resolved Born configuration $b=(\bar p_1,\bar p_2,\bar p_3)$. Radiation variables generate a four-parton state $\Phi_4$ or a five-parton state $\Phi_5$ while these three Born momenta remain fixed. The minimized geometric three-jettiness is $\tau_3(\Phi_m)$ on either state, as defined in Eq.~\eqref{eq:tau3}. As in the one-dimensional example, the contact is the sum of the coefficient of $\delta(\tau_3)$ in the EFT singular distribution and a finite radiation contribution.

At NLO ($k=1$) or NNLO ($k=2$), the singular cumulant differential in $b$ is
\begin{equation}
 \Sigma_{3,\mathrm{sing}}^{(k)}(b,t)
 =H_k(b)+\sum_{j=0}^{2k-1}\frac{a_{k,j}(b)}{j+1}\ln^{j+1}t.
\label{eq:contact-eft}
\end{equation}
Here $t$ is the upper limit on three-jettiness, $a_{k,j}(b)$ are the coefficients of its singular logarithms, and $H_k(b)$ is the constant term, including the finite virtual contribution. These are EFT inputs in the same Born variables, resolution definition, scale and normalization as the radiation calculation. Logarithms are referred to the dimensionless interval $[0,1]$. As in Eq.~\eqref{eq:contact-toy-matching}, $H_k$ must still be combined with the finite radiation integral to obtain the P2B contact.

The required EFT input includes the finite hard, jet and soft contributions. Numerical methods have been developed for two-loop soft and beam functions~\cite{Boughezal:2015eha,Bell:2018oqa,Bell:2022tmi,Bell:2023yso,Bell:2026pou}, while jet~\cite{Bruser:2018rad,Banerjee:2018ozf} and $N$-jettiness beam functions~\cite{Ebert:2020unb,Baranowski:2022vcn,Luo:2019szz,Ebert:2020yqt,Luo:2020epw,Ebert:2020qef} are known analytically through three loops. These ingredients must be combined with the hard function and conventions for the process under consideration.

We use the following four-dimensional radiation inputs. At NLO, $R_4(\Phi_4)$ is the four-parton tree coefficient. At NNLO, channel $a$ has a five-parton real contribution $R_a(\Phi_5)$ and a local counterterm $S_a(\Phi_5)$ whose measurement is evaluated on $\Phi_4=\widehat m_a\Phi_5$. The assigned real contributions obey $\sum_a R_a=R_5$, where $R_5$ is the complete five-parton tree coefficient; their definition is Eq.~\eqref{eq:nnlo-real-partition}. The remaining input $K_4(\Phi_4)$ is the renormalized four-parton one-loop coefficient plus its integrated NLO counterterms, finite at resolved four-parton kinematics.

The measure $\dd\mu_4^b$ integrates only the radiation coordinates $\xi$ of the inverse map $G_{4,h}(b,\xi)$, with $h$ labelling its branches. It includes the Jacobian and selected branch sum, so $\dd\Phi_4=\dd\Phi_3(b)\,\dd\mu_4^b$. In channel $a$, $\dd\mu_{5,a}^b=\dd\mu_4^b\dd\Phi_{\mathrm{ant},a}$ includes the extra-emission measure of Eq.~\eqref{eq:nnlo-five-measures}. Both generated states have the same prescribed Born point. The brackets $\langle X\rangle_\phi=[X(\phi)+X(\phi+\pi/2)]/2$ use the correlated antenna-azimuth pair needed for local spin cancellation in Appendix~\ref{app:nnlo-fixed-born}.

We now insert the auxiliary measurement $\tau_3^\zeta(\Phi_m)$, just as $x^\zeta$ was inserted in Eq.~\eqref{eq:contact-toy-laurent}. The two radiation integrals are
\begin{equation}
\begin{aligned}
 \mathcal Z_1(\zeta;b)&=\int\dd\mu_4^b\,R_4\tau_3^\zeta(\Phi_4),\\
 \mathcal Z_2(\zeta;b)&=\sum_a\int\dd\mu_{5,a}^b
       \left\langle R_a\tau_3^\zeta(\Phi_5)-S_a\tau_3^\zeta(\Phi_4)\right\rangle_\phi\\
 &\quad+\int\dd\mu_4^b\,K_4\tau_3^\zeta(\Phi_4),
\end{aligned}
\label{eq:contact-mellin}
\end{equation}
The real and counterterm measurements in the second line are evaluated on their respective states, within the same local pair. Expanding in the formal exponent $\zeta$ separates the poles in $\zeta$ generated by the singular radiation integrals from the constant radiation term $\mathcal A_k(b)=[\zeta^0]\mathcal Z_k(\zeta;b)$. The contacts are
\begin{equation}
\begin{aligned}
 c_{\widehat M}(b)&=H_1(b)+\mathcal A_1(b),\\
 c_{\mathcal P}^{(2)}(b)&=H_2(b)+\mathcal A_2(b).
\end{aligned}
\label{eq:contact-finite-part}
\end{equation}
The subscripts specify the projection prescription. Here $\widehat M$ is the NLO map of Appendix~\ref{app:fixed-born}, while $\mathcal P$ includes the outer map, NLO subtraction maps and real-channel assignment of Appendix~\ref{app:nnlo-fixed-born}. The plus sign has the same origin as in Eq.~\eqref{eq:contact-toy-matching}. The radiation integral above $t$ cancels the logarithms of Eq.~\eqref{eq:contact-eft} and leaves the constant $\mathcal A_k$.

The following construction represents these constants through finite differences and compensation intervals, generalizing Eq.~\eqref{eq:contact-toy-local}. The exact radiation kernels include all nonsingular terms. No resolved event is approximated by its small-jettiness limit, and neither a small numerical $\zeta$ nor an additional dimensionally regulated integral is required. The explicit NLO result is evaluated in the main text. The NNLO construction below includes the two-jet application and the proposed three-jet extension.

\subsection{Explicit NLO construction}

Set $Q=1$, with momenta expressed in units of $Q$, and use the inverse in Eqs.~\eqref{eq:app-inverse}--\eqref{eq:app-jacobian}. Its variable $\lambda$ rescales the Born spectators, while $\theta$ and $\phi$ are the polar and azimuthal angles of the two-body decay. Recall that $\chi_h=1$ when $h=(i,j)$ is the pair minimizing $d_{ij}=E_i+E_j-|\mathbf p_i+\mathbf p_j|$, and $\chi_h=0$ otherwise. Define $\delta=1-\lambda$ and $u=1-|\cos\theta|$. The inverse gives $d_{ij}=\delta$ and implies $\chi_h=0$ whenever $u<\delta$. Thus every four-parton phase-space configuration with $\chi_h=1$ can be parametrized, for $\sigma=\pm1$, by
\begin{equation}
 \delta=uv,\qquad \cos\theta=\sigma(1-u),\qquad
 0<u,v<1.
\label{eq:contact-nlo-coordinates}
\end{equation}
Here $\tau_3(\Phi_4)=d_{ij}=uv$ for $\chi_h=1$.

To verify $\chi_h=0$ for $u<\delta$, use the momenta in Eq.~\eqref{eq:app-inverse-daughters}. Here $e_1=\bar p_1^0$, $k=\lambda e_1$ and $\mathbf n$ is the direction of the Born leg replaced by the pair. If $u<\delta$, the lower-energy daughter $q$ has $E_q=(\delta+ku)/2<\delta$ and $\mathbf q\cdot\mathbf n=[(k+\delta)u-\delta]/2<0$. The spectator momenta sum to $-k\mathbf n$, so at least one spectator $r$ satisfies $\mathbf q\cdot\mathbf p_r>0$. Its pair with $q$ then obeys $d_{qr}=E_q+E_r-|\mathbf q+\mathbf p_r|\le E_q<d_{ij}$. Hence $(i,j)$ cannot minimize $d$, and $\chi_h=0$ there. Equation~\eqref{eq:contact-nlo-coordinates} therefore covers all configurations with $\chi_h=1$.

Include the exact matrix element, radiation measure at fixed $b$ and branch selection in $g$, defined by
\begin{equation}
\begin{aligned}
 R_4\dd\mu_4^b&=\sum_{h,\sigma}
           \frac{\dd u\,\dd v}{uv}\,\dd\phi\,g_{h,\sigma}(u,v,\phi;b),\\
 g_{h,\sigma}&=\chi_h\frac{e_1(1-uv)^3u^2v}{32\pi^3}R_4.
\end{aligned}
\label{eq:contact-nlo-density}
\end{equation}
The branch label $h$ also includes the assignments of the Born momenta to parton labels. Flavour and identical-particle factors are inherited from $R_4$. Use the common average over $\phi+n\pi/2$, $n=0,1,2,3$, in $g$ and its limits. Suppressing the branch and angular arguments, define $g_s(v)=g(0,v)$, $g_c(u)=g(u,0)$ and $g_0=g(0,0)$.

The finite radiation contribution is the two-variable generalization of the second line of Eq.~\eqref{eq:contact-toy-matching}. We subtract the soft and collinear limiting functions $g_s$ and $g_c$ and add back their common value $g_0$, which would otherwise be subtracted twice. Since the resolution variable is $uv$, the boundaries also contribute finite logarithmic terms. The result is
\begin{equation}
\begin{aligned}
 \mathcal A_1=\sum_{h,\sigma}\int\dd\phi\,\bigg\{
 &\int_0^1\frac{\dd u\,\dd v}{uv}
       [g-g_s-g_c+g_0]\\
 &+\int_0^1\frac{\dd v}{v}\ln v\,[g_s(v)-g_0]\\
 &+\int_0^1\frac{\dd u}{u}\ln u\,[g_c(u)-g_0]\bigg\}.
\end{aligned}
\label{eq:contact-nlo-boundary}
\end{equation}
The limiting functions follow from the soft and collinear limits of the tree amplitude. The double difference vanishes with a positive power in each unresolved coordinate; changes of the selected pair are confined to shrinking angular intervals near these limits. These properties make all three integrals finite. Equation~\eqref{eq:contact-nlo-boundary} can therefore be evaluated directly once these limiting functions are known.

Equation~\eqref{eq:contact-nlo-boundary} uses the limiting functions $g_s$, $g_c$ and $g_0$, just as Eq.~\eqref{eq:contact-toy-matching} uses $g_0$. To avoid using these limiting functions explicitly, we apply the finite rescaling in Eq.~\eqref{eq:contact-toy-rescaled} to $\mathcal Z_1$ in Eq.~\eqref{eq:contact-mellin}, using $\tau_3(\Phi_4)=uv$ and the density in Eq.~\eqref{eq:contact-nlo-density}. Taking the constant term will then give the two-variable counterpart of Eq.~\eqref{eq:contact-toy-local}. The resulting integral uses only interior values of $g$ and is more readily extended to higher orders, where overlapping limits make explicit soft and collinear limiting functions more involved. Choose $0<\rho<1$, set $L=\ln\rho$, and define $\Delta_u g(u,v)=g(u,v)-g(\rho u,v)$ and $\Delta_v g(u,v)=g(u,v)-g(u,\rho v)$. Applying both differences gives $g(u,v)-g(\rho u,v)-g(u,\rho v)+g(\rho u,\rho v)$. Changing variables in one shifted term gives, at $\operatorname{Re}\zeta>0$,
\begin{equation}
\begin{aligned}
 (1-\rho^{-\zeta})\int_0^1\frac{\dd u}{u}u^\zeta g(u,v)
 &=\int_0^1\frac{\dd u}{u}u^\zeta\Delta_u g(u,v)\\
 &\quad-\rho^{-\zeta}\int_\rho^1\frac{\dd u}{u}u^\zeta g(u,v).
\end{aligned}
\label{eq:contact-nlo-shift}
\end{equation}
This is the general-$\rho$ version of Eq.~\eqref{eq:contact-toy-rescaled}. Applying it in both $u$ and $v$ gives the denominator $(1-\rho^{-\zeta})^2$. Taking its constant term, as in Eq.~\eqref{eq:contact-toy-constant-weights}, now gives the polynomial weight
\begin{equation}
\begin{aligned}
 P_1(\mathcal L)&=[\zeta^0]\frac{e^{\mathcal L\zeta}}{(1-e^{-L\zeta})^2}\\
       &=\frac{\mathcal L^2}{2L^2}+\frac{\mathcal L}{L}+\frac{5}{12}.
\end{aligned}
\label{eq:contact-nlo-polynomial}
\end{equation}
Here $\mathcal L$ is the logarithm multiplying $\zeta$ in the numerator. It is $\ln(uv)$ for the double difference; each compensation contributes a factor $\rho^{-\zeta}$ and hence shifts $\mathcal L$ by $-L$. The second line follows from $(1-e^{-L\zeta})^{-2}=1/(L^2\zeta^2)+1/(L\zeta)+5/12+O(\zeta)$. The complete result is
\begin{equation}
\begin{aligned}
 \mathcal A_1=\sum_{h,\sigma}\int\dd\phi\,\bigg\{
 &\int_0^1\frac{\dd u}{u}\int_0^1\frac{\dd v}{v}
       P_1(\ln uv)\,\Delta_u\Delta_v g\\
 &-\int_\rho^1\frac{\dd u}{u}\int_0^1\frac{\dd v}{v}
       P_1(\ln uv-L)\,\Delta_v g\\
 &-\int_0^1\frac{\dd u}{u}\int_\rho^1\frac{\dd v}{v}
       P_1(\ln uv-L)\,\Delta_u g\\
 &+\int_\rho^1\frac{\dd u}{u}\int_\rho^1\frac{\dd v}{v}
       P_1(\ln uv-2L)\,g\bigg\}.
\end{aligned}
\label{eq:contact-nlo-interior}
\end{equation}
The first term cancels the singularities at both $u=0$ and $v=0$ through four interior evaluations. The next two restore the finite intervals left by each rescaling, and the last restores their overlap. All four terms are required. The exact change of variables in Eq.~\eqref{eq:contact-nlo-shift} ensures that this expression gives the same $\mathcal A_1$ as Eq.~\eqref{eq:contact-nlo-boundary}, now using only interior values of $g$. It is the two-variable counterpart of Eq.~\eqref{eq:contact-toy-local}. The remainders that vanish with positive powers of the unresolved coordinates make the differences integrable with the displayed logarithmic weights. The parameter $\rho$ can be chosen freely in $(0,1)$. It changes the finite integral representation, not the contact, and is not a cutoff; the exact compensations restore the full phase space.

To complete the contact, combine this finite radiation integral with the EFT coefficient of $\delta(\tau_3)$. In the same Born-differential normalization, the NLO singular three-jettiness spectrum is
\begin{equation}
\begin{aligned}
 \frac{\dd\Sigma_{3,\mathrm{sing}}^{(1)}}{\dd\tau_3}(b,\tau_3)
 &=H_1(b)\delta(\tau_3)
   +a_{1,0}(b)\left[\frac1{\tau_3}\right]_+\\
 &\quad+a_{1,1}(b)\left[\frac{\ln\tau_3}{\tau_3}\right]_+.
\end{aligned}
\label{eq:contact-nlo-eft-spectrum}
\end{equation}
The plus distributions use the interval $[0,1]$, as in Eq.~\eqref{eq:contact-eft}. Thus $H_1(b)$ is the coefficient multiplying $\delta(\tau_3)$ in the complete EFT prediction, with the hard, jet and soft contributions combined in this convention. The directly constructed NLO contact is
\begin{equation}
 \boxed{c_{\widehat M}(b)=H_1(b)+\mathcal A_1(b),}
\label{eq:contact-nlo-direct}
\end{equation}
where $\mathcal A_1$ is the explicit finite radiation integral in Eq.~\eqref{eq:contact-nlo-boundary}, or the interior expression in Eq.~\eqref{eq:contact-nlo-interior}. Both terms are densities at the same fixed Born point and use the same jettiness definition, scale and normalization.

Algorithm~\ref{alg:finite-contact} gives the NLO evaluation used in Sec.~\ref{sec:nlo-local-results}. It evaluates all four terms of Eq.~\eqref{eq:contact-nlo-interior} at common radiation coordinates, using indicators for the restored intervals. The calculation averages over four azimuthal orientations and uses independent samples at $\rho=0.1$ and $0.01$.

\begin{figure*}[t]
\centering
\begin{minipage}{0.97\textwidth}
\begingroup
\renewcommand{\lstlistingname}{Algorithm}
\begin{lstlisting}[basicstyle=\ttfamily\scriptsize,columns=fullflexible,keepspaces=true,breaklines=true,frame=tb,rulecolor=\color{black!25},xleftmargin=5pt,xrightmargin=5pt,caption={Local NLO contact evaluation used in Sec.~\ref{sec:nlo-local-results}. The density $g$ is defined in Eq.~\eqref{eq:contact-nlo-density}; the weight combines all four terms of Eq.~\eqref{eq:contact-nlo-interior} at the same sampled coordinates. The sampling density includes both the continuous radiation density and the probability of the chosen branch.},label={alg:finite-contact}]
INPUT: resolved Born momenta b; the four-parton tree coefficient R4;
       the complete EFT delta(tau_3) coefficient H1(b).
Use the common normalization and Born measure of Appendix D.
Choose rho = 0.1 or 0.01, as in the two independent production campaigns.
Set L = log(rho) and P1(log_tau) = log_tau^2/(2*L^2) + log_tau/L + 5/12.

For each sample in each independently scrambled Sobol batch:
    Draw (u,v,phi,h,sigma) with 0<u,v<1 and 0<=phi<2*pi.
    Use the frozen sampling density q, including the branch probability.
    h includes all six pair choices and six Born assignments;
    sigma = +/-1 gives the two polar signs.

    For i,j in {0,1}:
        g[i,j] = 0
        For n in {0,1,2,3}:
            us = rho^i*u; vs = rho^j*v; angle = phi+n*pi/2
            Generate four momenta at the same b, with
                1-lambda = us*vs and cos(theta) = sigma*(1-us).
            Recompute the selected-pair condition on this state.
            Evaluate the radiation density g_h,sigma(us,vs,angle;b),
                including the tree coefficient, Jacobian and selection.
            g[i,j] += g_h,sigma(us,vs,angle;b)/4

    log_tau = log(u*v)
    w = P1(log_tau)*(g[0,0]-g[1,0]-g[0,1]+g[1,1])
    If u >= rho: w -= P1(log_tau-L)*(g[0,0]-g[0,1])
    If v >= rho: w -= P1(log_tau-L)*(g[0,0]-g[1,0])
    If u >= rho and v >= rho: w += P1(log_tau-2*L)*g[0,0]
    w /= u*v*q
    Accumulate the complete weight in the batch mean.

Estimate A1(b) and its error from the independent batch means.
Return c_M(b) = H1(b) + A1(b).

Every scaled evaluation uses sufficient numerical precision; no large
weight or small radiation coordinate is discarded. rho rescales the
coordinates and leaves the full integration domain in the estimator.

For the distributions, also integrate over b with the physical Born
measurement and include the covariance with the P2B radiation term.
The direct limiting-function expression is evaluated on the same samples
for the contact comparison, including the shared-sample covariance.
\end{lstlisting}
\endgroup
\end{minipage}
\end{figure*}

\subsection{Reusing NLO subtraction for NNLO contact extraction}
\label{app:nnlo-contact-extension}

The contact construction can reuse any complete compatible extra-jet NLO calculation, including its real, virtual and integrated-subtraction contributions. In Fig.~\ref{fig:nnlo}, the radiation term uses the \texttt{EERAD3} antenna NLO input, while the two-jet contact uses our NLO P2B construction recursively. The antenna representation below provides an alternative NLO input for the three-jet extension. Both contact constructions use the same finite-part relation; the maps and compensations follow the chosen NLO representation.

For the two-jet application, let $\mathcal Z_2(\zeta)$ denote the three-jet NLO radiation integral with measurement $\tau_2^\zeta$, integrated over the Born orientation and normalized as in Eq.~\eqref{eq:nnlotwojet}. Inserting the three-jet NLO contact $H_1+\mathcal A_1$ cancels the projected real contribution against its counterpart in $\mathcal A_1$ before integration, giving
\begin{equation}
\begin{aligned}
 \mathcal Z_2(\zeta)
 &=[\zeta_1^0]\int\dd\Phi_4\,R_4
       \tau_3(\Phi_4)^{\zeta_1}\tau_2(\Phi_4)^\zeta\\
 &\quad+\int\dd\Phi_3\,H_1(\Phi_3)\tau_2(\Phi_3)^\zeta,\\
 \mathcal A_2&=[\zeta^0]\mathcal Z_2(\zeta),\qquad c_2=H_2+\mathcal A_2.
\end{aligned}
\label{eq:contact-twojet-recursion}
\end{equation}
Here $\tau_3$ is the globally minimized three-jettiness of Eq.~\eqref{eq:tau3}, $\tau_2=1-T$, and $H_1(\Phi_3)$ is the three-jet EFT $\delta(\tau_3)$ coefficient. The two-jet coefficient $H_2$ includes the hard, jet and soft contributions defined in Sec.~\ref{sec:nnlo}. The inner finite coefficient in $\zeta_1$ is taken first at fixed $\zeta$. As at NLO, both exponents are formal devices; the numerical calculation uses finite differences and their compensation integrals in four dimensions. Terms sharing an unresolved limit are combined before integration. Learned control functions then reduce the variance through Eq.~\eqref{eq:learned-contact-control}, leaving this contact formula unchanged.

For the three-jet extension, Eq.~\eqref{eq:contact-finite-part} defines $\mathcal A_2(b)=[\zeta^0]\mathcal Z_2(\zeta;b)$. The integration weight includes the four-parton term, the five-parton real--counterterm differences and all finite compensations at fixed Born momenta.

A useful reorganization transfers a finite part of an antenna counterterm between the two multiplicities. At a fixed four-parton state $x$, choose a dimensionless profile $\vartheta_a$ such that its antenna integral is finite, and define
\begin{equation}
 \Delta K_4(x)=\sum_a\int\dd\Phi_{\mathrm{ant},a}\,
              \langle\vartheta_a S_a\rangle_\phi.
\label{eq:contact-transfer}
\end{equation}
The profile, counterterm and measure use the same inverse map $p=G_a^{\mathrm{ant}}(x;r,s,\phi)$. Adding this contribution to the five-parton integrand and subtracting its integral from $K_4$ gives the exact identity
\begin{equation}
\begin{aligned}
 \mathcal Z_2(\zeta;b)
 &=\sum_a\int\dd\mu_{5,a}^b
   \big\langle R_a\tau_3^\zeta(\Phi_5)\\
 &\hspace{3em}-(1-\vartheta_a)S_a\tau_3^\zeta(\Phi_4)\big\rangle_\phi\\
 &\quad+\int\dd\mu_4^b\,[K_4-\Delta K_4]\tau_3^\zeta(\Phi_4).
\end{aligned}
\label{eq:contact-joint-source}
\end{equation}
The transfer and its compensation use the same four-parton state and together cover the full phase space. The actual $\tau_3(\Phi_5)$ and $\tau_3(\Phi_4)$ are evaluated on their respective states, with both azimuthal orientations included.

\subsection{Finite differences and exact compensations}

The finite-rescaling identity used at NLO also applies to the complete expression in Eq.~\eqref{eq:contact-joint-source}. In each radiation region, choose coordinates $z_i\in(0,1)$ that vanish in its soft or collinear limits, and denote the remaining variables and measures by $y$ and $\dd y$. Include the phase-space Jacobians, branch selections and channel weights in $g_\ell$. After extracting the logarithmic measure $\prod_i\dd z_i/z_i$, write
\begin{equation}
 f(\boldsymbol z,y;\zeta)=\sum_\ell g_\ell(\boldsymbol z,y)
                   \tau_{3,\ell}(\boldsymbol z,y)^\zeta,
 \qquad \tau_{3,\ell}=\prod_i z_i^{a_{i\ell}}U_\ell.
\label{eq:contact-chart}
\end{equation}
Here $\tau_{3,\ell}$ is three-jettiness on the state assigned to contribution $\ell$, including the transferred terms. The powers $a_{i\ell}\ge0$ describe how jettiness vanishes; $U_\ell$ contains its remaining dependence. Its logarithms contribute to the finite coefficient and must be kept. Additional zeros of $U_\ell$ require further coordinates. Each coordinate region retains its physical map and domain, including the complementary regions of any partition.

For each coordinate, put the distinct positive powers $a_{i\ell}$ in a list $\mathcal E_i$. Repeat a power $p+1$ times when its leading coefficient contains logarithms through degree $p$. For example, a quadratic logarithm requires three successive differences. Choose $0<\rho_i<1$ and define $\mathsf T_i f(z_i)=f(\rho_i z_i)$. The required finite differences and their polynomial weight are
\begin{equation}
\begin{aligned}
 \mathcal Q_i(\zeta)&=\prod_{a\in \mathcal E_i}(1-\rho_i^{-a\zeta}\mathsf T_i)
   =\sum_{j,A}c_{i,j,A}\rho_i^{-A\zeta}\mathsf T_i^j,\\
 D_i(\zeta)&=\prod_{a\in \mathcal E_i}(1-\rho_i^{-a\zeta}),\\
 P(\mathcal L)&=[\zeta^0]\frac{e^{\mathcal L\zeta}}{\prod_iD_i(\zeta)}.
\end{aligned}
\label{eq:contact-operator}
\end{equation}
Here $j$ counts rescalings, $A$ is the sum of selected powers, and $c_{i,j,A}$ is their signed multiplicity. At NLO, $\mathcal E_u=\mathcal E_v=\{1\}$ reproduces $P_1$ in Eq.~\eqref{eq:contact-nlo-polynomial}. Coordinates with no positive jettiness power retain the original NLO subtraction and require control of its remaining unresolved limits.

Each change of variables produces a finite interval, which is restored through the identity
\begin{equation}
\begin{aligned}
 D_i\int_0^1\frac{\dd z_i}{z_i}f
 &=\int_0^1\frac{\dd z_i}{z_i}\mathcal Q_i f\\
 &\quad+\sum_{j\ge1,A}c_{i,j,A}\rho_i^{-A\zeta}
                   \int_{\rho_i^j}^1\frac{\dd z_i}{z_i}f.
\end{aligned}
\label{eq:contact-compensation}
\end{equation}
Applying this identity to all rescaled coordinates includes every combination of full and restored intervals. Their overlaps are included, as in the four NLO terms of Eq.~\eqref{eq:contact-nlo-interior}. The constant coefficient is obtained algebraically, without evaluating a small numerical $\zeta$.

One possible parametrization maps every interval to $0<v_i<1$. Let $k_i=0$ denote the full interval and $k_i>0$ the restored interval $[\rho_i^{k_i},1]$, with $k_i=0,\ldots,|\mathcal E_i|$. Here $|\mathcal E_i|$ counts entries including repetitions, and $c_i=c_{i,j_i,A_i}$. For $k_i=0$, sum all terms in $\mathcal Q_i$; for $k_i>0$, retain those with $j_i=k_i$. The evaluation coordinate and integration factor are
\begin{equation}
 (z_i,\nu_i)=
 \begin{cases}
 (v_i\rho_i^{j_i},\ c_i/v_i),&k_i=0,\\
 (\rho_i^{k_i(1-v_i)},\ -c_i k_i\ln\rho_i),&k_i>0.
 \end{cases}
\label{eq:contact-block-map}
\end{equation}
For a given radiation region and interval combination, the resulting weight is
\begin{equation}
\begin{aligned}
 W_{\boldsymbol k}(\boldsymbol v,y)
 &=\sum_{\{j_i,A_i\}}\left(\prod_i\nu_i\right)\sum_\ell
       g_\ell(\boldsymbol z,y)\\
 &\qquad\times P\!\left(\ln\tau_{3,\ell}-\sum_iA_i\ln\rho_i\right).
\end{aligned}
\label{eq:contact-block-weight}
\end{equation}
At each rescaled point, regenerate the physical momenta, their mapped counterterms and all selection factors. The Born momenta remain fixed. All signed terms and azimuthal averages are formed before integration, with the transfer and its compensation treated consistently.

The cancellation of leading logarithms follows from the finite-difference identity
\begin{equation}
 (1-\rho_i^{-a\zeta}\mathsf T_i)^{p+1}
       [z_i^{a\zeta}p_p(\ln z_i)]=0,
\label{eq:contact-annihilation}
\end{equation}
where $p_p$ is a polynomial of degree $p$. The complete weight includes the finite interval terms and the logarithms of $U_\ell$.

\subsection{Combined integration and numerical status}

An explicit calculation requires the coordinates, endpoint powers and logarithms in every radiation region. The four- and five-parton weights, including all restored intervals and transfer compensations, must then be combined on common variables while preserving the local cancellation. Each term retains its map, Jacobian and Born projection. Lower-dimensional terms can be included by adding variables whose integrals equal one. If the combined density $\mathcal W_2$ on the six-dimensional unit cube is integrable, it gives
\begin{equation}
 \mathcal A_2(b)=\int_{(0,1)^6}\dd^6\boldsymbol x\,
                         \mathcal W_2(\boldsymbol x;b).
\label{eq:contact-joint-integral}
\end{equation}
For a normalized sampling density $q$ with support on the full integration domain, one complete sample has weight
\begin{equation}
 w_2(\boldsymbol x;b)=\frac{\mathcal W_2(\boldsymbol x;b)}{q(\boldsymbol x)},
 \qquad \boldsymbol x\sim q.
\label{eq:contact-joint-weight}
\end{equation}
The numerical mean and error are computed from this complete weight.

The proposed local extraction then takes the form
\begin{equation}
 \boxed{c_{\mathcal P}^{(2)}(b)=H_2(b)+\mathcal A_2(b).}
\label{eq:contact-nnlo-direct}
\end{equation}
Here $H_2$ is the EFT coefficient of $\delta(\tau_3)$ in the same Born variables, scale and normalization. Equations~\eqref{eq:contact-joint-source}--\eqref{eq:contact-joint-weight} describe one route to its radiation contribution using four-dimensional integrals over the full phase space.

Possible numerical implementations include finite rescalings, explicit limiting functions where available, and integration over angles and finite radiation ratios before sampling the remaining unresolved scales. All use the complete weight with its finite compensations.

Figures~\ref{fig:fixed-born-nnlo-local} and~\ref{fig:fixed-born-nnlo} test the radiation subtraction. Completing this three-jet contact construction and evaluating it for differential predictions is left for future work.

\subsection{Higher multiplicities and orders}

The finite-difference identity is independent of the number of Born legs. Appendix~\ref{app:njet-fixed-born} gives radiation maps at fixed Born momenta for any fixed massless final-state jet multiplicity. Extending contact extraction requires the complete combined weight and coordinates for radiation approaching each additional Born direction, together with the corresponding EFT singular distribution.

At N$^3$LO, P2B uses an extra-jet NNLO calculation and the third-order EFT singular distribution. Repeated finite differences can remove higher logarithmic powers. Their application would require maps preserving the NNLO input's local cancellation and a treatment of the simultaneous unresolved limits of three emissions. These extensions and their numerical study remain future work.

\clearpage

\end{document}